\documentclass[10pt]{article} 

\usepackage[utf8]{inputenc}
\usepackage{amsmath}
\usepackage{amsfonts}
\usepackage{amssymb}
\usepackage{mathtools}
\usepackage{listings}
\usepackage{mathrsfs}
\usepackage{times}
\usepackage{float}
\usepackage{color}
\usepackage{setspace}
\usepackage{color,soul}

\usepackage{amsmath,amsfonts,amsthm,eucal}
\usepackage{enumerate}
\usepackage[letterpaper,margin=3cm]{geometry}
\usepackage{float}
\usepackage[title]{appendix}
\usepackage{color}
\usepackage{tabularx}
\usepackage{siunitx}
\usepackage[tableposition=below]{caption}
\usepackage{array}

\usepackage[
pdfstartview=XYZ,
bookmarks=true,
colorlinks=true,
linkcolor=blue,
urlcolor=blue,
citecolor=blue,
pdftex,
bookmarks=true,
linktocpage=true,   
hyperindex=true
]{hyperref}

\usepackage{lipsum}
\usepackage{lineno}

\usepackage[FIGBOTCAP,TABTOPCAP,bf,tight]{subfigure}

\usepackage{natbib}
\setcitestyle{authoryear}

\usepackage[pdftex]{graphicx}
\usepackage{epstopdf}
\usepackage{booktabs} 
\usepackage{tikz,mathpazo}
\usetikzlibrary{shapes.geometric, arrows}
\usepackage{caption}

\newcommand{\tensor}[1]{\ensuremath{\boldsymbol{#1}}}

\DeclareMathOperator{\tr}{tr}

\DeclareMathOperator{\spn}{spn}

\theoremstyle{remark}
\newtheorem{rmk}{Remark}
\renewcommand{\vec}[1]{\ensuremath{\boldsymbol{#1}}}

\usepackage{mathtools}
\usepackage{ulem}
\usepackage{array}
\newcolumntype{H}{>{\setbox0=\hbox\bgroup}c<{\egroup}@{}}

\usepackage{paralist}
\usepackage{color}
\definecolor{darkblue}{rgb}{0.0, 0.0, 0.8}
\definecolor{darkred}{rgb}{0.8, 0.0, 0.0}
\definecolor{darkgreen}{rgb}{0.0, 0.8, 0.0}

\newcolumntype{M}[1]{>{\centering\arraybackslash}m{#1}}

\usepackage[ruled,linesnumbered]{algorithm2e}
\usepackage[noend]{algpseudocode}

\makeatletter
\let\@fnsymbol\@arabic
\makeatother

\date{}

\title{MD-inferred neural network monoclinic finite-strain hyperelasticity models  for $\beta$-HMX: Sobolev training and validation against physical constraints} 

\begin{document}

\author{
Nikolaos N. Vlassis \footnote{Department of Civil Engineering and Engineering Mechanics, Columbia University, New York, New York} 
\and   Puhan Zhao \footnote{Department of Chemistry, University of Missouri, Columbia, Missouri}
\and Ran Ma $^1$
    \and  Tommy Sewell  $^2$  
    \and WaiChing Sun  $^1$ \thanks{Corresponding author: \textit{wsun@columbia.edu} }
}

\maketitle
\begin{abstract}
We present a machine learning framework to train and validate neural networks to predict 
the anisotropic elastic response of the monoclinic organic molecular crystal $\beta$-HMX in the geometrical nonlinear regime. 
A filtered molecular dynamic (MD) simulations database is used to train the neural networks with a Sobolev norm 
that uses the stress measure and a reference configuration to deduce the elastic stored energy functional. 
To improve the accuracy of the elasticity tangent predictions originating from the learned stored energy, 
a transfer learning technique is used to introduce additional tangential constraints from the 
data while necessary conditions (e.g. strong ellipticity, crystallographic symmetry) for the
correctness of the model are either introduced as additional physical constraints or incorporated in the validation tests. 
Assessment of the neural networks is based on (1) the accuracy with which they reproduce the bottom-line constitutive responses predicted by MD, (2) detailed examination of their stability and uniqueness, and (3) admissibility of the predicted responses with respect to continuum mechanics theory in the finite-deformation regime.    
We compare the neural networks' training efficiency under different Sobolev constraints and assess the models' accuracy and robustness
against MD benchmarks for $\beta$-HMX.
\end{abstract}

\section{Introduction}

Plastic-bonded explosives (PBXs) are highly filled polymer composites in which crystallites of one or more energetic constituents are held together by a continuous polymeric binder phase.
The filler (i.e., explosive) mass fraction is typically 90\%-95\% and typically exhibits a wide range of crystallite sizes, spanning several orders of magnitude up to a maximum of a few hundred microns.
Detonation initiation in PBXs is often achieved by transmitting a mechanical shock wave into the explosive charge.
Shock passage leads to an abrupt increase in stress, strain, and temperature in the material.
In thermodynamic terms, the magnitude of the increase of these properties is given by the Hugoniot jump relations,
which yield the locus of thermodynamic states immediately behind the shock discontinuity as a function of the input shock strength
(with a parametric dependence on the initial thermodynamic state of the material).
However, except in the case of very strong shocks,
the stresses and temperatures achieved due to bulk hydrodynamic heating in 'perfect' crystal are insufficient to lead to prompt ignition of chemistry.
Rather, it is thought that additional energy localization mechanisms\textemdash such as pore collapse,
shear banding, and interfacial debonding and subsequent frictional heating \textemdash in the microstructure of the PBX
are required to achieve the necessary local thermodynamic states required for rapid, sustained chemistry.
These regions of locally high temperature, stress, and strain rate are known as \textit{hot spots} \citep{bowden1985initiation}.
If a given hot spot is sufficiently intense, chemistry will commence.
Although the initial chemical events, so-called \textit{primary reactions}, are typically endothermic,
subsequent \textit{secondary reactions} will follow leading to large, localized heat release and formation of small-molecule products.
This results in thermal and stress pulses that propagate into the surrounding material.
If the spatial density of such hot spots in a sample is sufficiently high,
interactions among them will lead to accelerating chemistry culminating in detonation initiation.

The elastic properties of the constituents in a PBX play an important role in determining the states on the Hugoniot locus.
The most obvious connection is their appearance in the reactant equation of state (EOS). For a useful summary, see \citet{hooks2015elasticity}.
The isotropic EOS can be built around the isothermal compression curve,
typically by fitting $V=V(P)$ to the 3rd-order Birch-Murnaghan (B-M) equation of state or some other convenient functional form at room temperature or zero kelvin.
For the B-M EOS, the fitting variables are the bulk modulus $K$ and the initial pressure derivative $K^\prime$.
More sophisticated models account for crystal elastic anisotropy by incorporating the full elastic tensor.
The advantage is a higher fidelity description of the elastic response,
but doing so for a material under shock conditions requires knowledge of the pressure and temperature dependence of the elastic coefficients,
which in most cases is only available from simulations \citep{pereverzev2020elastic}.
Furthermore, the possibility of coupling between the volumetric and deviatoric responses may make it difficult to frame a proper inverse problem for experiments \citep{borja2013plasticity, bryant2018mixed, ma2021atomistic}. 

The substance octahydro-1,3,5,7-tetranitro-1,3,5,7-tetrazocine (HMX) is the energetic constituent in many PBXs. HMX exhibits several crystal polymorphs [Cady]. The thermodynamically stable form on the 300 K isotherm, for pressures between 0 and approximately 30 GPa is known as $\beta$-HMX, for which the crystal structure is monoclinic with a unit cell containing two molecules \cite{cady1963crystal}. Numerous theoretical studies of HMX physical properties and thermo-mechanical response to shocks have been reported; we do not discuss them here, but  \citet{das_2021} provides a recent entry point into that literature. All MD simulations discussed below were performed for $\beta$-HMX in the $P2_{1}/n$ space group setting.

Previous work, such as \citet{pereverzev2020elastic}), has obtained pressure- and temperature-dependent elastic coefficients 
by applying small strain increment to a sample at thermal equilibrium at the desired thermodynamic state
and determining the corresponding stress and elasticity tangential tensor at that state.
Another feasible alternative that we consider here is to assume that the finite strain elasticity of $\beta-$HMX is that of a \textit{Green elastic material} or hyper-elastic material \citep{marsden1994mathematical, ogden1997non}. 
In this approach we postulate that (1) the state of the stress in the current configuration can be solely determined by the state of the deformation of the current configuration relative to one choice of a reference configuration such as the crystal lattice vectors at (300 K, 1 atm) and (2) there exists an elastic stored energy functional of which the derivative with respect to the strain measure is the energy-conjugated stress measure.
Comparing to the former approach, which tabulates the elasticity tensor at prescribed states for a given pressure and temperature, the hyperelasticity approach 
has several distinct advantages. First, the prediction of the elastic strain energy, stress measure, and elastic tangential stress are all bundled together into one scalar-valued tensor function, instead of separate calculations for stress and elastic tangent that might not be consistent with each other.
Second, unlike the more widely used tabular approach, the hyperelasticity model does not require pressure as an input to predict elastic constitutive responses and hence enables consistency easily. Finally, by assuming the existence of such an elastic stored energy, the stability, and uniqueness of the constitutive responses 
as well as other attributes such as convexity, material frame indifference, and symmetry can be more easily analyzed mathematically \citep{ogden1997non, borja2013plasticity}. 

Nevertheless, with a few exceptions, such as \citet{holzapfel2009planar, holzapfel2004anisotropic, latorre2015anisotropic},
the majority of hyperelasticity models are limited to isotropic materials or
materials of simple symmetry such as transverse isotropic and orthotropic. 
Hyperelastic models for materials of lower symmetry such as monoclinic or triclinic are less common \citep{clayton2010nonlinear}. 
This can be attributed to the fact that the strain and the stress for anisotropic materials are not necessarily co-axial, 
and handcrafting a mathematical expression for the energy functional that leads to accurate predictions of stress and tangent, therefore, becomes a challenging task. 

To overcome this technical barrier, we introduce a transfer learning approach that generates a neural network model for the hyperelastic response of $\beta$-HMX from molecular dynamics (MD) simulations. Our new contributions, to the best knowledge of the authors, are listed below: 
\begin{enumerate}
\item Traditional supervised learning approaches often employ objective/loss functions that match the stress-strain responses \citep{ghaboussi_knowledge-based_1991, lefik2003artificial, heider2020so, frankel2019predicting}, the elastic stored energy \citep{le2015computational,teichert2019machine}, or matching the energy, stress, or elastic tangent fields \citep{vlassis2020geometric,vlassis2021sobolev} with the raw data considered as the ground truth. 
This direct approach, however, is not suitable for MD data where the change of one state to another will lead to fluctuation that makes direct Sobolev training not productive \citep{czarnecki2017sobolev}. To overcome this problem, we introduce a pre-training step in which the data are pre-processed through a filter and the underlying non-fluctuating patterns are extracted to train the neural network models. 
\item We introduce a transfer learning approach where the additional desirable attributes (e.g. frame invariance) and necessary conditions for the correctness of the constitutive laws (e.g. material symmetry) can be enforced 
with a simple re-training. 
\item We also introduce a post-training validation procedure where the focus is not only on predicting stress-strain responses
but on the desirable properties of the elastic tangential operator.
To compare to the previous literature that employs measures in the geometrical linear regime to measure anisotropy,
we introduce a reverse mapping from \citet{cuitino1992material} that generates the infinitesimal small-strain tangent from the finite strain counterpart.
With these metrics available, we can examine the convexity and strong ellipticity 
of the learned function and also evaluate whether predicted constitutive responses
exhibit the same \text{evolution} of anisotropy as the MD benchmark while ensuring that the filtering process does not lead to non-physical responses at the continuum level.
The accuracy of the model is assessed by comparing MD-simulated and learned stresses as functions of strain,
and by comparing the pressure-dependent tangent stiffness from the learned model against explicit predictions of the elastic tensor
reported recently \citep{pereverzev2020elastic} for $\beta$-HMX states on the \SI{300}{\kelvin} hydrostatic isothermal compression curve.
The latter comparison, in particular, provides an incisive test of the accuracy of the learned functional,
as this information was not used explicitly as part of the training set.
\end{enumerate}

The rest of the paper is organized as follows. We first provide a brief account of the database generation procedure,
including pertinent details of the MD simulations, the procedure to generate stress-strain data from the MD predictions,
and the procedure to filter out the high-frequency responses (Section~\ref{sec:database}).
We briefly review the setup of our hyperelastic model (Section~\ref{sec:hyperelasticity}) and
then outline the major ingredients for the supervised learning of the hyperelastic energy functional,
including the Sobolev training, the Hessian sampling techniques for controlling the higher-order derivatives
and the way to incorporate the physical constraints in the training procedure (Section~\ref{sec:sobolev_constraints}).
This section is followed by the validation procedure that tests the attributes of the learned hyperelasticity models
with physical constraints not included in the training problems (Section~\ref{sec:post_training_validation}).
The results of the numerical experiments are reported in Section~\ref{sec:results} followed by concluding remarks in Section~\ref{sec:conclusion}. 

As for notations and symbols, bold-faced and blackboard bold-faced letters denote tensors (including vectors which are rank-one tensors); 
the symbol '$\cdot$' denotes a single contraction of adjacent indices of two tensors 
(e.g.,\ $\vec{a} \cdot \vec{b} = a_{i}b_{i}$ or $\tensor{c} \cdot \tensor{d} = c_{ij}d_{jk}$); 
the symbol `:' denotes a double contraction of adjacent indices of tensor of rank two or higher
(e.g.,\ $\mathbb{C} : \vec{\varepsilon}$ = $C_{ijkl} \varepsilon_{kl}$); 
the symbol `$\otimes$' denotes a juxtaposition of two vectors 
(e.g.,\ $\vec{a} \otimes \vec{b} = a_{i}b_{j}$)
or two symmetric second-order tensors 
[e.g.,\ $(\tensor{\alpha} \otimes \tensor{\beta})_{ijkl} = \alpha_{ij}\beta_{kl}$]. 
We also define identity tensors: $\tensor{I} = \delta_{ij}$, $\mathbb{I} = \delta_{ik}\delta_{jl}$, and $\bar{\mathbb{I}} = \delta_{il}\delta_{jk}$, where $\delta_{ij}$ is the Kronecker delta.
We denote the Eulerian coordinate as $\{x_{1}, x_{2}, x_{3} \}$ and the corresponding three orthonogonal basis vectors as $\vec{e}_{1}$, $\vec{e}_{2}$ and $\vec{e}_{3}$ accordingly. 
As for sign conventions, unless specified, the directions of the tensile stress and dilative pressure are considered as positive.

\section{Database generation via molecular dynamics simulations} \label{sec:database}
In this section, we discuss the specifics of the MD simulation setup used to generate the database used for the hyperelastic energy functional discovery. We provide a theoretical background for the simulations as well as details on the system setup. We demonstrate the output results for the simulations and describe the post-processing procedure to render them suitable for our machine learning algorithms.

Training data for the neural networks are obtained by computing the Cauchy stress tensor for isothermal samples as functions of imposed tensorial strains. The strains used correspond variously to uniaxial compression or tension, pure shear, and combination strains. The imposed strains are restricted to states below the threshold for mechanical failure of $\beta$-HMX as predicted by the MD. By learning the underlying free-energy functional, we can extract the hyperelastic response from second-order and higher-order strain derivatives.

\subsection{Force field}

The MD simulations were performed using LAMMPS \cite{plimpton_1995} in conjunction with a modified version of the all-atom, fully flexible, non-reactive force field originally developed for HMX by Smith and Bharadwaj (S-B). 
\citet{smith_1999,bedrov_2000,kroonblawd_2016,mathew_2018,chitsazi_2020} Intramolecular interactions in the S-B force field are modeled using harmonic functions for covalent bonds, three-center angles, and improper dihedral ("wag") angles; and truncated cosine expansions for proper dihedrals. Intermolecular non-bonded interactions between atoms separated by three or more covalent bonds (\textit{i.e.}, 1-4 and more distant intramolecular atom pairs) are modeled using Buckingham-plus-charge (exponential-6-1) pair terms. Here and in Refs.~\cite{zhao_2020,kroonblawd_2020,das_2021}, a steep repulsive pair potential was incorporated between non-bonded atom pairs to prevent `overtopping' of the exponential-6-1 potential at short non-bonded separations $R$, which can occur under shock-wave loading due to the global maximum in the potential at distances of approximately 1~{\AA} with a divergence to negative infinity as $R \rightarrow 0$. This is accomplished by superposing on the Buckingham potential a Lennard-Jones 12-6 potential in a way such that the $R^{-12}$ repulsive core strictly prevents overtopping while having practically no effect on the potential, and therefore on the interatomic forces, for non-bonded distances more than approximately 1~{\AA}. Evaluation of dispersion and Coulomb pair terms was computed using the particle-particle particle-mesh (PPPM) k-space method \citep{hockney_1988} with a cutoff value of 11~{\AA} and with the PPPM precision set to $10^{-6}$. 


\subsection{MD Simulation cell setup}

\begin{figure}[h!]
\centering
\includegraphics[width=0.65\textwidth]{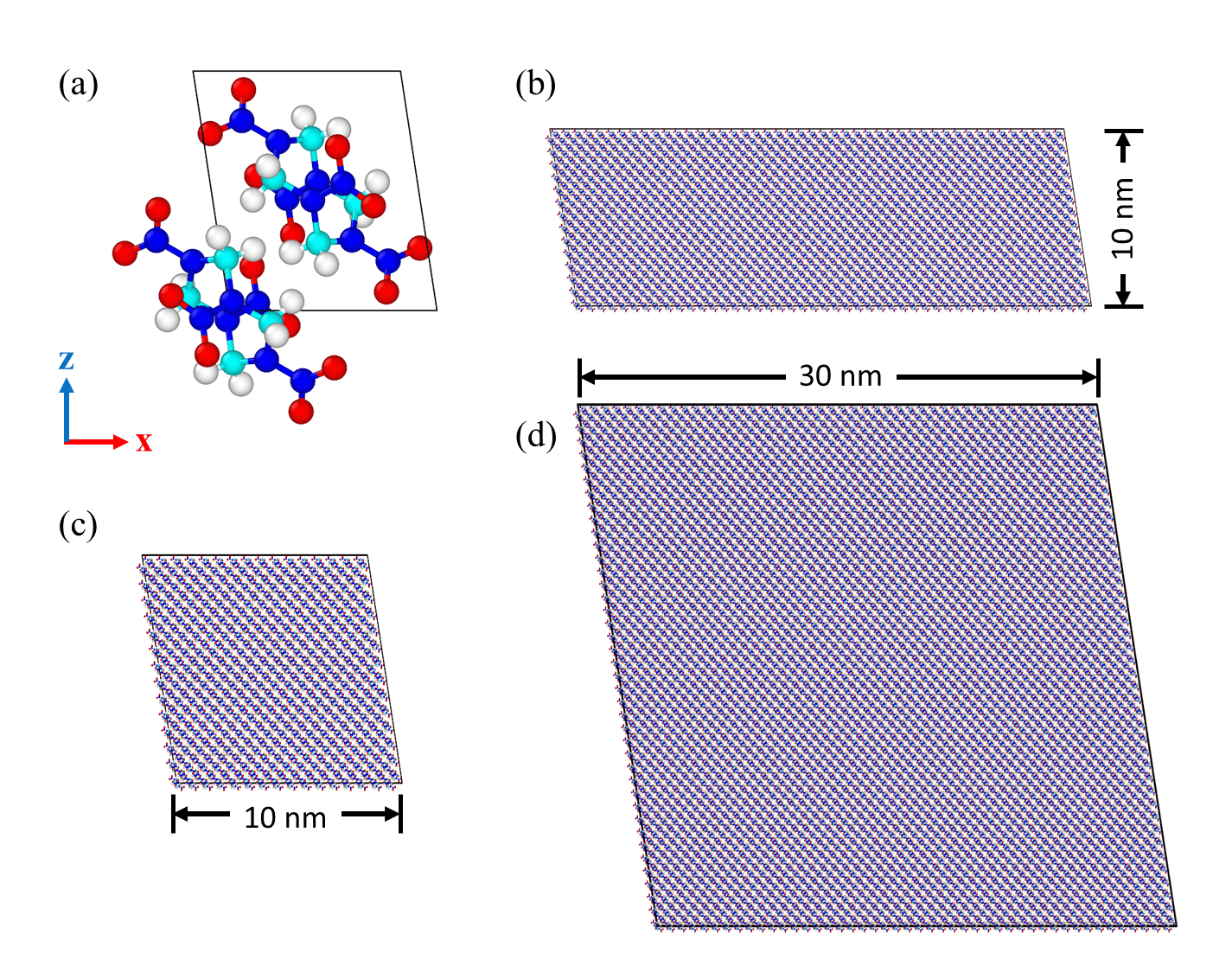}
\caption{\label{fig:mdcell} Unit cell of $\beta$-HMX (panel (a)) and snapshots of representative simulation cells for (b) uniaxial compression and tension, (c) shear deformation, and (d) biaxial compression. Cyan for carbon, navy for nitrogen, red for oxygen, and white for hydrogen.}
\end{figure}

Three-dimensionally periodic (3-D) primary simulation cells were generated starting from the unit-cell lattice parameters for $\beta$-HMX (P2$_{1}$/n space group setting) predicted by the force field (at 300~K and 1~atm), by simple replication of the unit cell in 3-D space. This results in a monoclinic-shaped primary simulation cell. The mapping of the crystal frame to the Cartesian lab frame is $\mathbf{a}$ $\|$ $\hat{\mathbf{x}}$, $\mathbf{b}$ $\|$ $\hat{\mathbf{y}}$, and $\mathbf{c}$ in the \textit{+z} space. Starting primary cell sizes for the uniaxial compression and uniaxial tension cases were approximately 30~nm parallel to the strain direction and approximately 10~nm transverse to it; those for pure shear deformation were approximately 10~nm $\times$ 10~nm $\times$ 10~nm; and those for biaxial compression were approximately 30~nm $\times$ 30~nm $\times$ 30~nm. Figure~\ref{fig:mdcell} depicts a unit cell of $\beta$-HMX and snapshots of representative simulation cells prior to the beginning of deformation. Table~\ref{tab:systemsize} contains details of the system sizes used.\par

\begin{table}
	\begin{center}
	\caption{System sizes for uniaxial compression and tension, pure shear deformation, and biaxial compression production simulations.}
	\label{tab:systemsize}
	\begin{tabular}{c c c c c c}	
	\hline\hline
	Simulation & $\rm{L}_{x}$ (nm) & $\rm{L}_{y}$ (nm) & $\rm{L}_{z}$ (nm) & Number of Molecules \\ [0.5ex]
	\hline
	Compression/Tension along $\hat{\mathbf{x}}$ & 30.3 & 10.5 & 10.6 & 12,880 & \\

	Compression/Tension along $\hat{\mathbf{y}}$ & 10.5 & 30.3 & 10.6 & 12,992 & \\
	
	Compression/Tension along $\hat{\mathbf{z}}$ & 10.5 & 10.5 & 30.4 & 12,800 & \\
	
	Shear deformation & 10.5 & 10.5 & 10.6 & 4,480 & \\
	
	Biaxial compression & 30.3 & 30.3 & 30.4 & 106,720 & \\
	\hline\hline
	\end{tabular}
	\end{center}
\end{table}

\subsection{Simulation details}
MD trajectories were propagated using the velocity Verlet integrator in LAMMPS \citep{verlet_1967,swope_1982}.
Primary cells constructed as described in the preceding paragraph were thermally equilibrated in the isochoric-isothermal (NVT) ensemble at 300~K by initially selecting atomic velocities from the 300~K Maxwell distribution followed by 20~ps of trajectory integration.
Temperature control was achieved using the Nos{\'e}-Hoover thermostat \citep{nose_1984,hoover_1985} as implemented in LAMMPS with the damping parameter set to \SI{50.0}{\fs}.
A 0.2~fs time step was used for the thermal equilibration.\par 

Fifteen isothermal MD production simulations, comprising three apiece for uniaxial compression, uniaxial tension, and biaxial compression, and six for pure shear (\textit{i.e.}, positive and negative shear directions for three distinct shear cases) were performed at $T = 300$~K using NVT integration in conjunction with the LAMMPS \textit{fix deform} command. The integration time step was 0.20~fs and the thermostat damping parameter was set to 20.0~fs. The system potential energy, temperature, pressure, Cauchy stress-tensor components, and primary cell lattice vectors were recorded at 10~fs intervals for subsequent analysis.\par

For the uniaxial compression and tension simulations, strain was applied parallel to the long direction of the primary cell
while holding both the transverse cell lengths and the tilt factors constant.
The strain rate was set to the constant value $\pm$ 0.1/100~ps, applied uniformly at each time step.
The uniaxial simulations were performed for 300~ps, resulting in a total strain of 0.3 for those cases.\par

For the shear simulations, the system was deformed along one of the three tilt factors (\textit{i.e.}, xy, xz, and yz) while the cell edge lengths were maintained at constant values. A constant strain rate of $\pm$ 0.1/100~ps was applied for 300~ps, resulting in total positive or negative shear strains of 0.3.\par

For the biaxial compression simulations, the primary cell was compressed along two axes simultaneously in the lab frame (\textit{i.e.}, $\mathbf{x}$ and $\mathbf{y}$, $\mathbf{y}$ and $\mathbf{z}$, or $\mathbf{x}$ and $\mathbf{z}$) while holding the third cell length and the tilt factors constant. The strain rate was set to $\pm$ 0.05/100~ps along both directions. Trajectory integration was performed for 300~ps resulting in a strain of 0.15 along \textit{each} of the two affected directions.\par

\subsubsection{MD results}

\begin{figure*}[h!]
\centering
\hspace{-1cm}\includegraphics[width=.8\textwidth ,angle=0]{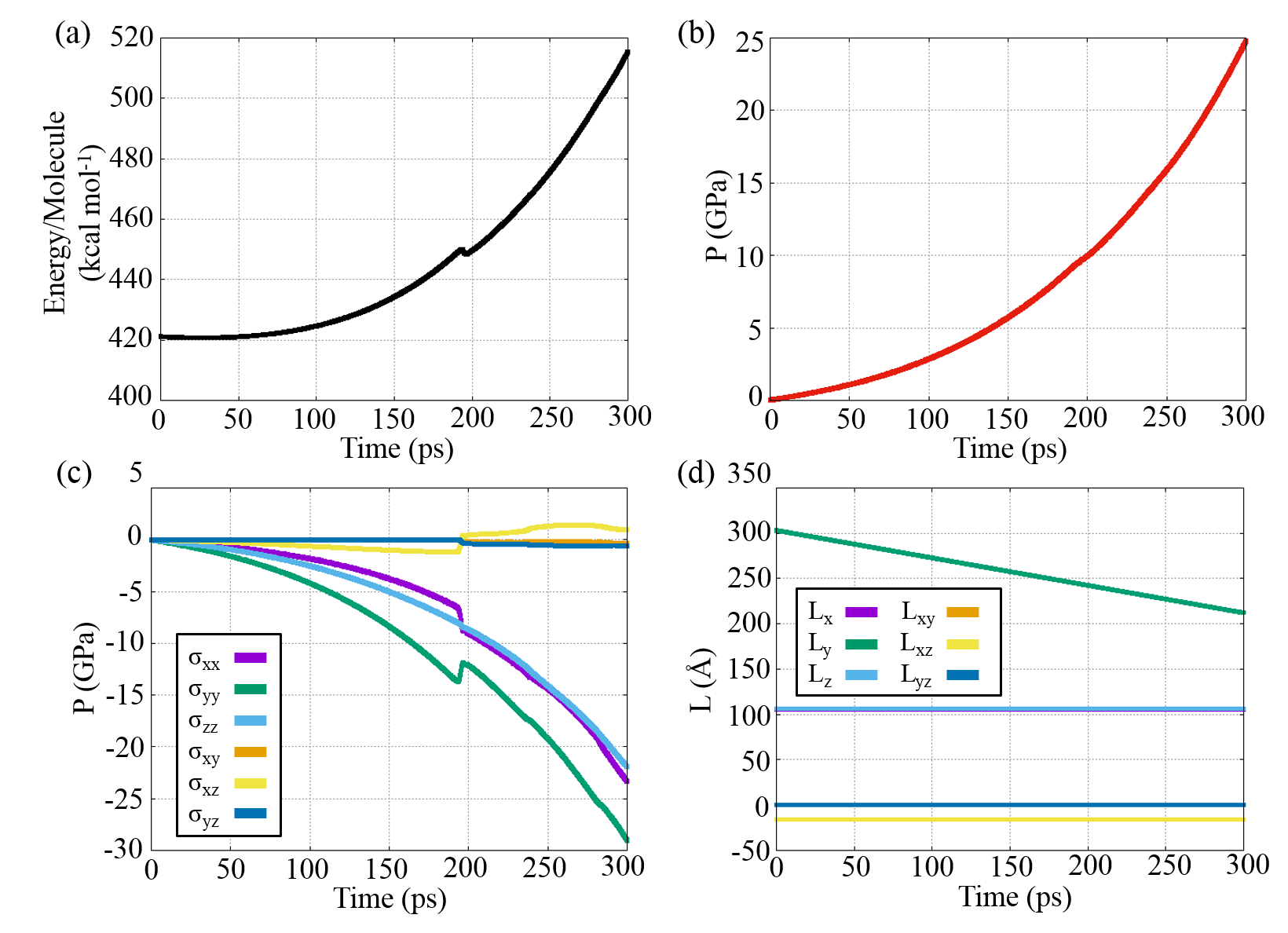} 
\caption{\label{fig:comp010} From MD, system (a) potential energy, (b) pressure, (c) Cauchy stress-tensor components,
and (d) lattice vectors vs.~time for uniaxial compression along $\hat{\mathbf{y}}$.}
\label{fig:data_curves}
\end{figure*}

Figure~\ref{fig:comp010} contains the system potential energy, pressure, Cauchy stress-tensor components,
and lattice vectors vs.~time for the case of uniaxial compression along $\hat{\mathbf{y}}$.
The effects of deformation are clearly evident in the potential energy and stress-tensor components (panels (a) and (c)),
where it can be seen that the sample yields at $t \approx \SI{190}{\ps}$.
Data for times up to approximately 10~ps before failure were used for further analysis using machine learning.\par
The Cauchy stress is obtained from the standard LAMMPS command and the expression can be found there (cf. \cite{lammps}).

\subsection{Filtering MD simulation data} \label{sec:filtering}

The raw data from the MD simulations are not expected to be smooth, due to thermal fluctuations. These fluctuations may depend on the thermostat employed and the size of the system. 
This temperature fluctuation, however, is not supposed to be captured by the hyperelasticity energy functional,
which is only designed to capture the macroscopic constitutive responses. 

To deal with the MD data, we can either introduce a regularization process during the machine learning training or we can simply filter out the Gaussian noise that might otherwise affect the convexity and therefore the stability of the hyperelasticity model. 

While one can filter the Cauchy stress tensor on a component-by-component basis,
such a strategy may lead to a filtered Cauchy stress that depends on the coordinate system.
Thus, this strategy should be avoided. 
While there are potentially more sophisticated techniques for filtering tensorial and multi-dimensional data (e.g. \citet{muti2005multidimensional}),
here we introduce a spectral decomposition on the Cauchy stress such that 
\begin{equation}
\tensor{\sigma} =\sum_{a=1}^{3} \sigma_{a} \; \vec{n}^{a} \otimes \vec{n}^{a}.
\end{equation}
Following this step, a 1D moving average filter is applied to each of the eigenvalues of the Cauchy stress
and to the Euler angles that represent the orthogonal basis vector\textemdash $\vec{n}^{a}$. 
To remove the noise, we used a 1D uniform filter on the data series that works similar to a rolling-average window.
The temporal length of the filter window is equal to that of 300 MD observations. This length of the filter window 
is selected after a manual trial-and-error such that we may 
suppress the noise of the tensorial time series without greatly distorting the global recorded constitutive response.
Note that a highly fluctuated stress data may increase the difficulty of Sobolev training the hyperelasticity energy
functional but also affect the stability of the constitutive responses at the continuum scale. Hence, this preliminary step 
is necessary. 

\begin{figure}[h!]
\newcommand\siz{.32\textwidth}
\centering
\begin{tabular}{M{.33\textwidth}M{.33\textwidth}M{.33\textwidth}}
\hspace{-1cm}\includegraphics[width=.32\textwidth ,angle=0]{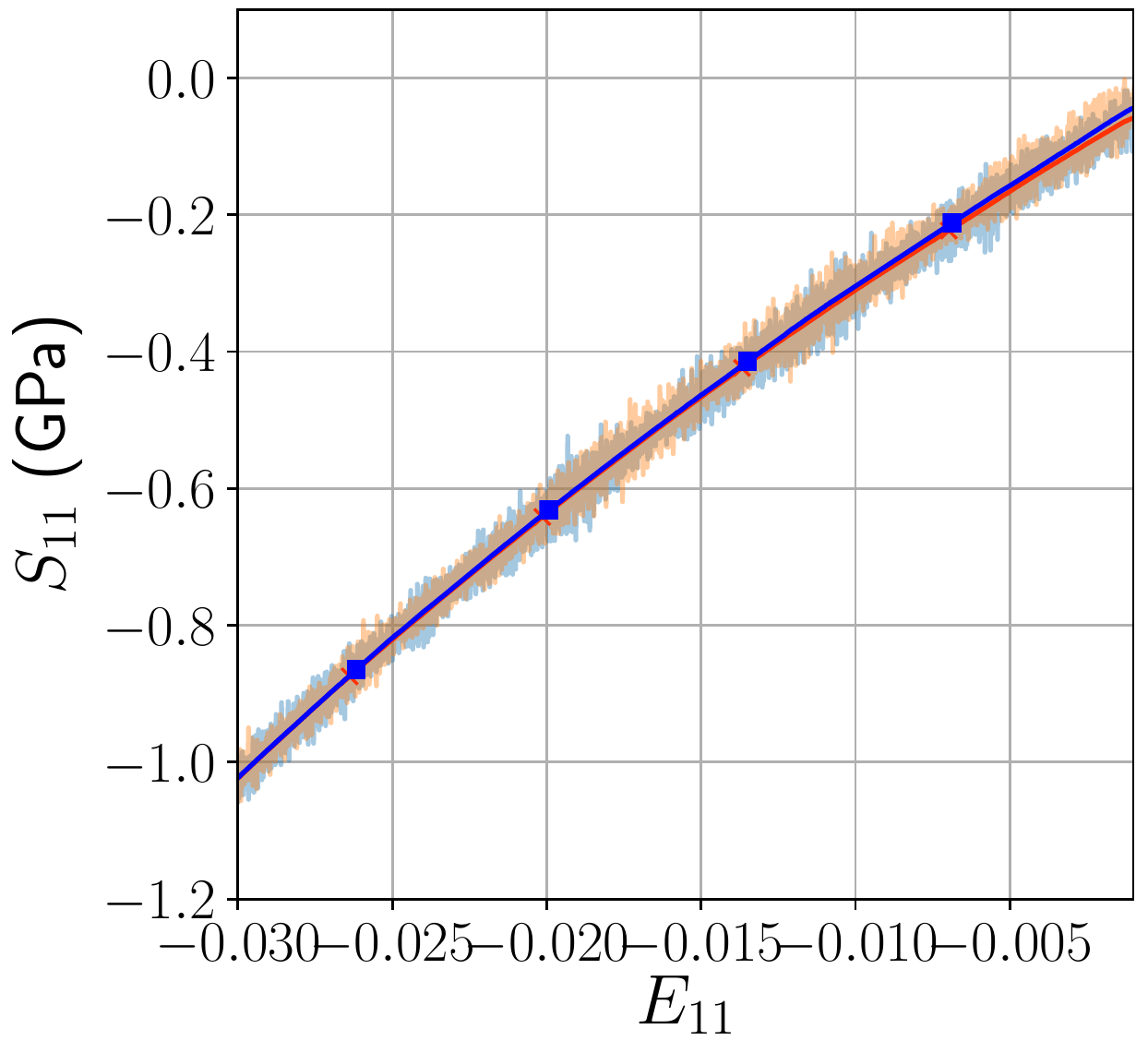} &
\hspace{-1cm}\includegraphics[width=.33\textwidth ,angle=0]{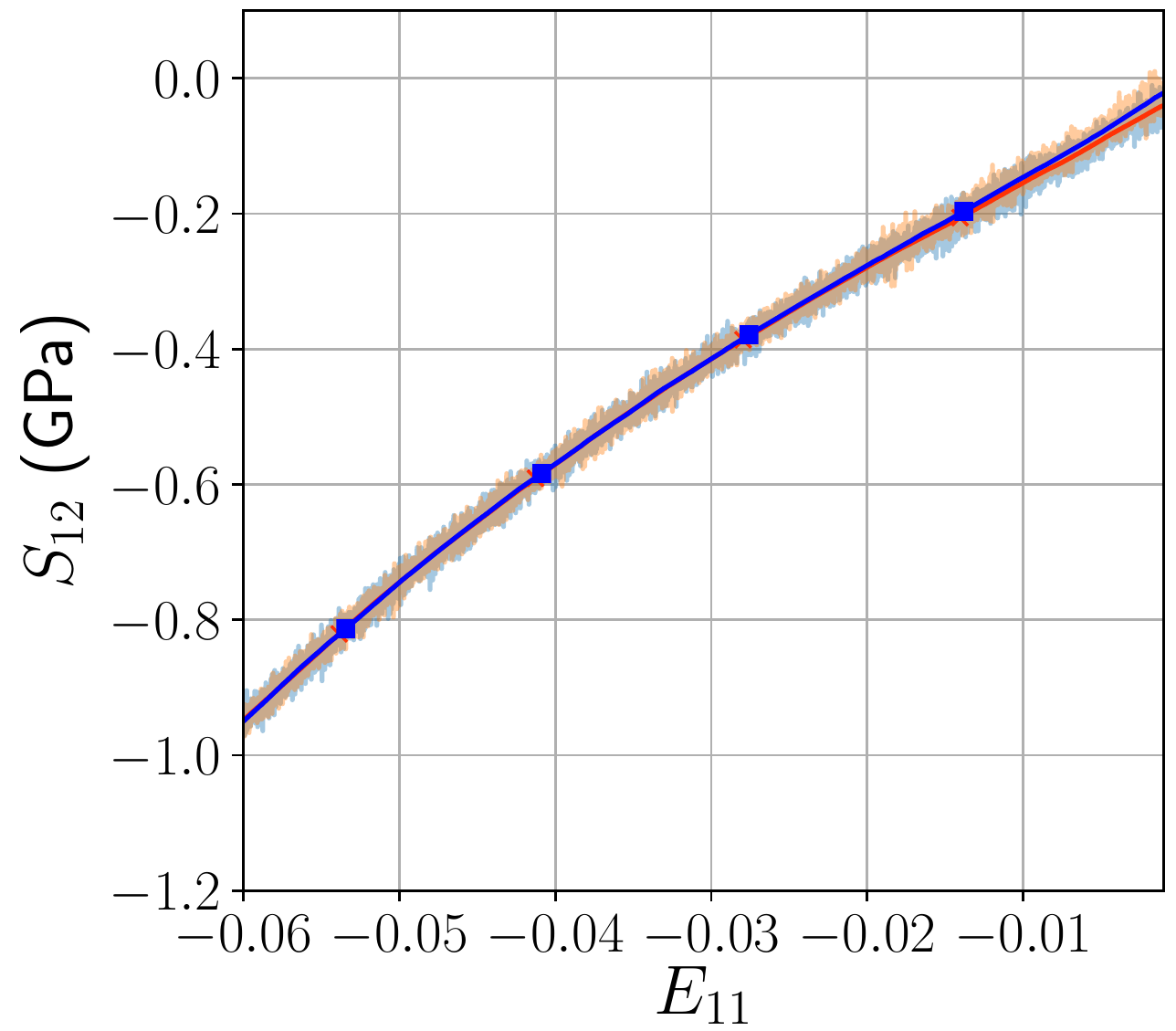} &
\hspace{-1cm}\includegraphics[width=.22\textwidth ,angle=0]{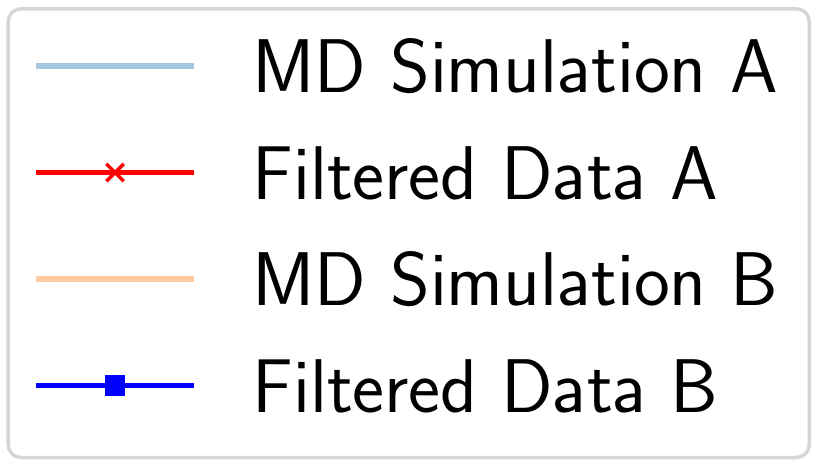} \\     
\end{tabular}
\caption{Filtering of MD simulation data with a uniform filter for a compression test along the $x$ axis.
The filtering is performed twice for two MD simulations with different thermostat coupling parameters and thus different oscillations.}
\label{fig:data_denoising}
\end{figure}

To examine whether the filter introduces significant bias to the filter data, we apply our filtering procedure to two MD simulations with the same strain path
but initiated from different initial conditions. The filtered and unfiltered constitutive responses are compared for both cases, as shown in 
Fig.~\ref{fig:data_denoising}. The two MD simulations demonstrate different fluctuation patterns but the filtered responses are 
very close 
The uniform filter used to process the data appears to capture almost identical behaviors for both simulations.

\section{Finite strain hyperelastic neural network functional for $\beta$-HMX}
\label{sec:hyperelasticity}
In this work, we will approximate a finite strain hyperelastic energy functional for $\beta$-HMX
using a feed-forward neural network architecture trained with a modified Sobolev training loss function that incorporates
additional physical constraints via a transfer learning technique. The following assumptions and setup have been made: 
\begin{enumerate}
\item There exists one stress-free configuration for the $\beta$-HMX for which the stored elastic energy is zero.
This configuration constitutes the reference configuration for the deformation mapping.  
\item We assume that all the data used in the training are purely elastic with no path dependence. 
\item Thermo-mechanical and rate-dependence effects on the elasticity are neglected. 
\item A filter is used to reduce the high-frequency responses. 
\end{enumerate}

The stored energy functional $\bar{\psi}$ can be written as a function of the deformation gradient $\tensor{F}$.
The first Piola-Kirchhoff stress $\tensor{P}$ is conjugate to
the deformation gradient $\tensor{F}$ and can be obtained from the following relation, 
\begin{equation}
\tensor{P}(\tensor{F}) = \frac{\partial \bar{\psi}(\tensor{F}) }{\partial \tensor{F}}.
\label{eq:1stpiola}
\end{equation}
Notice that a necessary condition for this energy functional to be correct is the material-frame indifference. 
Here the deformation gradient is not sensitive to rigid-body translation. However, to ensure the 
the $SO(3)$ equivalence, the machine learning generated energy functional must satisfy the following constraint, 
\begin{equation}
\bar{\psi}(\tensor{F}) = \bar{\psi}(\tensor{QF}), \quad
\forall \tensor{Q} \in SO(3)
\label{eq:frame_invariance_F}
\end{equation}

A possible way to bypass the need to introduce additional constraints in the loss function is to 
to derive the energy functional as a function of the Green strain tensor $\tensor{E}$ for which:
\begin{equation}
\tensor{E}^\prime = \frac{1}{2}(\tensor{C}^\prime - \tensor{I}) = \frac{1}{2}(\tensor{F}^{\prime T} \cdot \tensor{F}^\prime - \tensor{I}) =  \frac{1}{2}(\tensor{F}^{T} \cdot \tensor{Q}^T \cdot \tensor{Q} \cdot \tensor{F} - \tensor{I}) 
= \frac{1}{2}(\tensor{F}^{T} \cdot \tensor{F} - \tensor{I}) =  \frac{1}{2}(\tensor{C} - \tensor{I}) = \tensor{E},
\label{eq:green_strain_tensor_rotation}
\end{equation}
so we then acquire an equivalent expression:
\begin{equation}
\bar{\psi}(\tensor{F}) = \psi(\tensor{E}).
\label{eq:equivalent_expression}
\end{equation}
The second Piola-Kirchhoff stress $\tensor{S}$ is conjugate to the Green strain $\dot{\tensor{E}}$,
which is derived as:
\begin{equation}
\tensor{S}(\tensor{E}) = \frac{\partial \psi(\tensor{E}) }{\partial \tensor{E}}.
\label{eq:2ndpiola}
\end{equation}

In addition to the frame invariance, another major benefit of expressing the energy functional in terms of the Green strain tensor is that
the resultant stress measure is symmetric and the elastic tangential operator possesses both major and minor symmetries.
These symmetries may reduce the dimension of the input parametric space 9 to 6 and hence simplify the training. 
Furthermore, while $\tensor{C}$ and $\tensor{E}$ can both be used as the input for the inherently frame-indifferent energy functional that yields $\tensor{S}$ as the first derivative, $\tensor{E} = \tensor{0}$ implies the energy functional becomes zero. 
Meanwhile, training $\bar{\psi}(\tensor{F})$ as the learned function can be more convenient for implicit PDE solver for large deformation where 
the tangent corresponding to $\tensor{P}-\tensor{F}$ is required to solve the linearized system of equation. 

As such, we will train two hyperelasticity functionals, one takes the deformation gradient as input and another one takes the Green strain tensor as input respectively and we will compare the results obtained from numerical experiments. 
The relationships among elasticity tangential tensors corresponding to different stress-strain conjugate pairs will also be discussed in Section \ref{sec:postanalysis}.

\section{Stress-based Sobolev training for stored-energy function}
\label{sec:sobolev_constraints}

We introduce a neural network training technique that constructs the hyperelasticity energy functional using solely the stress data and a single reference configuration where $\tensor{F} = \tensor{I}$.
Recall that 
a feed-forward neural network can be trained to approximate 
an energy functional $\psi$ that takes the Green-Lagrange deformation tensor $\boldsymbol{E}$ as input. 
This energy function is parametrized by weights $\boldsymbol{W}$ and biases $\boldsymbol{b}$. 
The supervised learning that minimizes the inner product of the difference between the true $\psi$ and the approximated $\hat{\psi}$ for $N$ samples can be written as

\begin{equation}
    W^{\prime}, b^{\prime}=\underset{W, b}{\operatorname{argmin}}\left(\frac{1}{N} \sum_{i=1}^{N}\left\|\psi_{i}-\hat{\psi}_{i}\right\|_{2}^{2}\right),
\label{eq:L2_loss}
\end{equation}
where $\psi_{i} = \psi(\tensor{E}_{i})$ and $\hat{\psi}_{i} = \hat{\psi}(\tensor{E}_{i})$ accordingly. 
While this approach could reduce the discrepancy of the predicted and true free energy values\textemdash if available,
it would not necessarily improve the performance of the stress predictions. 
However, true energy functional value samples were not available from the MD simulations, thus,
we designed a variation of the free energy functional loss function that only uses stress data.
From this point forward, we refer to free energy simply as energy.

\subsection{Sobolev constraints for the hyperelastic energy functional}
To introduce a hyperelasticity model suitable to incorporate into numerical solvers for boundary value problems, 
the accuracy, stability, robustness, smoothness, and uniqueness of the hyperelasticity responses are all important 
to consider. Unlike neural networks that directly generate stress predictions,
a hyperelasticity model must be sufficiently smooth and differentiable to avoid discontinuity
in the predicted stress and elastic tangent \citep{vlassis2020geometric, vlassis2021sobolev, le2015computational}.

Consider the stored-energy functional solely constructed via
(1) a reference configuration where the Green strain tensor equals to $\tensor{E}_{0}$,
and (2) the Cauchy stress measured in the MD simulations. 
The corresponding loss function reads,

\begin{equation}
    W^{\prime}, b^{\prime}=
    \underset{\boldsymbol{W}, b}{\operatorname{argmin}}
    \left(\left\|\psi_{0}-\hat{\psi}_{0}\right\|_{2}^{2}+ w_{\tensor{S}} \left\|\frac{\partial \psi_{0}}{\partial \boldsymbol{E}_{0}}-\frac{\partial \hat{\psi}_{0}}{\partial \boldsymbol{E}_{0}}\right\|_{2}^{2}+\frac{w_{\tensor{S}} }{N} \sum_{i=1}^{N}\left\|\frac{\partial \psi_{i}}{\partial \boldsymbol{E}_{i}}-\frac{\partial \hat{\psi}_{i}}{\partial \boldsymbol{E}_{i}}\right\|_{2}^{2}\right),
\label{eq:incomplete_sobolev}
\end{equation}
where $\psi_{0}=  \psi(\tensor{E}_{0})$ and $\hat{\psi}_{0}=  \psi(\tensor{E}_{0})$ are the true and approximated values of the energy functional at strain $\boldsymbol{E}_0$, $N$ is the number of non-trivial stress data, and $w_{\tensor{S}}$ is the weighting factor for the multi-objective optimization.  
In this work, we use the configuration at (\SI{300}{\kelvin}, \SI{1}{atm}) as the reference and assume this configuration is stress-free. 

The corresponding loss function for the energy conjugate pair $\tensor{P}-\tensor{F}$ hyperelastic model is:
\begin{equation}
    W^{\prime}, b^{\prime}=
    \underset{\boldsymbol{W}, b}{\operatorname{argmin}}
    \left(\left\|\psi_{0}-\hat{\psi}_{0}\right\|_{2}^{2}+ w_{\tensor{P}} \left\|\frac{\partial \psi_{0}}{\partial \boldsymbol{F}_{0}}-\frac{\partial \hat{\psi}_{0}}{\partial \boldsymbol{F}_{0}}\right\|_{2}^{2}+\frac{w_{\tensor{P}} }{N} \sum_{i=1}^{N}\left\|\frac{\partial \psi_{i}}{\partial \boldsymbol{F}_{i}}-\frac{\partial \hat{\psi}_{i}}{\partial \boldsymbol{F}_{i}}\right\|_{2}^{2}\right),
\label{eq:incomplete_sobolev_PF}
\end{equation}
where $\psi_{0}=  \psi(\tensor{F}_{0})$ and $\hat{\psi}_{0}=  \psi(\tensor{F}_{0})$ are the true and approximated values of
the energy functional at the reference configuration $\tensor{F}_0$ at (\SI{300}{\kelvin}, \SI{1}{atm}).

\subsection{Transfer learning to enforce frame invariance}
\label{sec:frame_invariance}

A hyperelastic model described by the conjugate pair $\tensor{P}-\tensor{F}$ tensors is expected to satisfy the frame invariance conditions described in Eq.~\eqref{eq:frame_invariance_F}.
To ensure that the frame invariance is preserved during training, we re-use a previously trained neural network but 
modifying the loss function by introducing a number $M$ of random rotations $\tensor{Q}^{m}, i=1,2,...,M$ and penalizing the violation of the objectivity by adding the following weighted objectives: 


\begin{equation}
\begin{split}
w_{\psi} \frac{1}{M} \sum_{m=1}^{M} \left\| \psi( \tensor{Q}^{m} \tensor{F}  ) - \psi(\tensor{F}) \right\| + 
w_{\tensor{P}} \frac{1}{M} \sum_{m=1}^{M}\left\|
\tensor{P}(\tensor{Q}^{m}  \tensor{F} )  - \tensor{Q}^{m} \tensor{P}(\tensor{F}) 
   \right\|_{2}^{2} \\
+
w_{\mathbb{C}} \frac{1}{M} \sum_{m=1}^{M}\left\|
\tensor{A} (\tensor{Q}^{m}  \tensor{F} )  -   \tensor{Q}^{m}  \tensor{Q}^{m}  \tensor{A}(\tensor{F})
   \right\|_{F}^{2} .
\end{split}
\label{eq:frame_invariance_constraint}
\end{equation}

\subsection{Transfer learning to enforce crystal symmetries}
\label{sec:crystal_symmetries}

\begin{figure}[!htb]
\centering
\includegraphics[width=0.4\textwidth]{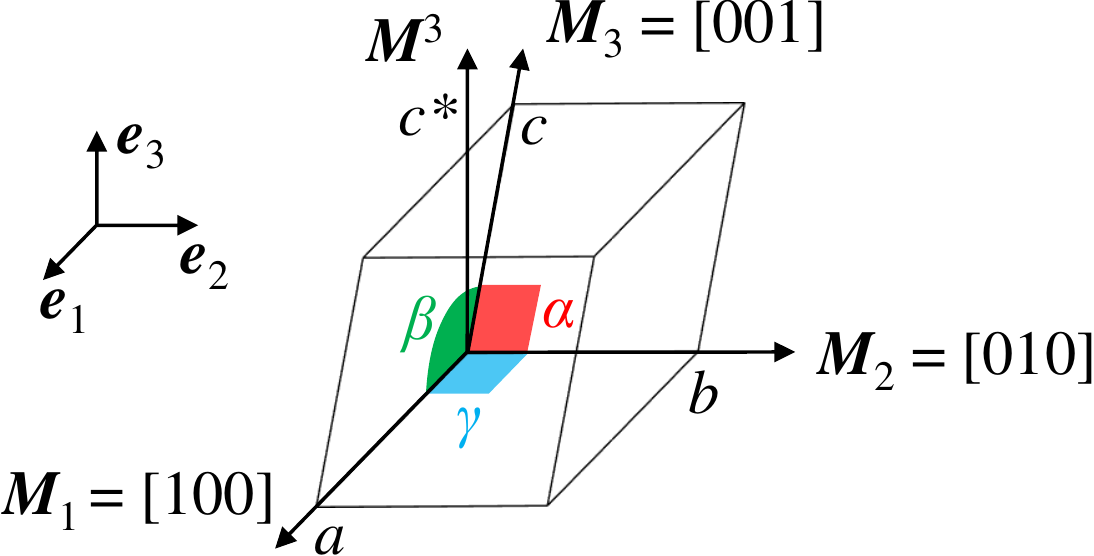}
\caption{
Monoclinic unit cell of $\beta$-HMX in the $P2_1/n$ space group setting.
The lattice constants are $a = \SI{6.53}{\angstrom}$, $b = \SI{11.03}{\angstrom}$,
$c = \SI{7.35}{\angstrom}$, $\alpha = \gamma = \SI{90}{\degree}$,
and $\beta = \SI{102.689}{\degree}$ (at \SI{295}{\kelvin}) \citep{eiland1954crystal}.
The vectors $\vec{e}_1$, $\vec{e}_2$, and $\vec{e}_3$ denote
the basis vectors of the global Cartesian coordinate system.
}
\label{fig:unit_cell}
\end{figure}

The monoclinic unit cell of the single crystal $\beta$-HMX in the $P2_1/n$ space group setting is shown in Figure \ref{fig:unit_cell}.
The covariant crystal basis vectors $\vec{M}_1$, $\vec{M}_2$, and $\vec{M}_3$ represent
the crystal axis in the reference configuration,
with corresponding contravariant basis vectors $\vec{M}^1$, $\vec{M}^2$, and $\vec{M}^3$ such that
$\vec{M}^i \cdot \vec{M}_j = \delta^i_j$.
Furthermore, the covariant crystal basis vectors in the current configuration are denoted
as $\vec{m}_1$, $\vec{m}_2$, and $\vec{m}_3$, where $\vec{m}_i = \tensor{F} \vec{M}_i$.
A general form of the deformation gradient $\tensor{F}$ that maintains the monoclinic unit cell reads:
\begin{equation}
\label{eq:F_restriction}
V_F = \lbrace \tensor{F} \vert \tensor{F} = \tensor{RU},
\tensor{U} = \sum_{i = 1}^3 ( a_i \vec{M}_i \otimes \vec{M}^i )
+ a_4 ( \vec{M}_1 \otimes \vec{M}^3 + \vec{M}_3 \otimes \vec{M}^1 ),
\tensor{R} \in \textrm{SO(3)}, a_j \in \mathbb{R} \rbrace.
\end{equation}

Under an imposed deformation gradient under the constraint in Equation \eqref{eq:F_restriction},
the symmetry group of the monoclinic unit cell in the current configuration reads
\begin{equation}
V_Q = \lbrace \tensor{Q} \vert \tensor{Q} =
\exp \left[ k \pi \frac{\spn( \vec{m}_2 )}{\Vert \vec{m}_2 \Vert} \right], \quad
\vec{m}_2 = \tensor{F} \vec{M}_2, \quad \tensor{F} \in V_F, \quad k \in \mathbb{Z} \rbrace.
\end{equation}
Here, the infinitesimal rotation map and the finite rotation map are defined as \citep{simo1989stress}
\begin{equation*}
\spn (\vec{\theta}) = - \tensor{\varepsilon} \cdot \vec{\theta}, \quad
\exp \left[ \spn (\vec{\theta}) \right] = \tensor{I} + \frac{\sin(\theta)}{\theta} \spn (\vec{\theta})
+ \frac{1 - \cos(\theta)}{\theta^2} \spn (\vec{\theta})^2,
\end{equation*}
where $\tensor{\varepsilon}$ is the permutation tensor
and $\theta = \Vert \vec{\theta} \Vert$ is the rotation angle.

Due to material symmetry, the first elasticity tensor $\mathbb{C}^{PF}$ has
the following symmetric property:
\begin{equation}
 \mathbb{C}^{PF}  ( \tensor{F \tensor{Q}}) =  \mathbb{C}^{PF}  (\tensor{F}), \quad
\forall \tensor{F} \in V_F, \tensor{Q} \in V_Q,
\end{equation}
where the tensor components are expressed in the global Cartesian frame for convenience.

To ensure that the crystal symmetry is preserved, we can again re-use
the previous trained function \eqref{eq:incomplete_sobolev_PF} and \eqref{eq:frame_invariance_constraint},
and modify the loss function by introducing $M$ number of
rotations $\tensor{Q}^{k}_\text{sym} \in V_{Q}, k=1,2,...,M$ to penalize the violation of
the material symmetry by adding the following weighted objectives: 
\begin{equation}
\begin{split}
\sum_{k=1}^{M} \left( w_{\psi} \frac{1}{N}\sum_{i=1}^{N} \left\| \psi( \tensor{F} \tensor{Q}^{k}_{\text{sym}} ) - \psi(\tensor{F}) \right\| + 
w_{\tensor{P}} \frac{1}{N} \sum_{i=1}^{N}\left\|
 \tensor{P} (\tensor{F} \tensor{Q}^{k}_{\text{sym}} )  -   \tensor{P} (\tensor{F}) \tensor{Q}^{k}_{\text{sym}}
   \right\|_{2}^{2} \right. \\
\left. +
w_{\mathbb{C}} \frac{1}{N} \sum_{i=1}^{N}\left\|
 \tensor{A} (\tensor{F} \tensor{Q}^{k}_{\text{sym}} )  -  \tensor{A} (\tensor{F}) \tensor{Q}^{k}_{\text{sym}} \tensor{Q}^{k}_{\text{sym}}
   \right\|_{F}^{2} \right).
\end{split}
\label{eq:symmetry_constraint}
\end{equation}


\section{Post-training validation of the predicted elastic tangential operators} \label{sec:postanalysis}
\label{sec:post_training_validation}

In this section, we introduce numerical tests to determine whether the predicted constitutive responses 
are thermodynamically admissible, preserve the symmetry, and lead to unique and stable elastic responses. 
A subset of these criteria are required to constitute a correct constitutive law (e.g. material frame invariance),
while others such as the convexity and the strong ellipticity are not necessary conditions 
but are desirable properties for stability and uniqueness of the boundary value problem. 
While in principle many of these physics constraints/laws can be incorporated into the loss function in the supervised learning process,
putting all the constraints explicitly into the loss function is not necessarily always ideal,
as the multiple constraints may alter the landscape of the loss function and thus complicate the search for the optimal energy functional \citep{mavrotas2009effective}.

As such, our goal is to introduce a suite of {\it necessary} conditions which the learned hyperelasticity constitutive law must fulfill.
These necessary conditions, along with the fact that the hyperleasticity constitutive law must be capable of generating predictions within a threshold error,
are necessary but not sufficient to guarantee the safety of using the machine learning model for high-consequence high-risk predictions (such as those for explosives). 

\subsection{Mapping between finite and infinitesimal kinematics} \label{sec:mapping}
To examine the admissibility of the hyperelasticity model and compare the finite strain model with other published results based on the infinitesimal strain assumption,
the connections among the tangents of different energy-conjugate pairs are provided below for completeness. 
Here our first goal is to obtain an underlying small-strain tangent of the finite-strain counterpart
by using the logarithmic and exponential mappings,
such that the elasticity tensors predicted here and those from the literature can be compared. 
Recall that the logarithmic elastic strain $\tensor{\epsilon}$ can be defined as \citep{cuitino1992material},  
\begin{equation}
\tensor{\epsilon} = \ln \tensor{U}^{2} = \frac{1}{2} \ln \tensor{C}, \quad, 
\label{eq:eps}
\end{equation}
where $\tensor{U}$ is the right-stretch tensor and $\tensor{C}$ is the right Cauchy-Green strain tensor.
The small-strain elastic tensor $\mathbb{C}^{\tensor{\sigma}-\tensor{\epsilon}}$ can be obtained from the chain rule,
\begin{equation}
\mathbb{C}^{\tensor{\sigma}-\tensor{\epsilon}} = \frac{\partial \tensor{\sigma}}{\tensor{\partial \epsilon}} = \frac{\partial \tensor{\sigma}}{\partial \tensor{S}} :  \frac{\partial \tensor{S}}{\partial \tensor{E}} :  \frac{\partial \tensor{E}}{\partial \tensor{\epsilon}} = \frac{1}{2} \frac{\partial \tensor{\sigma}}{\partial \tensor{S}} :  \frac{\partial \tensor{S}}{\partial \tensor{E}} :  \frac{\partial \tensor{C}}{\partial \tensor{\epsilon}}  =
 \frac{1}{2} \frac{\partial \tensor{\sigma}}{\partial \tensor{S}} :  \frac{\partial \tensor{S}}{\partial \tensor{E}} :  \frac{\partial \exp\tensor{2 \epsilon}}{\partial \tensor{\epsilon}}, 
\label{eq:chainrule}
\end{equation}
where $\tensor{\sigma} = J^{-1}\tensor{F} \cdot \tensor{S} \cdot \tensor{F}^{T}$ is the Cauchy stress.
To compute the small-strain elasticity tensor, one first rewrites Eq. \eqref{eq:eps} in an infinite series representation,
\begin{equation}
\tensor{C} = \exp 2 \tensor{\epsilon} = \sum^{\infty}_{n=0} \frac{1}{n!} (2 \tensor{\epsilon})^{n}.
\end{equation}
As such, the Cartesian component of the derivative $\partial \tensor{C} / \partial {\tensor{\epsilon}}$ reads \citep{miehe1998comparison}
 \begin{equation}
[ \frac{\partial \tensor{C}}{\partial \tensor{\epsilon}} ]_{ijkl}= \sum^{\infty}_{n=1} \frac{2^{n}}{n!}  \sum^{\infty}_{m=1} [\tensor{\epsilon}^{m-1}_{ik} ][\tensor{\epsilon}^{n-m}_{lj} ].
 \end{equation}

The first tangential tensor $\mathbb{C}^{\tensor{P}-\tensor{F}}$ can be related to the second derivative of the hyperelastic energy functional $\psi(\tensor{E})$,
\begin{equation}
\mathbb{C}^{\tensor{P}-\tensor{F}} =  \frac{\partial \tensor{P}}{\partial \tensor{F}}
= \frac{\partial \tensor{S}}{\partial \tensor{E}}  \cdot \tensor{F} \cdot \tensor{F} \cdot \tensor{g} + \tensor{S} \otimes \tensor{\delta}, 
\label{eq:cpf}
\end{equation}
where $\tensor{g}$ is the metric tensor.
This expression is derived from \citet{marsden1994mathematical} (see page 215),
where we simply use the chain rule to link the tangents $\partial \tensor{S} \slash \partial \tensor{E} $ with $\partial \tensor{S}/\partial \tensor{C}$.
Note that this tensor corresponds to the first Piola-Kirchhoff stress and the deformation gradient, and does not possess minor symmetry.  

In both Eq. \eqref{eq:chainrule} and Eq. \eqref{eq:cpf},
the derivative $\partial \tensor{S}/ \partial \tensor{E}$ is obtained from the neural elastic stored energy,
while the rest of the terms can be obtained via either analytical solution or automatic differentiation. 

\subsection{Strong ellipiticity}
\label{sec:strong_ellipticity}
While many works are dedicated to training elastic constitutive laws \citep{ghaboussi1998autoprogressive, pernot1999application, le2015computational, hoerig2018data, fuhg2021local, huang2020learning, vlassis2020geometric, vlassis2021sobolev}, 
surprisingly few among these analyze the stability and uniqueness 
of the learned neural network constitutive laws or provide any evidence of the well-posedness for the trained model. 
Recent work by \citep{klein2021polyconvex} address this issue by enforcing polyconvexity via invariants (cf. \citet{hartmann2003polyconvexity}).

Consider $\tensor{A}$ to be the acoustic tensor corresponding to $\mathbb{C}^{PF}$ and that
$\mathbb{C}^{PF}$ is the elastic tangential operator for the energy conjugate pairs $(\tensor{P}, \tensor{F})$, that is, 

\begin{equation}
\tensor{A}(\vec{N}) = \vec{N} \cdot \mathbb{C}^{PF}   \cdot \vec{N} 
\label{eq:acoustic_tensor}
\end{equation}
The Legendre-Hadamard condition requires that for any pair of vectors $\vec{N}$ and $\vec{m}$, the following condition holds:
\begin{equation}
\vec{m} \cdot \tensor{A} \cdot \vec{m} \geq 0,
\end{equation}
where $\vec{N}$ is a Lagrangian unit vector and $\vec{m}$ is an Eulerian vector. Because we assume that
$\beta$-HMX is a Green-elastic material, the necessary and sufficient conditions for strong ellipticity are (cf. \citet{ogden1997non} page 392)

\begin{eqnarray}
A_{ii}(\vec{N}) & > & 0, \quad i \in \{ 1, 2, 3\} \label{eq:a1} \\
A_{ii}(\vec{N})A_{jj} (\vec{N})- A_{ij}(\vec{N})^{2}  & > & 0, \quad j \neq i \in \{ 1, 2, 3\}  \label{eq:a2} \\
\det A(\vec{N}) & > & 0 \label{eq:a3}
\end{eqnarray}
for any $\vec{N} \in \mathbb{R}^{3}$. 
Notice that the material response is nonlinear and the ellipticity may vary according to the Eulerian vector $\vec{m}$.
A simple way to ensure the conditions \eqref{eq:a1}-\eqref{eq:a3} are satisfied is to create the worst-case scenario, that is, find the 
\textit{infimum}, and the unit vectors $\vec{N}$ that minimize $A_{ii}(\vec{N})$, $A_{ii}(\vec{N})A_{jj} (\vec{N})- A_{ij}(\vec{N})^{2}$,
and $\det A(\vec{N})$ accordingly and check whether the three terms remain positive. 
Depending on the parameterization, the corresponding minimization problems can be written as
\begin{eqnarray}
f(q) & = & A_{ii}(\vec{N}(q)), \quad \underset{q}{\operatorname{argmin}} \; f(q),  \quad \vec{N}(q) \in S^{2}\label{eq:m1} \\
g(q) & = & A_{ii}(\vec{N}(q))A_{jj} (\vec{N}(q))- A_{ij}(\vec{N}(q))^{2}, \quad 
\underset{q}{\operatorname{argmin}}\;  g(q),  \quad \vec{N}(q) \in S^{2}  \label{eq:m2} \\
d(q) & = & \det A(\vec{N}(q)), \quad \underset{q}{\operatorname{argmin}} \; d(q),  \quad \vec{N}(q) \in S^{2}, \label{eq:m3}
\end{eqnarray}
where $q$ represents a parametrization of the unit vector $\vec{N}(q)$.
\citet{mota2016cartesian} provide a comprehensive review of how different parameterizations,
namely the spherical, stereographic, projective and tangent parameterizations,
may lead to different local mininizers of the acoustic tensor in the parametric space.
For spherical parameterization, a unit vector $\vec{N}$ is an element of the unit sphere $S^{2}$ which can be parameterized by the spherical coordinates,
that is, the polar angle $\phi \in [0, \pi]$ and the azimuthal angle $\theta \in [0, \pi]$:
\begin{equation}
\vec{N}(\phi, \theta) = \sin \phi \cos \theta \vec{e}_{1} + \sin \phi \sin \theta \vec{e}_{2} + \cos \phi \vec{e}_{3},
\label{eq:sample_N}
\end{equation}
where $\{\vec{e}_{1}, \vec{e}_{2}, \vec{e}_{3} \}$ is the the orthogonal basis for $\mathbb{R}^{3}$. 

To ensure stability for any given admissible deformation, 
we must ensure that Eqs. \eqref{eq:a1}-\eqref{eq:a3} are valid for any $\tensor{F}$. 
While this can be, in principle, determined analytically for hand-crafted energy functionals, 
the expression of the neural network energy functional would likely be too complicated to analyze.  
As such, we again resort to constructing a test to check the hypothesis that the material demonstrates strongly ellicipticity,
via an attempt to find the minima, that is, 
\begin{eqnarray}
f'(q, \tensor{F}) & = & A_{ii}(\vec{N}(q), \tensor{F}), \quad \underset{q, \tensor{F}}{\operatorname{argmin}} \; f'(q, \tensor{F}),  \quad \vec{N}(q) \in S^{2}, \tensor{F} \in GL^{+}(3) \label{eq:um1} \\
g'(q, \tensor{F}) & = & A_{ii}(\vec{N}(q), \tensor{F})A_{jj} (\vec{N}(q), \tensor{F})- A_{ij}(\vec{N}(q), \tensor{F})^{2}, \quad \nonumber \\
& &\underset{q, \tensor{F}}{\operatorname{argmin}}\;  g'(q),  \quad \vec{N}(q) \in S^{2},   \tensor{F} \in GL^{+}(3)\label{eq:um2}  \\
d'(q, \tensor{F}) & = & \det A(\vec{N}(q), \tensor{F}), \quad \underset{q, \tensor{F}}{\operatorname{argmin}} \; d'(q, \tensor{F}),  \quad \vec{N}(q) \in S^{2}, \tensor{F} \in GL^{+}(3).\label{eq:um3}
\end{eqnarray}
It is impossible to test all the possible deformation gradients in the MD simulations while maintaining the path independence of the constitutive responses,
so we instead construct a test where we only consider 
a range of possible deformation gradients and search for the minima within this range. 

The numerical strong ellipticity test is conducted via the following three steps. 
\begin{enumerate}
\item We create two sets of point clouds in the parametric space with uniform spacing,
$V_{q} = \{q_{1}, q_{2}, q_{3}, ....\}$ and $V_{\tensor{F}} = \{ \tensor{F}_{1}, \tensor{F}_{2}, \tensor{F}_{3}, ...\}$,
and select the combination of $(q, \tensor{F})$ that minimizes $f'$, $g'$, $d'$.
If there exist other $(q, \tensor{F})$ combinitions that yield a value sufficiently close to the minimum (say within 5\% difference),
then the additional coordinates will be stored as the candidate position(s) for the gradient-free search.
This treatment is to ensure that more local optimal points can be identified and compared and to avoid the issues exhibited in \citet{mota2016cartesian}.
\item We then use the candidate position determined from the previous step as the starting point and 
apply a gradient-free optimizer via the third-party gradient-free optimizer library (cf. \citet{gfo2020}) to examine whether we can find new coordinates
for which the functions $f'(q, \tensor{F})$, $g'(q, \tensor{F})$, and 
$d'(q, \tensor{F})$ are smaller than the candidate position(s) identified in Step 1. 
\item If Eqs. \eqref{eq:a1} - \eqref{eq:a3} are not violated in the worst case obtained from the Step 2,
then we consider the neural network functional to have passed the strong ellipticity test. 
\end{enumerate}

 \subsection{Convexity and growth conditions}
 \label{sec:convexity_growth}
In nonlinear elasticity in the finite strain regime, convexity is not necessary and can be over-restrictive for physical phenomena
that involve instability or buckling \citep{clayton2010nonlinear}.
Nevertheless, the convexity condition has to be satisfied
to predict stable elastic responses under large deformation.
The convexity condition can be stated as (cf. \citep{ogden1997non}), 

\begin{equation}
\psi(\tensor{F}') - \psi(\tensor{F}) - \tr ( \tensor{P} \cdot (\tensor{F}  - \tensor{F}')) \geq 0
\label{eq:convexity}
\end{equation}

Because convexity is not a requirement for realistic simulations (although it might be expected for HMX), we do not incorporate this criterion in the training of the neural network.
However, the uniqueness and stability of the elasticity 
model are not only important for predicting realistic elastic responses but crucial if the model will be deployed as 
the underlying elasticity model for crystal plasticity and damage models. 

Another important condition to prevent degenerated elastic behavior is from \citet{rosakis1994relation} which requires
\begin{equation}
\psi(\tensor{F}) \rightarrow \infty \quad \text{as} \quad \det \tensor{F} \rightarrow 0^{+}.
\label{eq:degenerated}
\end{equation}
Recall that $\det \tensor{F} \rightarrow 0$ only happen if the distance between two material points
that was non-zero in the reference configuration vanishes in the current configuration. 
Note that it is unlikely a material would remain elastic if the volumetric deformation is extremely large. 
furthermore, enforcing these constraints explicitly in the loss function is difficult due to the infinity. 
Nevertheless, the constraint may provide a helpful indicator of the admissibility of the 
machine learning extrapolated predictions.
As a result, we suggest a post-training validation test where we generate the response for deformation gradients with $\det \tensor{F}$ approaching zero and observe whether the resultant energy is monotonically increasing.

\subsection{Material Anisotropy}

A predictive elasticity model must preserve the overall crystal symmetry while capturing how the
anisotropy of the elasticity tensor evolves under arbitrary deformation.
The degree of anisotropy of the elastic response can be measured by various metrics available in the literature
(cf. \citep{li1987single, kube2016elastic, ranganathan2008universal}).
Many of these anisotropy metrics (or indices) are intended for components of the elasticity tensor.
Typically, the distinction between the secant and tangential elastic tensors is not taken into account. 
This can be confusing for materials undergoing finite deformation where both material and geometrical non-linearities play important roles in the anisotropy of the constitutive response.
More importantly, the impacts of the former and latter types of non-linearity should be distinguished properly such that a meaningful evaluation can be conducted.

\subsubsection{Ledbetter and Migliori general anisotropy index}
\label{sec:anisotropic_index}
Here, we use the idea from previous work due to \citet{ledbetter2006general},
where the ratio between the maximum and minimum shear-wave speed is used to define an anisotropy measure. 
Interestingly, this method can also be used to detect instability as the vanishing of the slowest wave speed
is accompanied by divergence of the Ledbetter-Migliori index. 

This measure can be easily extended to the finite strain regime by replacing the infinitesimal elasticity tangent
with the elasticity tensor corresponding to the first Piola-Kirchhoff stress and deformation gradient \citep{ogden1997non}. 
This idea can be summarized into the following steps. 
\begin{enumerate}
\item Generate as many unit vectors $\vec{N}$ as possible.
\item Solve the Christoffel equation for each unit vector $\vec{N}$, that is, 
\begin{equation}
 \det \left( \vec{N} \cdot \mathbb{C}(\tensor{F}) \cdot \vec{N} - \rho v^{2} \tensor{I} \right) = 0
\end{equation}
\item Pick the largest solution $v_{2}$ and the smallest solution $v_{1}$.
Then, the anistropy index is simply 
\begin{equation}
A_{I} = v_{2}^{2} / v_{1}^{2}
\label{eq:anisotropic_index}
\end{equation}
\end{enumerate}

Here, instead of a Monte Carlo search, we can leverage the search formulated in Section \ref{sec:strong_ellipticity} to
obtain the smallest eigenvalue $v_{1}$ and largest eigenvalue $v_{2}$ of the acoustic tensor.
Again, the optimization is conducted by using a uniformly spaced point cloud to search for the initial guess,
then a gradient-free optimizer is used to find the normal vectors that maximize and minimize $v$.

\section{Results} 
\label{sec:results}

In this section, we discuss the performance of neural network models for discovering the hyperelastic energy functional from the $\beta$-HMX MD simulation data.
We describe the training setup of the networks and compare the performance of the architectures.
We then demonstrate the predictive capabilities of the models against the present MD simulation data and elastic constants taken from the literature
for the same MD force field used here.
Finally, we investigate the energy functional models in terms of how well they satisfy desired properties from the hyperelasticity literature.  

\subsection{Training performance and learning capacity}
\label{sec:network_training}

In this section, we discuss the performance of the neural network architectures for the Sobolev constraints described in Section~\ref{sec:sobolev_constraints}. 
We first demonstrate how we trained the neural networks to generate a hyperelastic energy functional data from the MD simulation data. 
We use two different architectures to discover the hyperelastic energy functional for $\beta$-HMX.
The first architecture is based on the energy conjugate pair $\tensor{S}-\tensor{E}$ (Model $\mathcal{M}_{1}$). 
The input and output variables are symmetric tensors and, thus, can be described by six components.
The second architecture is based on the energy conjugate pair $\tensor{P}-\tensor{F}$ (Model $\mathcal{M}_{2}$). 
In addition, we also re-train Model $\mathcal{M}_{2}$ with 
an additional material frame indifference constraint  (Eq.~\eqref{eq:frame_invariance_constraint} )in the loss function (model $\mathcal{M}_{3}$). 
As the difference in the predictions obtained from  Models $\mathcal{M}_{2}$ and $\mathcal{M}_{3}$ is minor, 
we did not enforce the Eq.~\eqref{eq:frame_invariance_constraint} explicitly in the the last model we trained (Model $\mathcal{M}_{4}$).  
Instead, only monoclinic symmetry is enforced as an additional term for the weighted loss function in the re-training step to ensure  that
the material symmetry is preserved. 

\begin{table}[h]
\centering
\caption{Summary of the trained models.}
\label{tab:AbbrevModels}       
{
\begin{tabular}{p{0.8cm}p{10.5cm}}
\hline\noalign{\smallskip}
Model & Description\\
\noalign{\smallskip}\hline\\[-3mm]
$\mathcal{M}_{1}$ & Energy conjugate pair $\tensor{S}-\tensor{F}$ model trained via the loss function described in Eq.~\ref{eq:incomplete_sobolev}.  \\[2mm]
$\mathcal{M}_{2}$ & Energy conjugate pair $\tensor{P}-\tensor{F}$ model trained via  the loss function described in Eq.~\ref{eq:incomplete_sobolev_PF}. \\[3mm]
$\mathcal{M}_{3}$ & Energy conjugate pair $\tensor{P}-\tensor{F}$ model trained with pre-trained model $\mathcal{M}_{2}$ and additional loss function Eq.~\eqref{eq:frame_invariance_constraint} to enforce material frame indifference. \\[3mm]
$\mathcal{M}_{4}$ & Energy conjugate pair $\tensor{P}-\tensor{F}$ model trained pre-trained model $\mathcal{M}_{2}$ and additional loss function Eq.~\eqref{eq:symmetry_constraint} to enforce monoclinic symmetry.  \\
\noalign{\smallskip}\hline
\end{tabular}
}
\end{table} 

\begin{figure}[h!]
\newcommand\siz{.32\textwidth}
\centering
\begin{tabular}{M{.33\textwidth}M{.33\textwidth}M{.33\textwidth}}
\hspace{-1cm}\includegraphics[width=.33\textwidth ,angle=0]{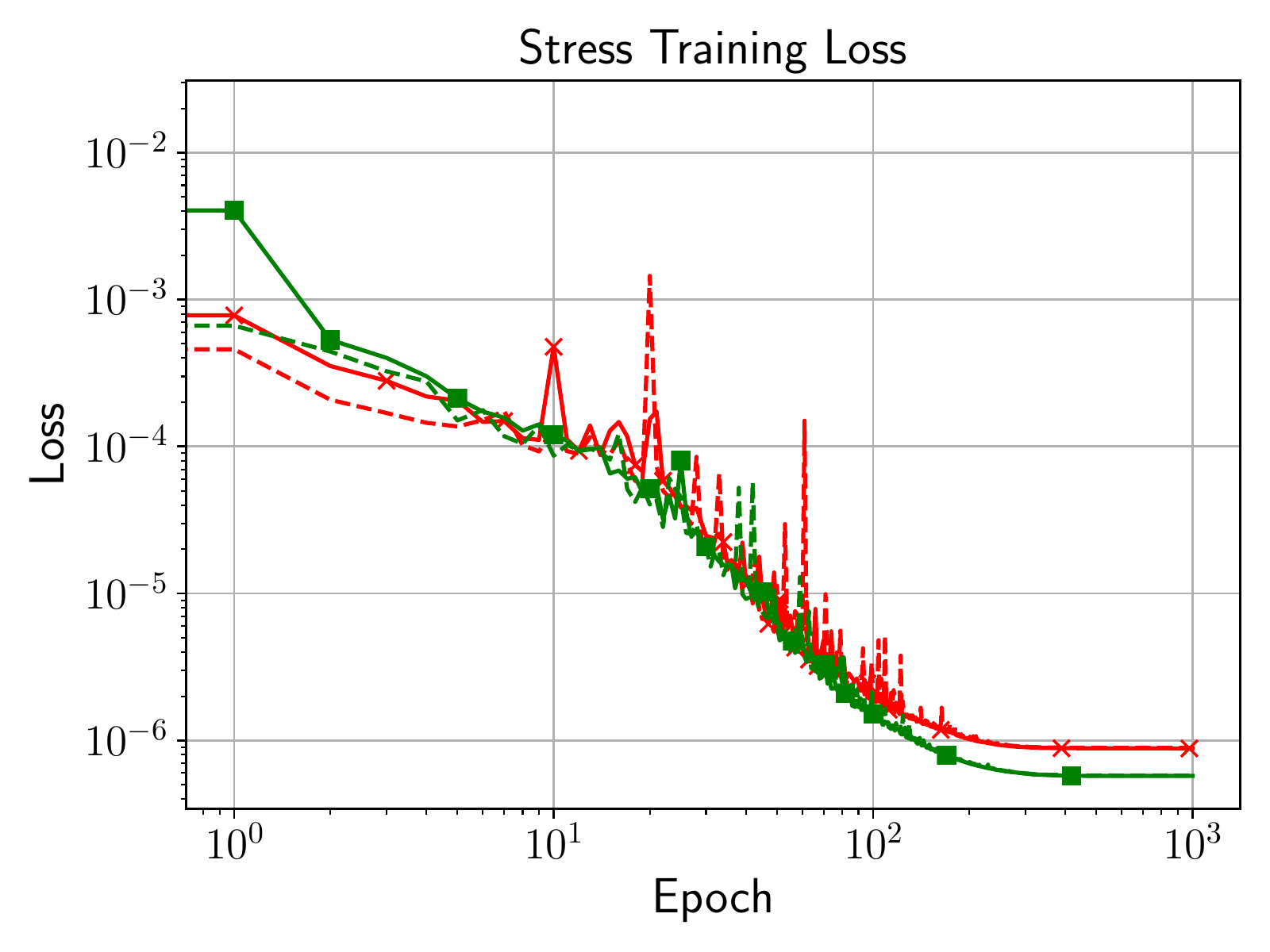} &
\hspace{-1cm}\includegraphics[width=.33\textwidth ,angle=0]{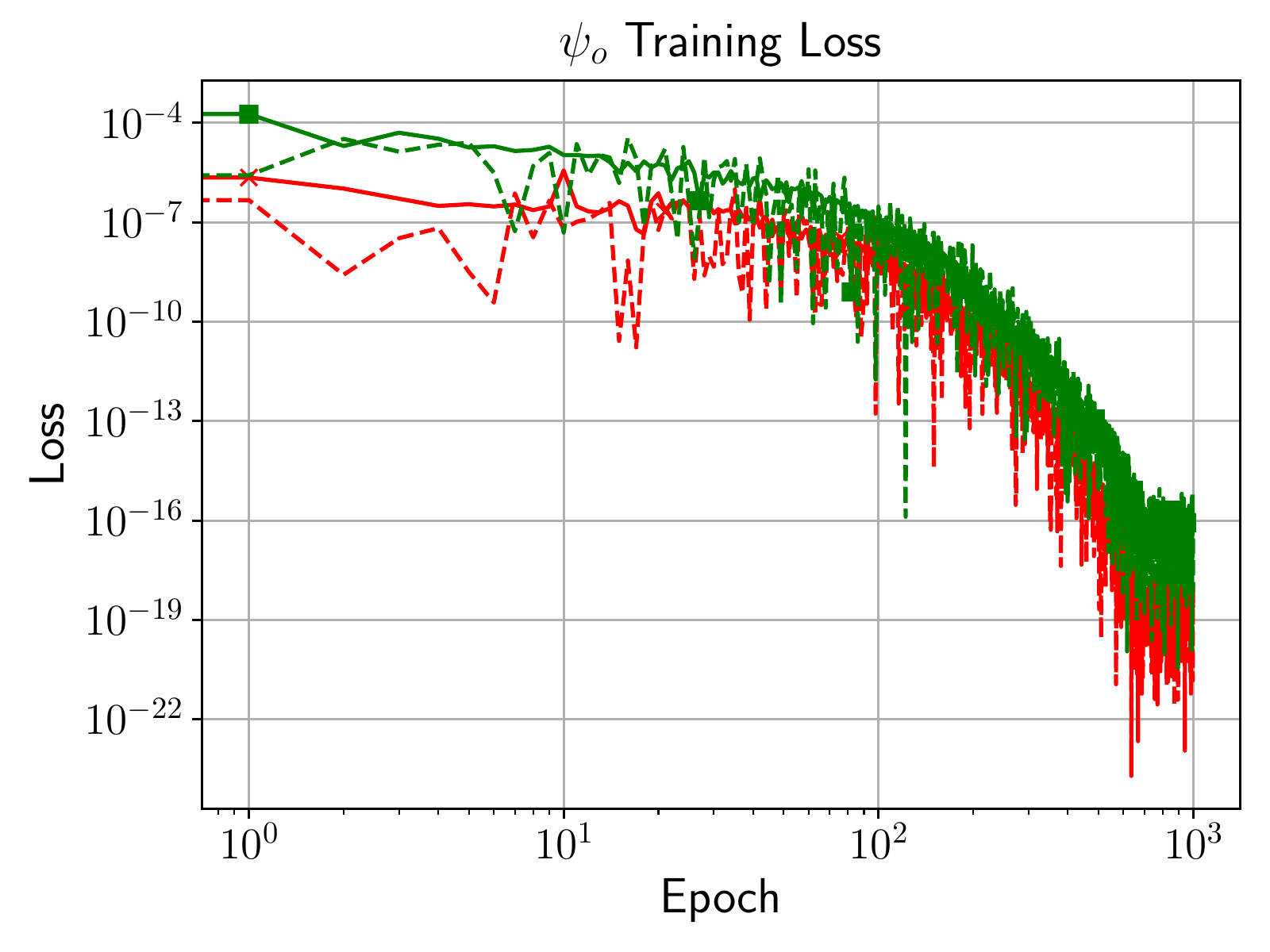} &
\hspace{-1cm}\includegraphics[width=.33\textwidth ,angle=0]{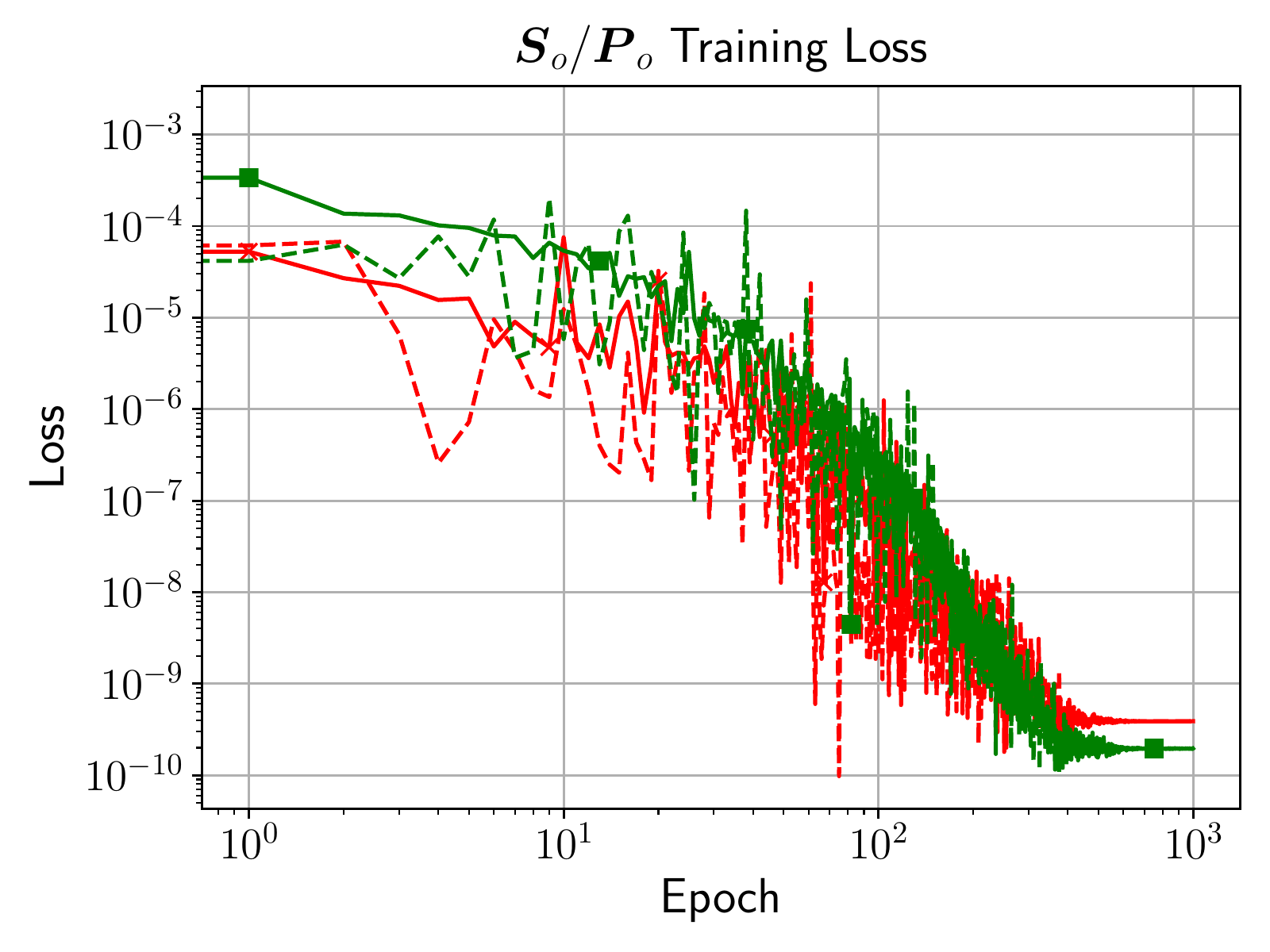} \\     

(a) & (b) & (c) \\
\end{tabular}
\includegraphics[width=.6\textwidth ,angle=0]{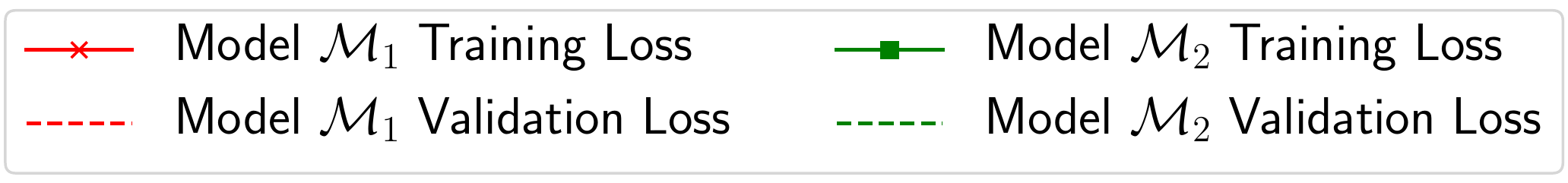}
\caption{Comparison of the training loss curves for the energy conjugate pair $\tensor{S}-\tensor{E}$ model ($\mathcal{M}_{1}$) and
the energy conjugate pair $\tensor{P}-\tensor{F}$ model ($\mathcal{M}_{2}$) for (a) the stress,
(b) the energy, and (c) stress value at the state of zero strain.}
\label{fig:training_loss}
\end{figure}

The energy functional neural networks have a feed-forward architecture consisting of a hidden dense layer (100 neurons / ReLU),
followed by two multiply layers (cf. \citet{vlassis2021sobolev}),
then another hidden dense layer (100 neurons / ReLU), and finally an output dense layer (Linear).
The training and validation procedures of the neural network are implemented in Python with machine learning libraries 
Keras \citep{chollet2015keras} and Tensorflow \citep{tensorflow2015}.
The kernel weight matrix of the layers was initialized with a Glorot uniform distribution and the bias vector with a zero distribution.
The models were trained on 163400 MD simulation data points and validated on 70030 data points.
All the models were trained for 1000 epochs with a batch size of 512, using the Nadam optimizer \citep{dozat2016incorporating}
initialized with default values in the Keras library.

The loss function training curves for the architectures $\mathcal{M}_{1}$ and $\mathcal{M}_{2}$ are demonstrated in Fig.~\ref{fig:training_loss}.
The two architectures appear to have similar accuracy so they will be used interchangeably below.
The predictive capabilities of $\mathcal{M}_{1}$ and $\mathcal{M}_{2}$ are further demonstrated in Section~\ref{sec:md_simulation_predictions}.

\begin{figure}[h!]
\newcommand\siz{.32\textwidth}
\centering
\begin{tabular}{M{.43\textwidth}M{.43\textwidth}}
\hspace{-1cm}\includegraphics[width=.35\textwidth ,angle=0]{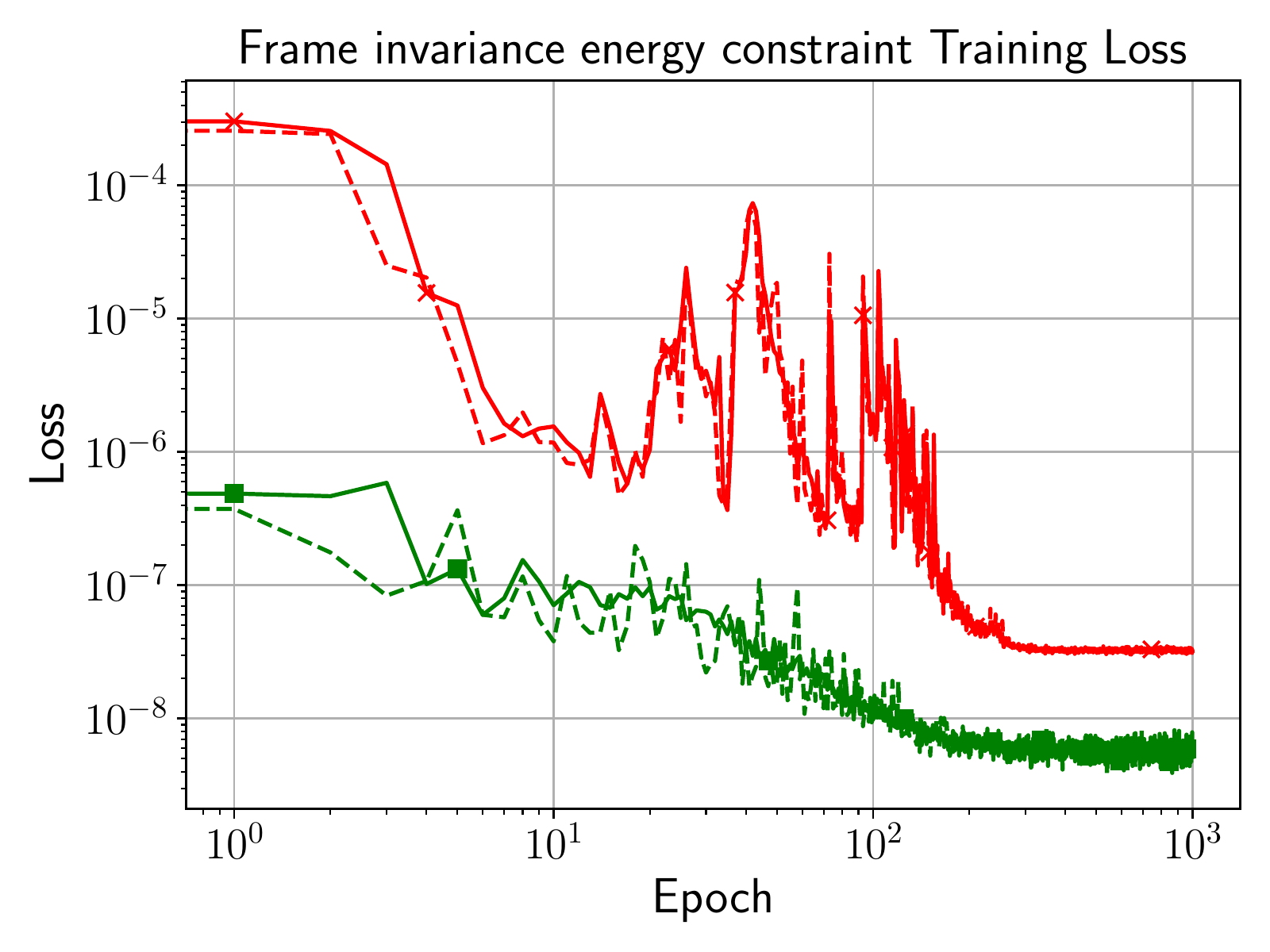} &
\hspace{-1cm}\includegraphics[width=.35\textwidth ,angle=0]{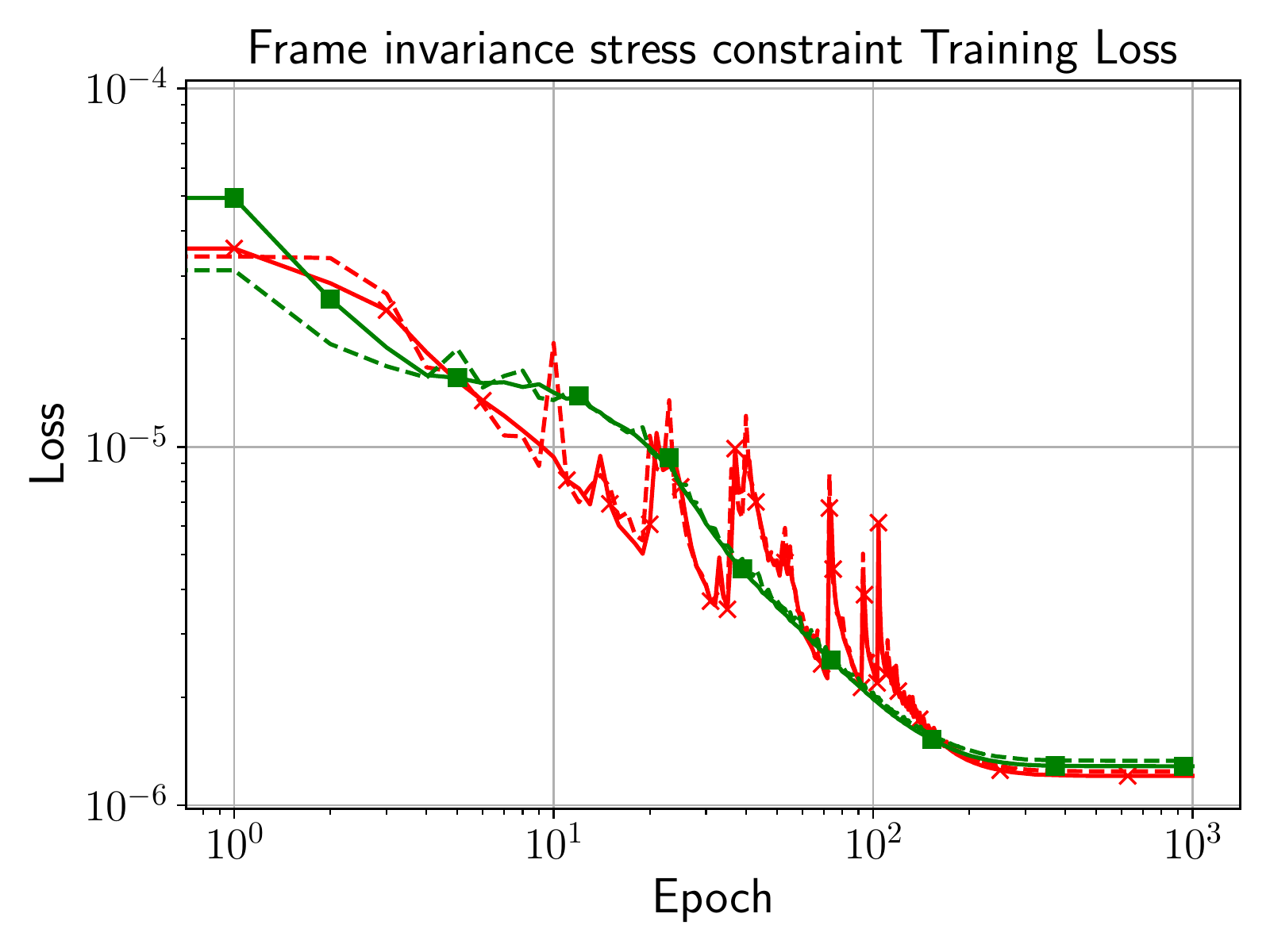} \\     

(a) & (b) \\
\end{tabular}
\includegraphics[width=.6\textwidth ,angle=0]{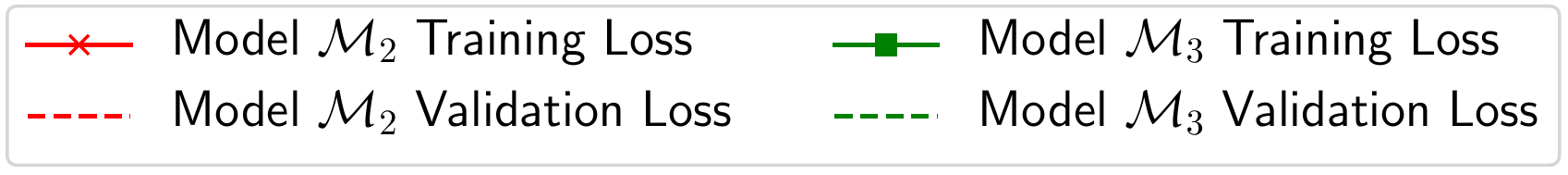}
\caption{Comparison of the training loss curves for (a) the energy and (b) stress frame invariance constraints for the energy conjugate pair $\tensor{P}-\tensor{F}$ model ($\mathcal{M}_{2}$) without any additional constraints in the loss function and the energy conjugate pair $\tensor{P}-\tensor{F}$ model ($\mathcal{M}_{3}$) trained with the additional frame invariance constraint loss function Eq.~\eqref{eq:frame_invariance_constraint}.}
\label{fig:frame_invariance_training_loss}
\end{figure}

To check and, if necessary, enforce the frame invariance of the neural network hyperelastic models as described in Section~\ref{sec:frame_invariance},
we conduct a transfer learning experiment by retraining the neural network model $\mathcal{M}_{2}$.
We first train the energy conjugate pair $\tensor{P}-\tensor{F}$ model ($\mathcal{M}_{2}$) for 1000 epochs
without any frame invariance constraints in the loss function (i.e., Eq.~\eqref{eq:incomplete_sobolev_PF}). 
We record the frame invariance metrics during training by applying random rotation $\tensor{Q}$ tensors on the input deformation gradient tensors
and examine whether the material response is frame invariant;
that is, whether the predicted energy remains the same before and after rotation and whether the predicted stress tensor rotates accordingly. 
The trained model $\mathcal{M}_{2}$ is then retrained with the additional frame invariance constraints in Eq.~\eqref{eq:frame_invariance_constraint} for another 1000 epochs (model $\mathcal{M}_{3}$). 
The comparison of the training curves for $\mathcal{M}_{2}$ and $\mathcal{M}_{3}$ is shown in Fig.~\ref{fig:frame_invariance_training_loss}. 
Model $\mathcal{M}_{2}$ appears to already satisfy well the frame invariant properties,
with the additional constraints of model $\mathcal{M}_{3}$ mostly improving the frame invariance energy constraints.

\begin{figure}[h!]
\newcommand\siz{.32\textwidth}
\centering
\begin{tabular}{M{.43\textwidth}M{.43\textwidth}}
\hspace{-1cm}\includegraphics[width=.35\textwidth ,angle=0]{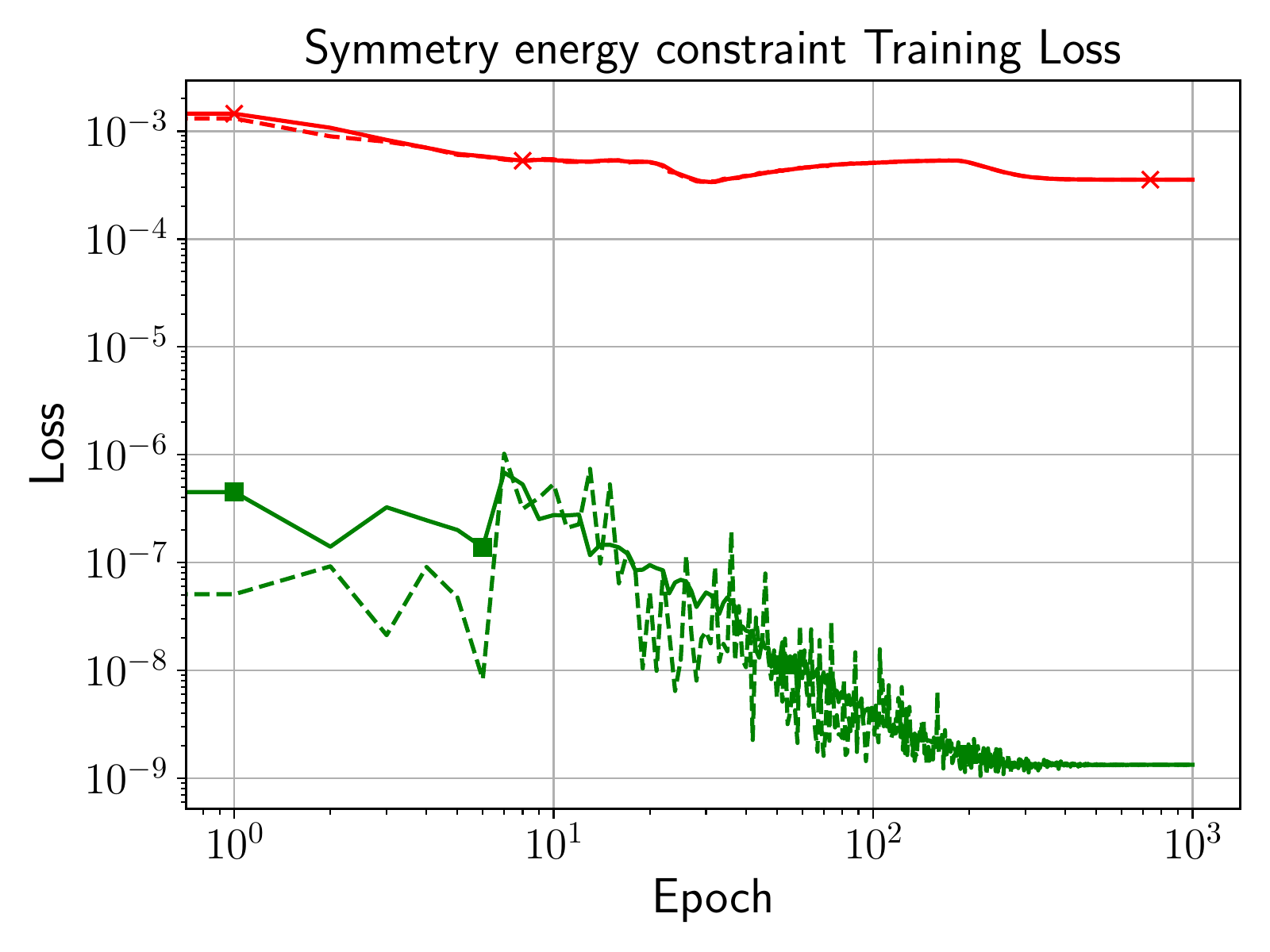} &
\hspace{-1cm}\includegraphics[width=.35\textwidth ,angle=0]{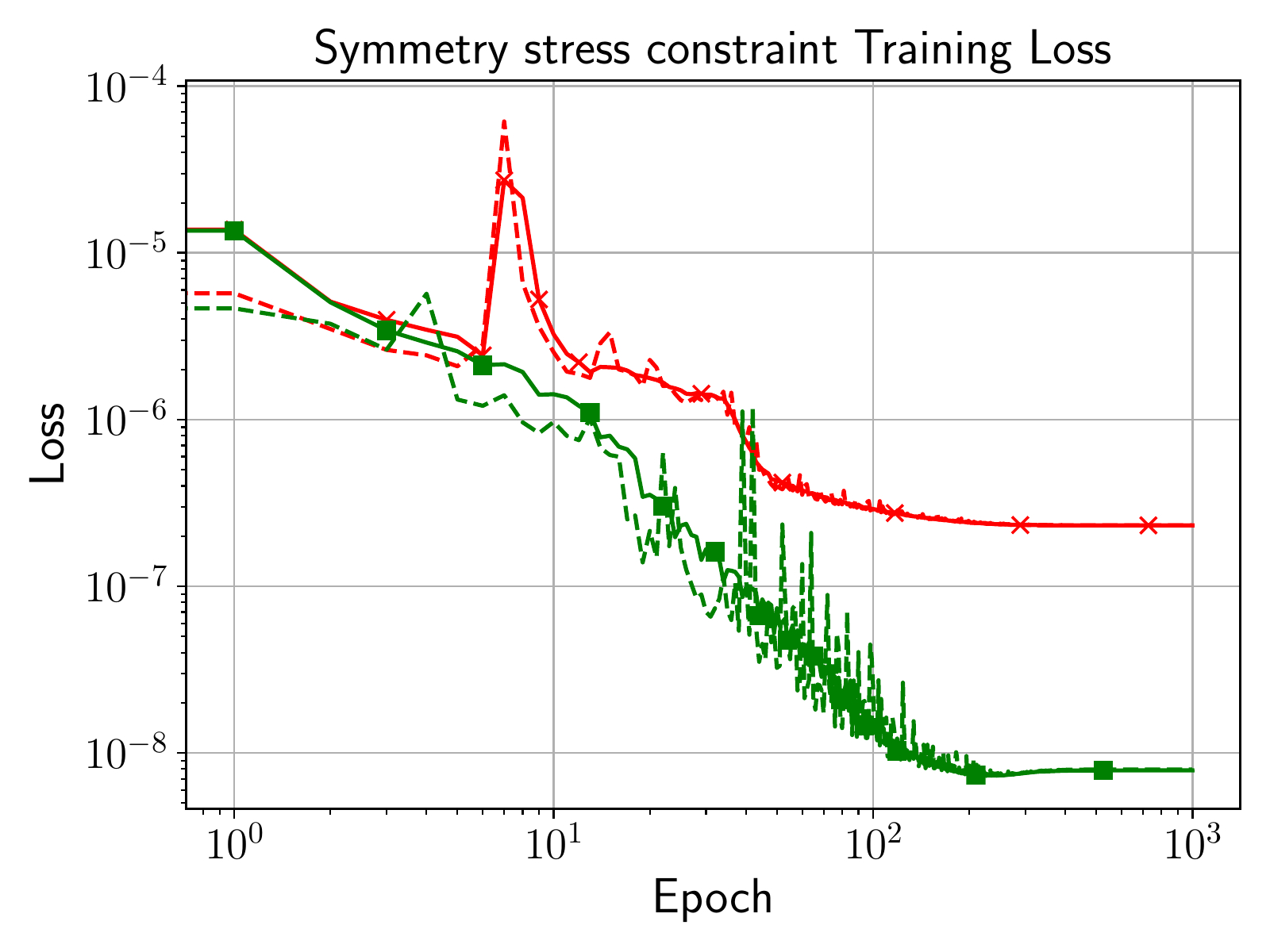} \\     

(a) & (b) \\
\end{tabular}
\includegraphics[width=.6\textwidth ,angle=0]{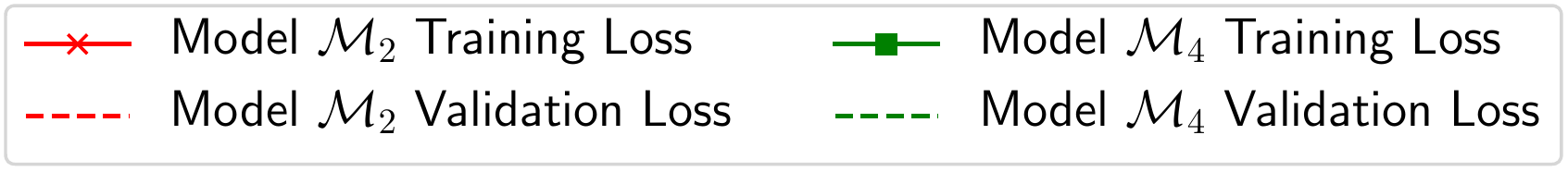}
\caption{Comparison of the training loss curves for (a) the energy and (b) stress symmetry constraints for the energy conjugate pair $\tensor{P}-\tensor{F}$ model ($\mathcal{M}_{2}$) without any symmetry constraints in the loss function and the energy conjugate pair $\tensor{P}-\tensor{F}$ model ($\mathcal{M}_{4}$) trained with the additional symmetry-constraint loss function Eq.~\eqref{eq:symmetry_constraint}.}
\label{fig:symmetry_training_loss}
\end{figure}

We also perform a transfer learning experiment by retraining the neural network model $\mathcal{M}_{2}$ to ensure it retains the observed $\beta$-HMX crystal symmetries as described in Section~\ref{sec:crystal_symmetries}.
We first train the energy conjugate pair $\tensor{P}-\tensor{F}$ model ($\mathcal{M}_{2}$) for 1000 epochs without any symmetry constraints in the loss function and record the symmetry metrics during training.
By applying a rotation $\tensor{Q}_\text{sym}$ on the input deformation gradient tensors, we check for the material response to retain the expected monoclinic symmetry behavior.
The check includes the constraints up to the first-order derivatives of the network.
The trained model $\mathcal{M}_{2}$ is then retrained with the additional symmetry constraints in Eq.~\eqref{eq:symmetry_constraint} for another 1000 epochs (model $\mathcal{M}_{4}$). 
The results for the two training experiments are shown in Fig.~\ref{fig:symmetry_training_loss},
where the additional symmetry constraints appear to be improving both the energy and the stress symmetry constraints.

\rmk{\label{rem:scaling}\textbf{Rescaling of the training data}. 
As a pre-processing step, we have normalized all data to avoid the vanishing or 
exploding gradient problem that may occur during the back-propagation 
process \citep{bishop1995neural}. The $X_i$ sample of a measure $X$ is scaled to a unit interval via,

\begin{equation}
\overline{X_i}:= \frac{X_i - X_{\text{min}}}{X_{\text{max}}-X_{\text{min}}} ,
\label{eq:scaled_feature_X}
\end{equation} 
where $\overline{X_i}$ is the normalized sample point. $X_{\text{min}}$ and $X_{\text{max}}$ are the minimum and maximum values of the measure $X$ in the training data set such that all different types of data used in this paper (e.g. strain, stress, etc) are all normalized within the range $[0,1]$. 
}

\subsection{Validation of the constitutive responses}
In this section, we validate the neural network predicted constitutive response against MD simulation data as well as $\beta$-HMX elastic coefficients from the literature.
We also monitor the learned physical properties for the trained models, such as the strong ellipticity, the energy growth, and the anisotropy index.

\subsubsection{Validation against unseen MD simulations}
\label{sec:md_simulation_predictions}
We validate the predictive performance of the learned models against \textbf{unseen} MD simulation loading paths. 
The neural network architectures considered in this section are the energy conjugate pair $\tensor{S}-\tensor{E}$ model ($\mathcal{M}_{1}$) 
and the energy conjugate pair $\tensor{P}-\tensor{F}$ model ($\mathcal{M}_{2}$).

\begin{figure}[h!]
\newcommand\siz{.32\textwidth}
\centering
\begin{tabular}{M{.33\textwidth}M{.33\textwidth}M{.33\textwidth}}
\hspace{-1cm}\includegraphics[width=.33\textwidth ,angle=0]{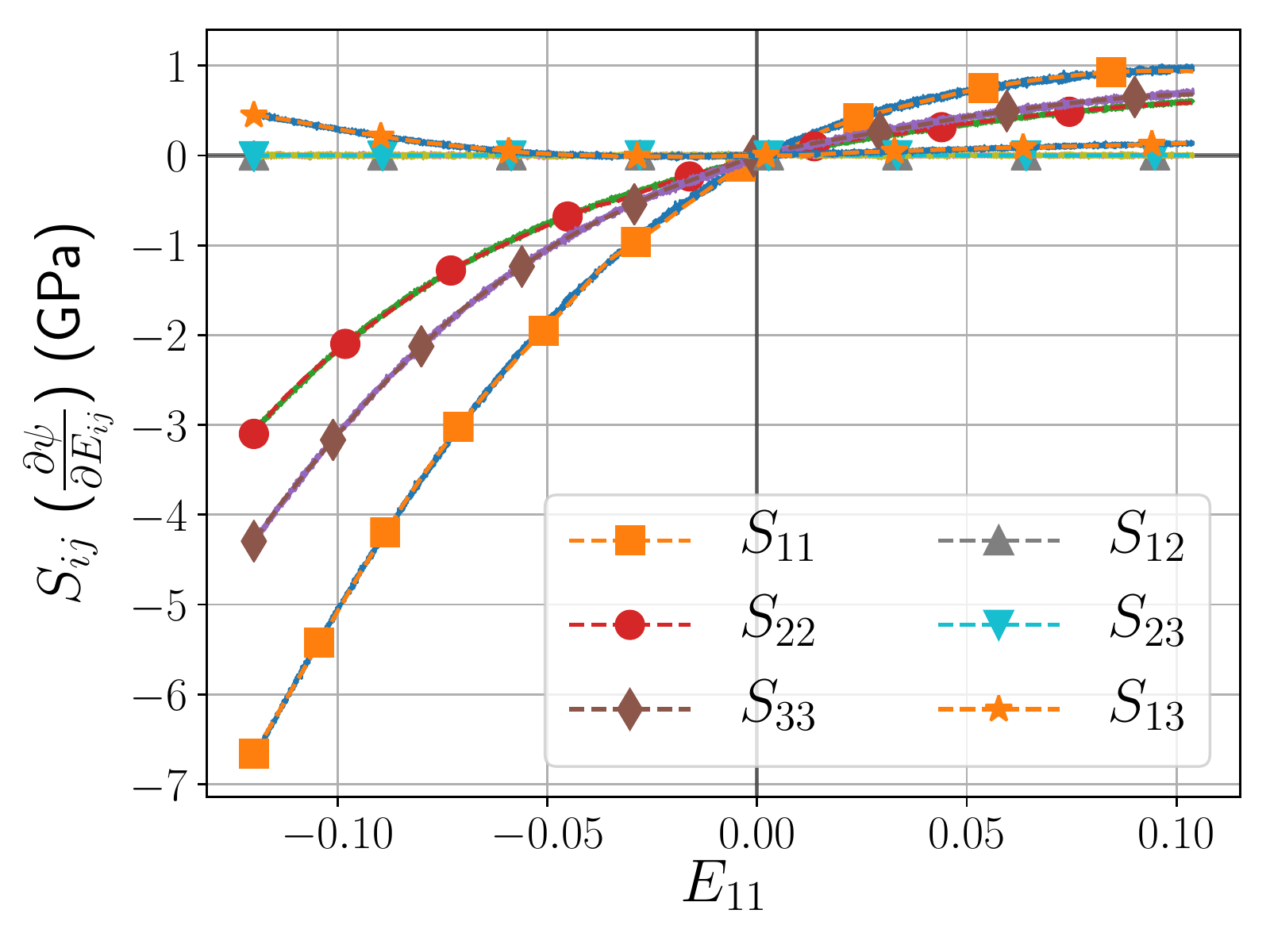} &
\hspace{-1cm}\includegraphics[width=.33\textwidth ,angle=0]{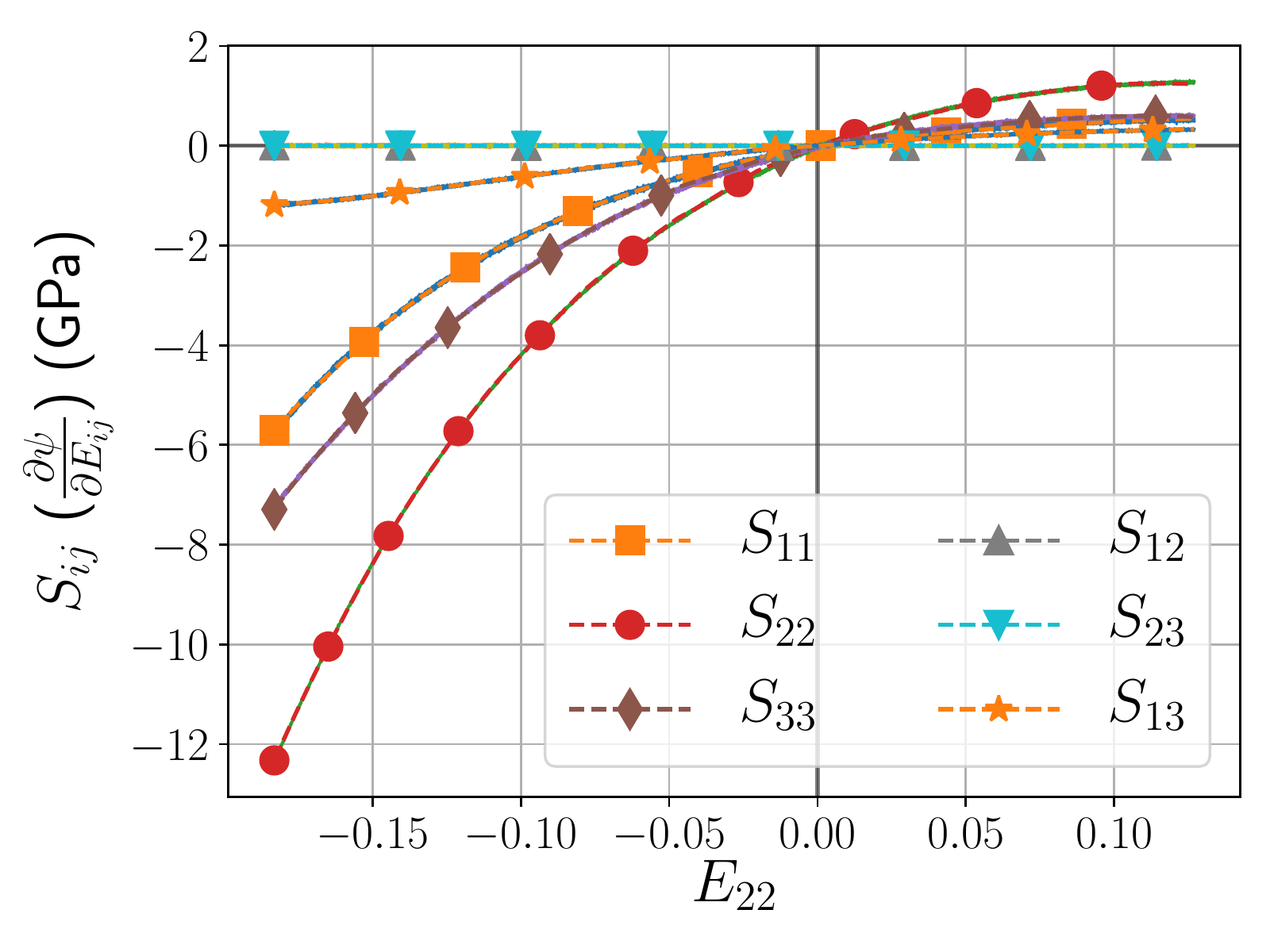} &
\hspace{-1cm}\includegraphics[width=.33\textwidth ,angle=0]{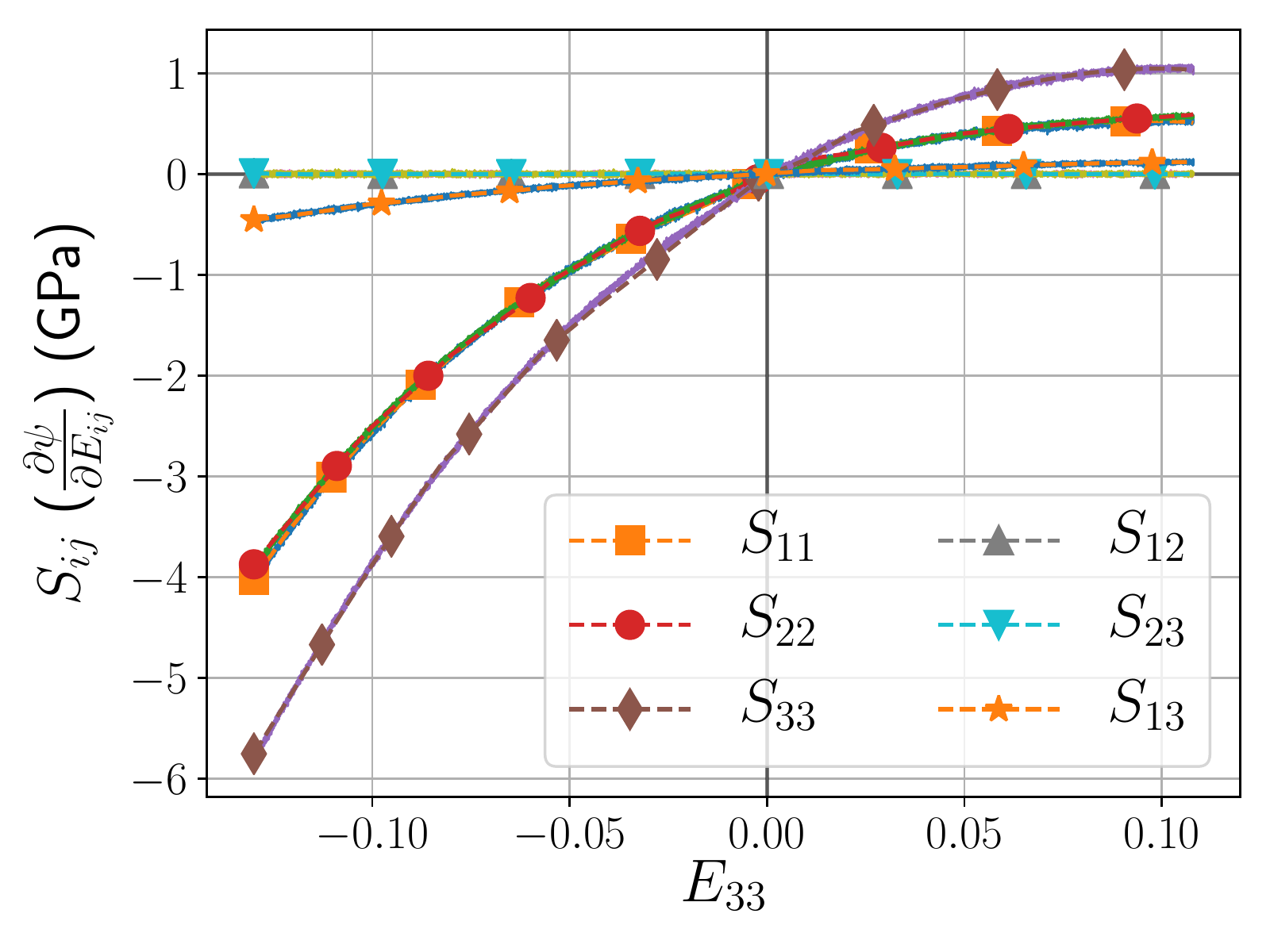} \\     

(a) & (b) & (c) \\
\end{tabular}

\caption{Comparison of the predicted stress response against three uniaxial MD simulations for the energy conjugate pair $\tensor{S}-\tensor{E}$ model ($\mathcal{M}_{1}$). (a) Uniaxial compression and extension along the $x_{1}$ axis. (b) Uniaxial compression and extension along the $x_{2}$ axis. (c) Uniaxial compression and extension along the $x_{3}$ axis.
}.
\label{fig:axial_tests}
\end{figure}

\begin{figure}[h!]
\newcommand\siz{.32\textwidth}
\centering
\begin{tabular}{M{.33\textwidth}M{.33\textwidth}M{.33\textwidth}}
\hspace{-1cm}\includegraphics[width=.33\textwidth ,angle=0]{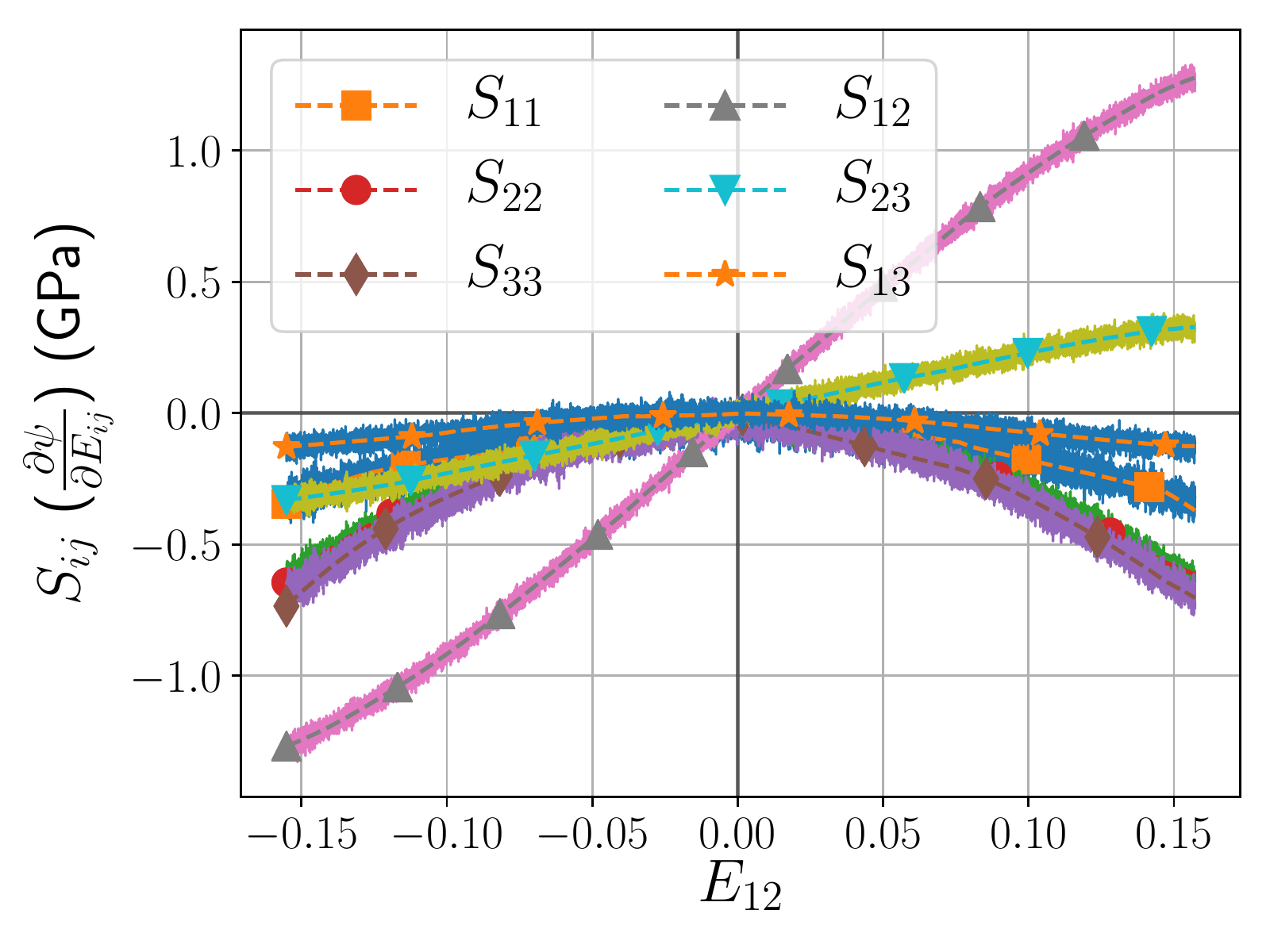} &
\hspace{-1cm}\includegraphics[width=.33\textwidth ,angle=0]{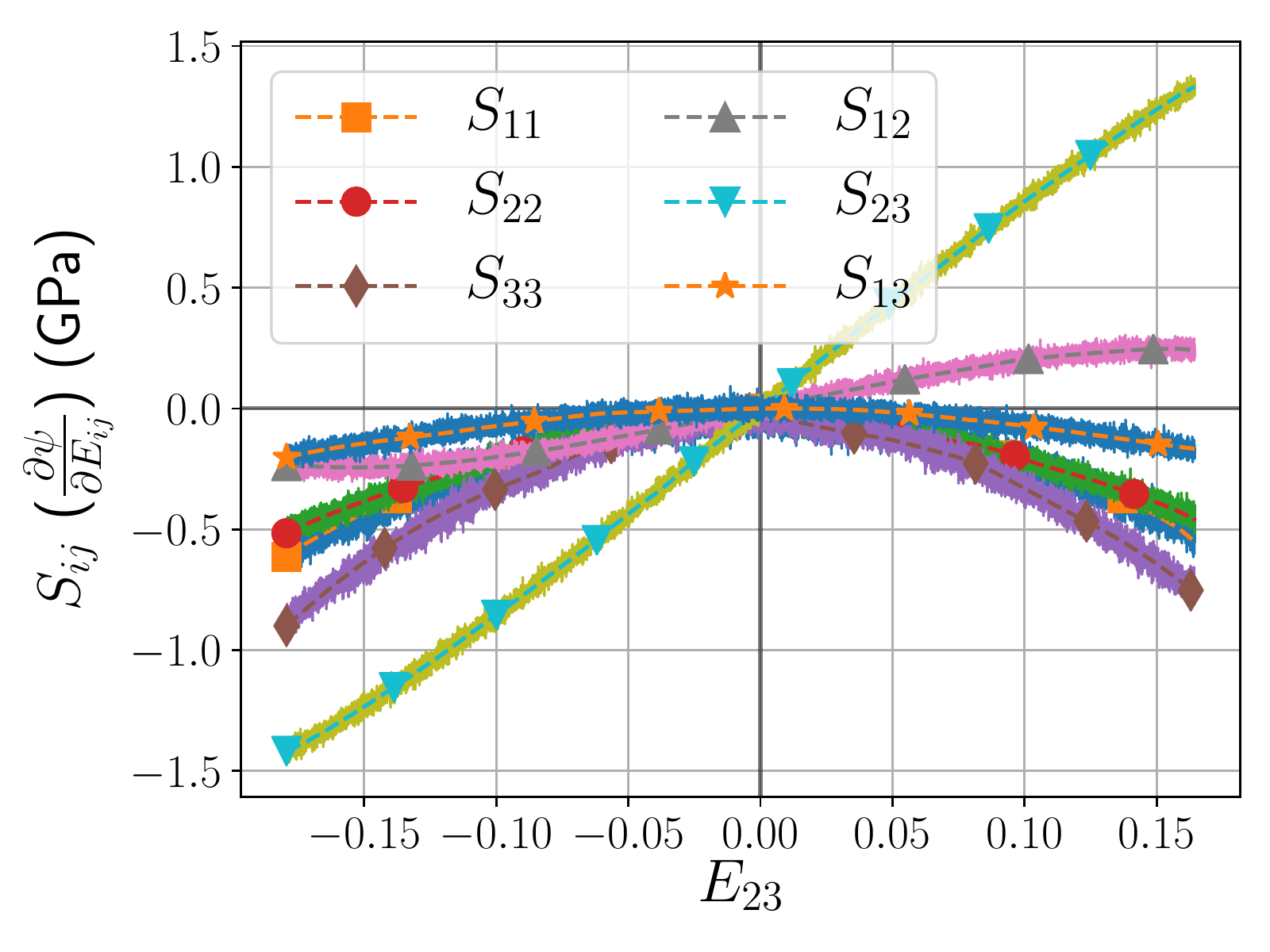} &
\hspace{-1cm}\includegraphics[width=.33\textwidth ,angle=0]{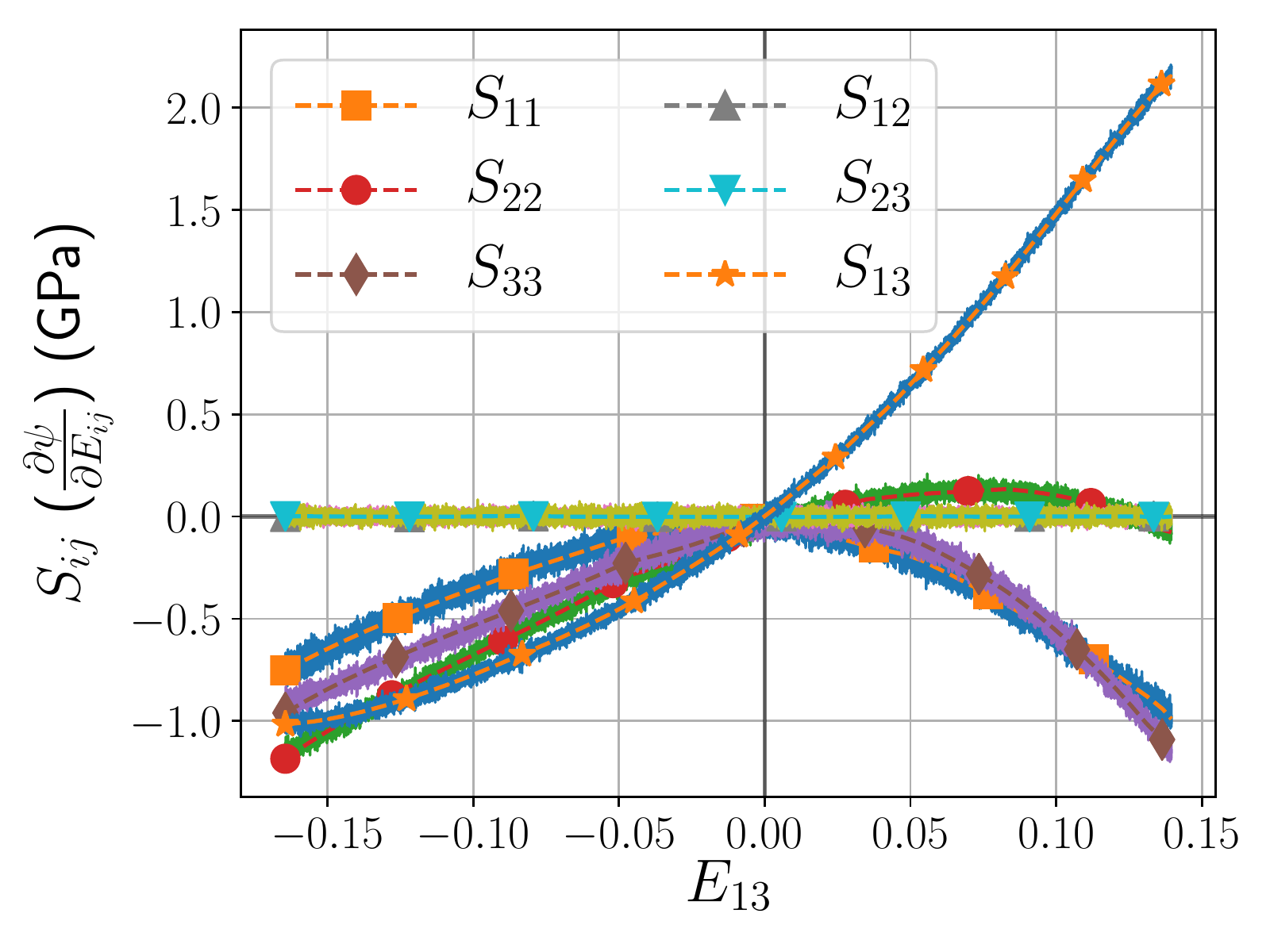} \\     

(a) & (b) & (c) \\
\end{tabular}

\caption{Comparison of the predicted stress response against three shear MD simulations for the energy conjugate pair $\tensor{S}-\tensor{E}$ model ($\mathcal{M}_{1}$). (a) Shear tests for positive and negative directions along the $\vec{e}_{1} \otimes \vec{e}_{2}$ direction. (b) Shear tests for positive and negative directions along the $\vec{e}_{2} \otimes \vec{e}_{3}$ direction. (c) Shear tests for positive and negative directions along the $\vec{e}_{1} \otimes \vec{e}_{3}$ direction.}
\label{fig:shear_tests}
\end{figure}

\begin{figure}[h!]
\newcommand\siz{.32\textwidth}
\centering
\begin{tabular}{M{.33\textwidth}M{.33\textwidth}M{.33\textwidth}}
\hspace{-1cm}\includegraphics[width=.33\textwidth ,angle=0]{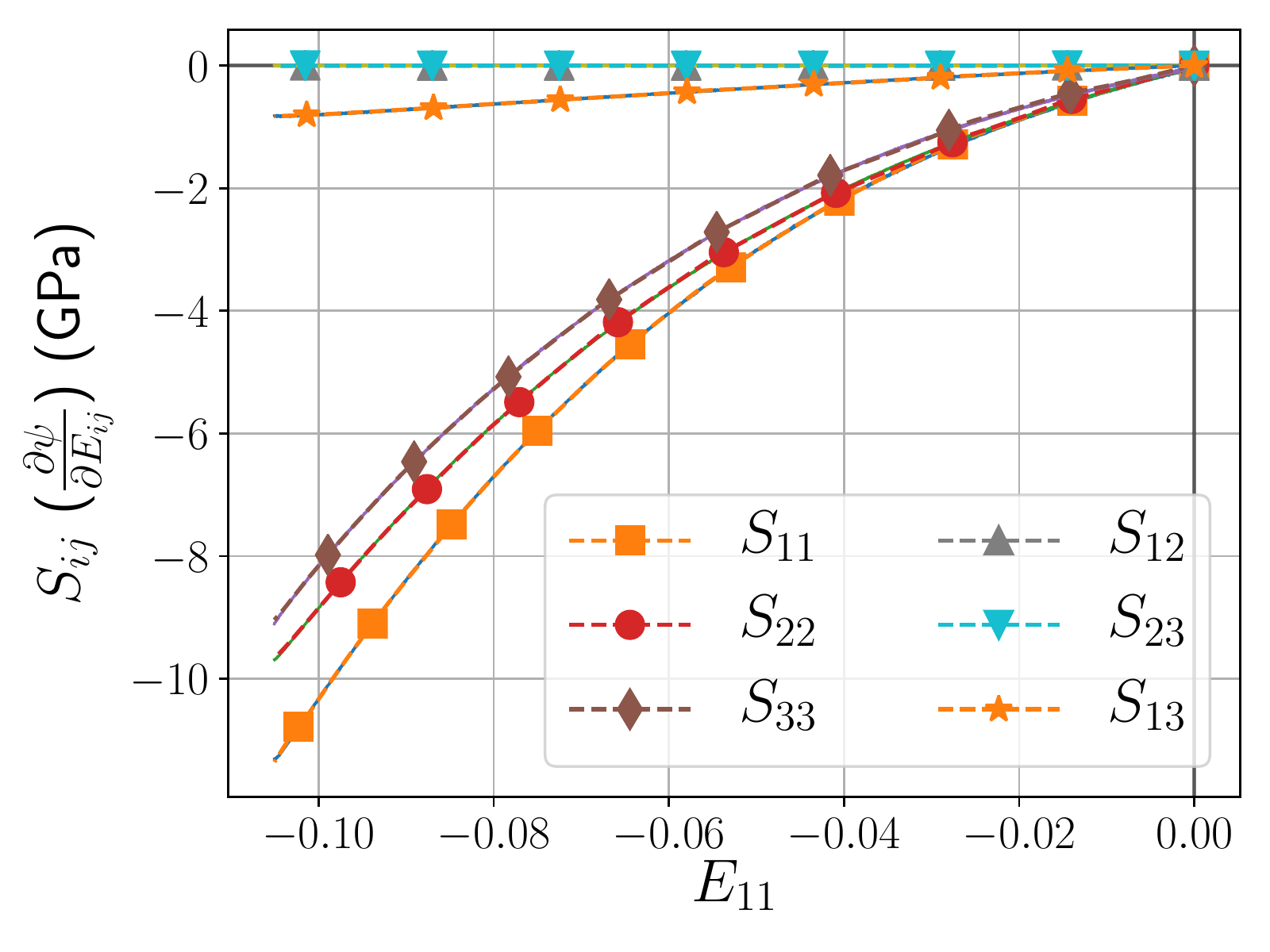} &
\hspace{-1cm}\includegraphics[width=.33\textwidth ,angle=0]{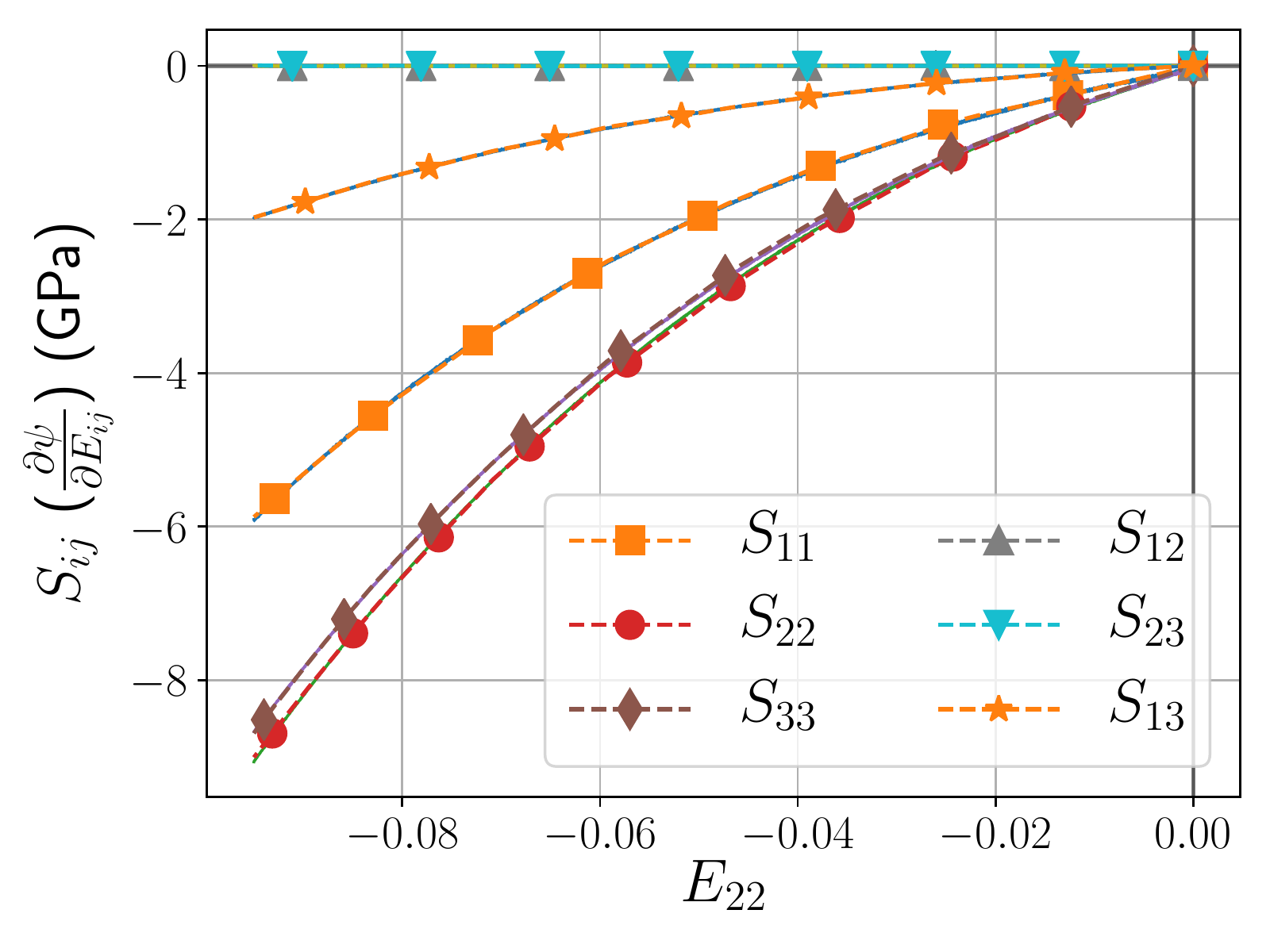} &
\hspace{-1cm}\includegraphics[width=.33\textwidth ,angle=0]{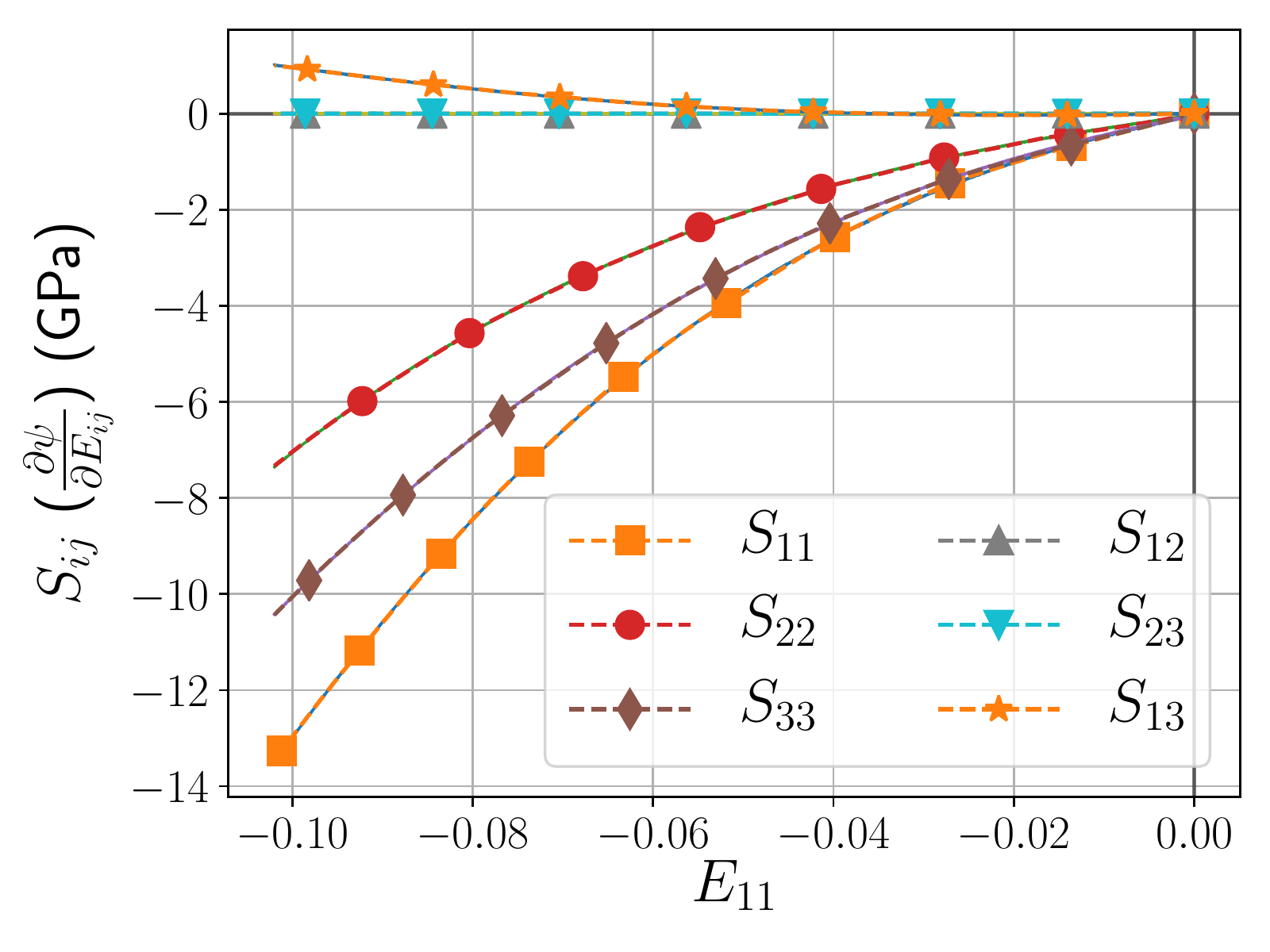} \\     

(a) & (b) & (c) \\
\end{tabular}
\caption{Comparison of the predicted stress response against three biaxial MD simulations for the energy conjugate pair $\tensor{S}-\tensor{E}$ model ($\mathcal{M}_{1}$). (a) Biaxial compression along the $x_{1}$ and $x_{2}$ axes. (b) Biaxial compression along the $x_{2}$ and $x_{3}$ axes. (c) Biaxial compression along the $x_{1}$ and $x_{3}$ axes.}
\label{fig:biaxial_compression}
\end{figure}

\begin{figure}[h!]
\newcommand\siz{.32\textwidth}
\centering
\begin{tabular}{M{.33\textwidth}M{.33\textwidth}M{.33\textwidth}}
\hspace{-1cm}\includegraphics[width=.33\textwidth ,angle=0]{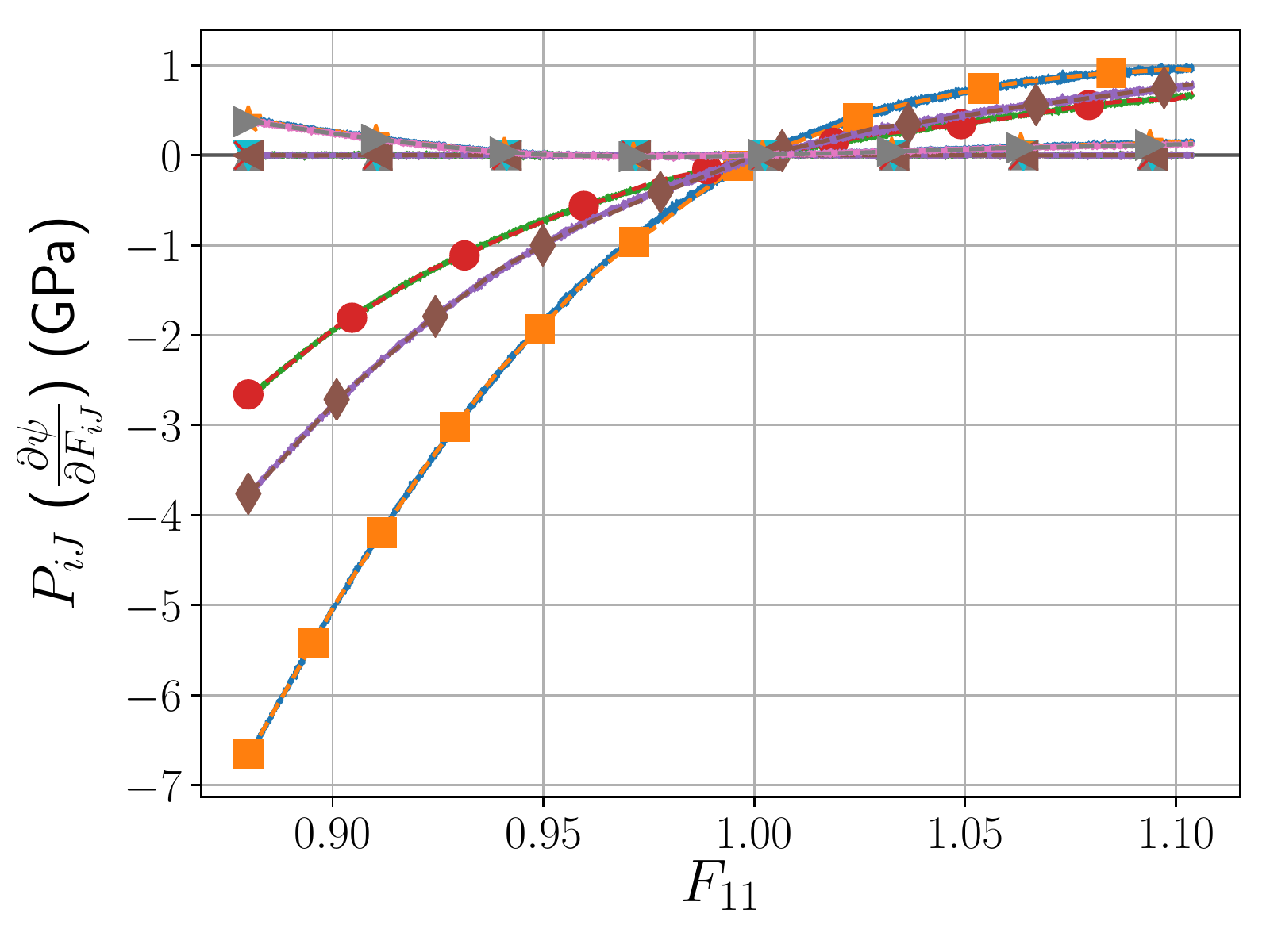} &
\hspace{-1cm}\includegraphics[width=.33\textwidth ,angle=0]{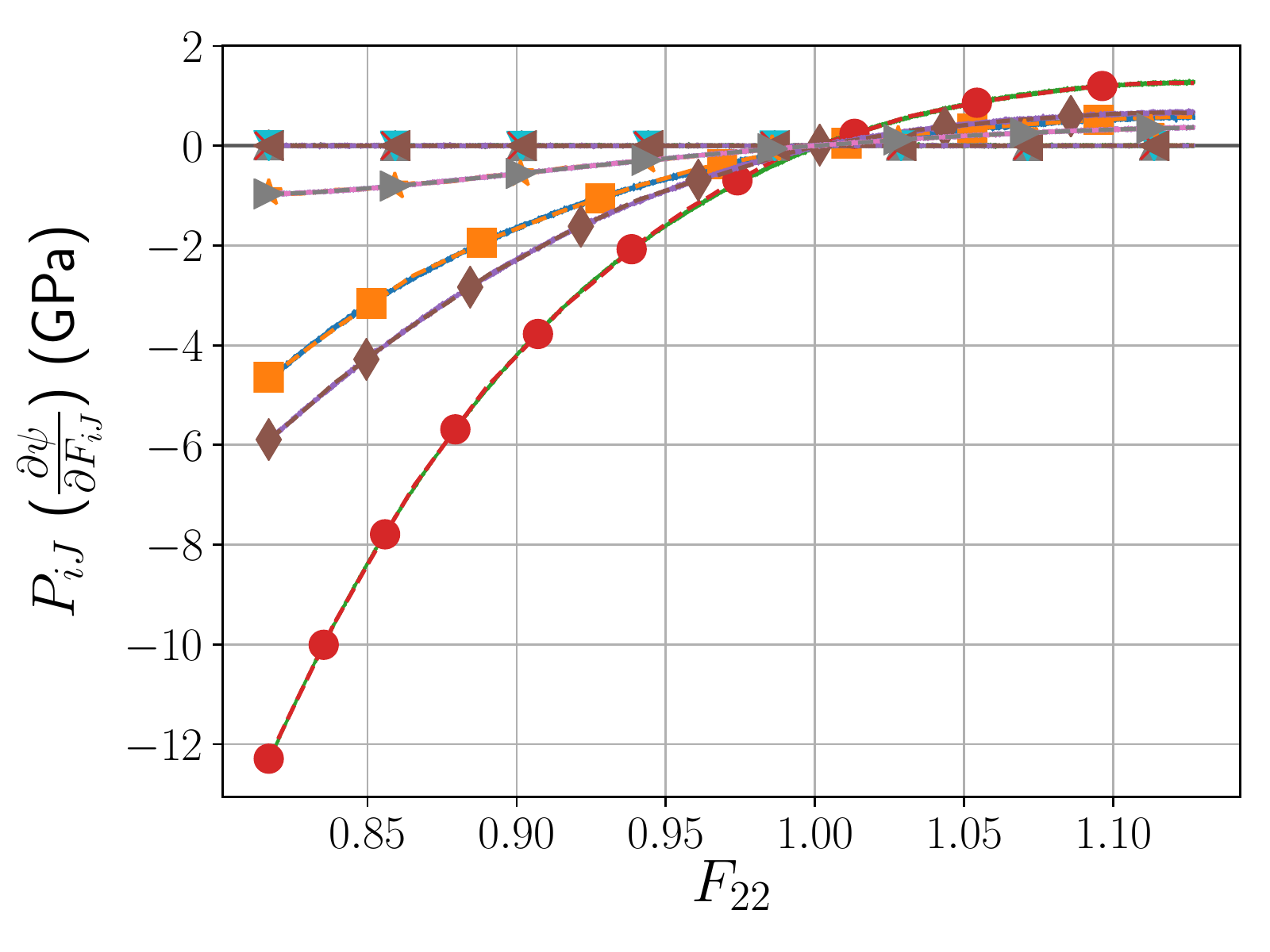} &
\hspace{-1cm}\includegraphics[width=.33\textwidth ,angle=0]{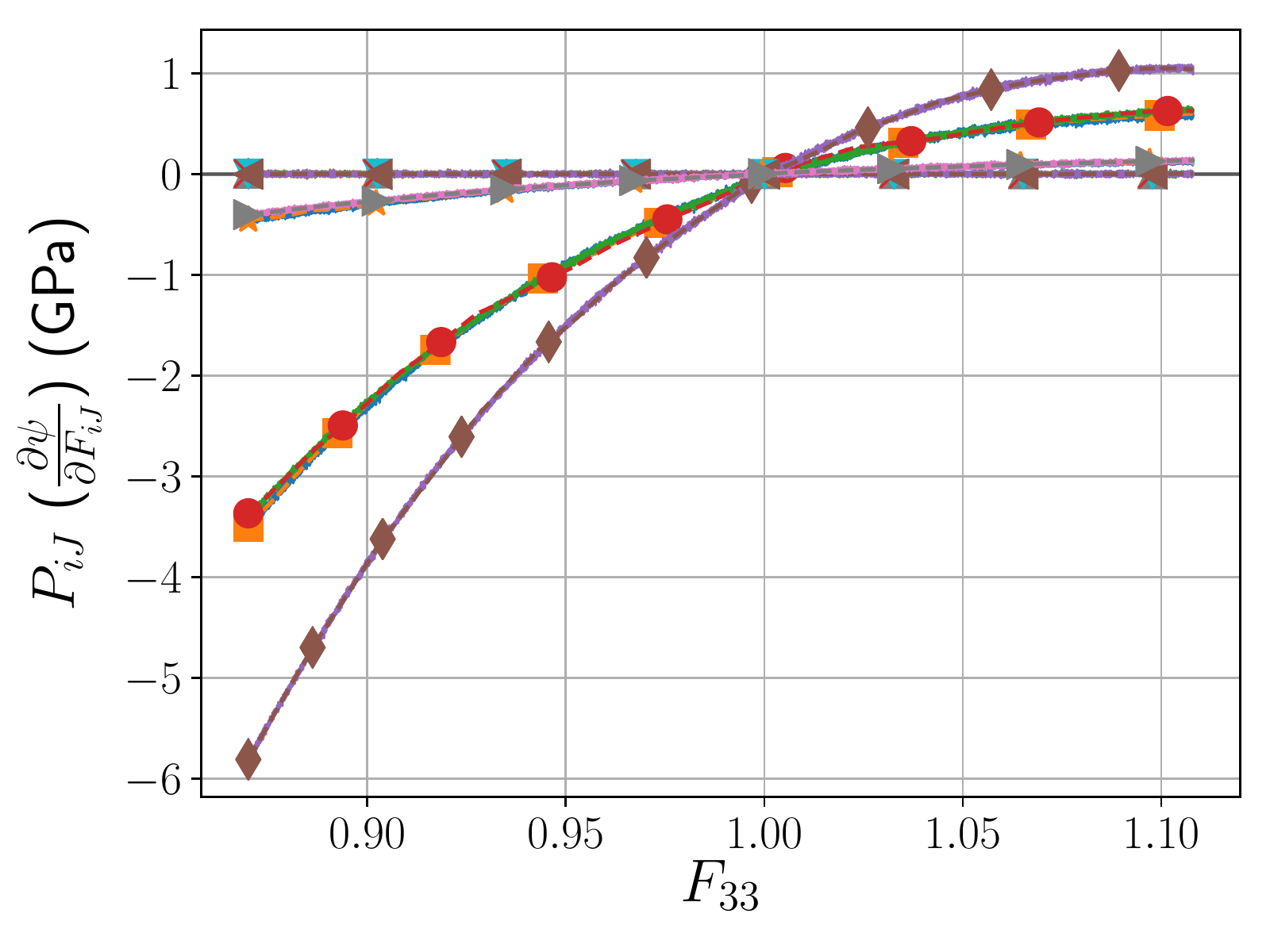} \\     

(a) & (b) & (c) \\

\hspace{-1cm}\includegraphics[width=.33\textwidth ,angle=0]{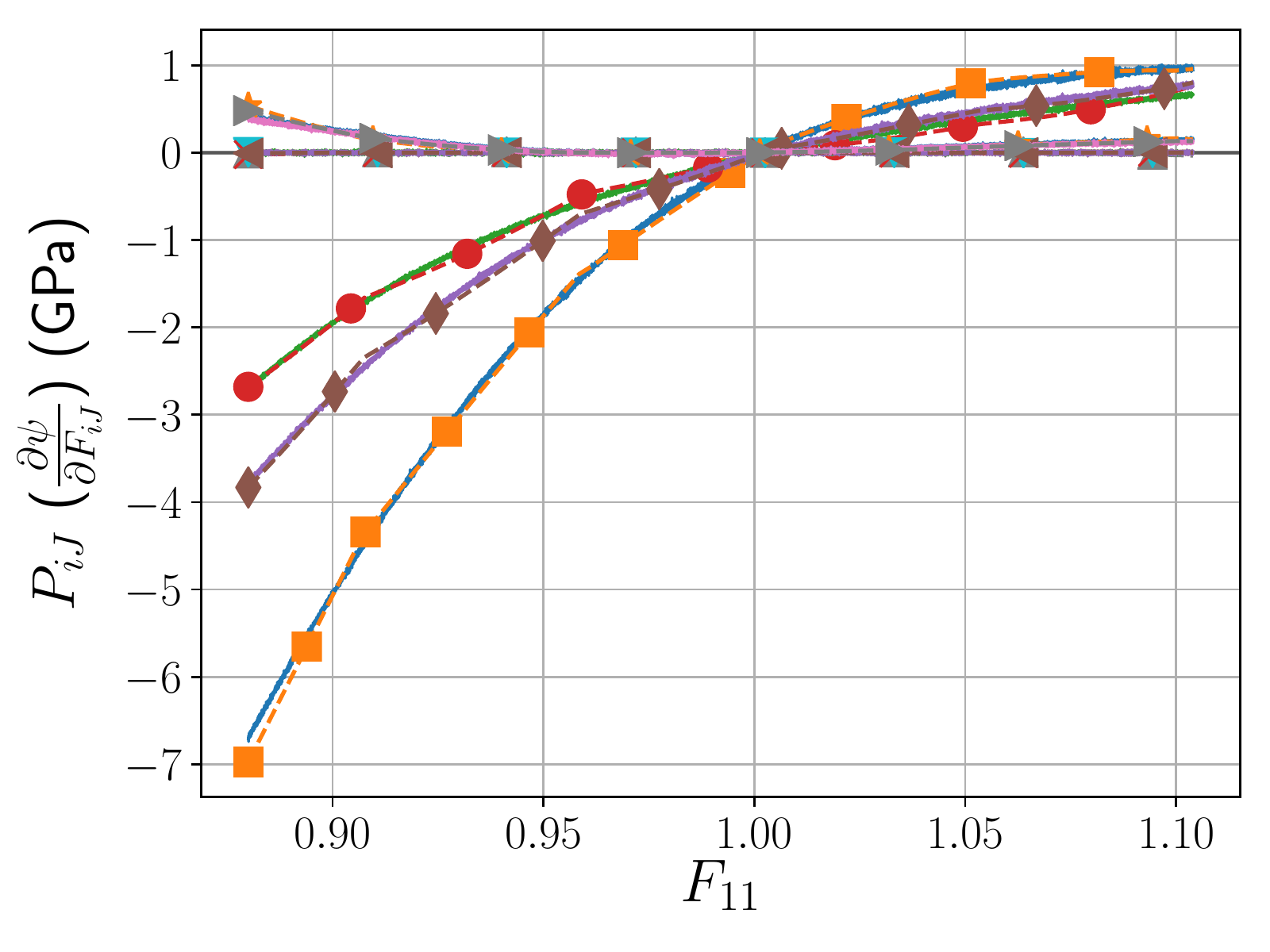} &
\hspace{-1cm}\includegraphics[width=.33\textwidth ,angle=0]{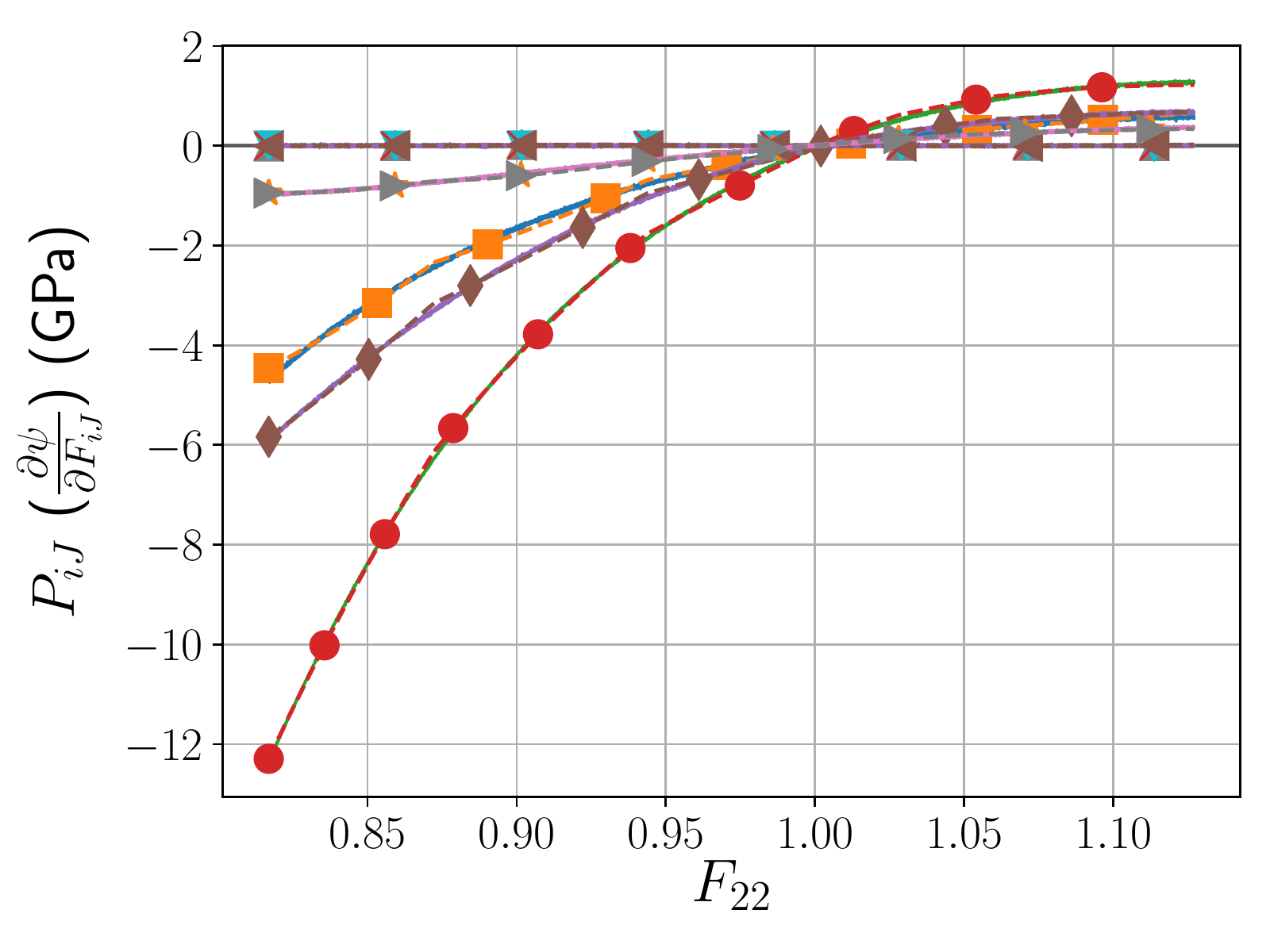} &
\hspace{-1cm}\includegraphics[width=.33\textwidth ,angle=0]{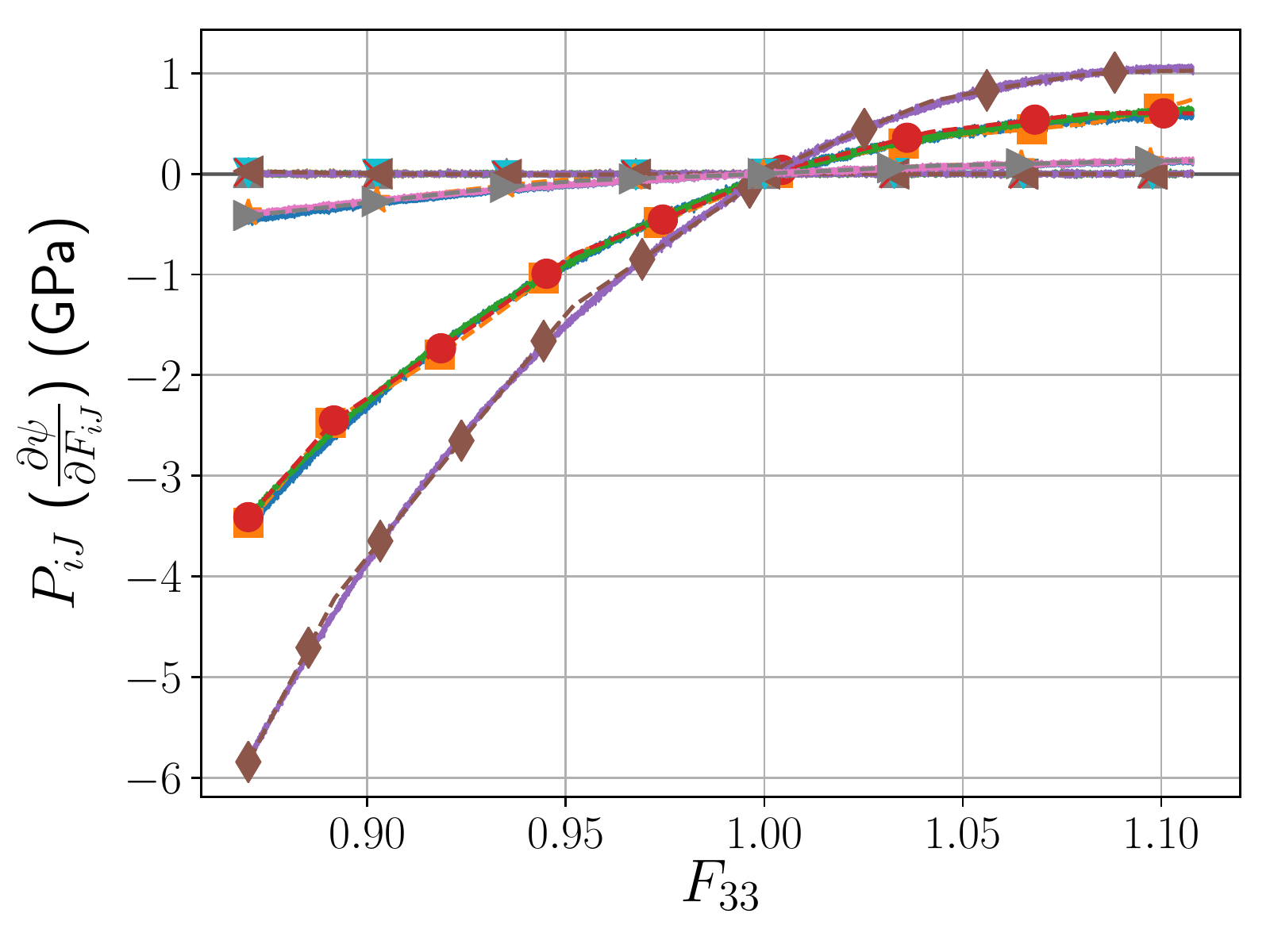} \\     

(d) & (e) & (f) \\

\end{tabular}
\hspace{2cm}\includegraphics[width=.3\textwidth ,angle=0]{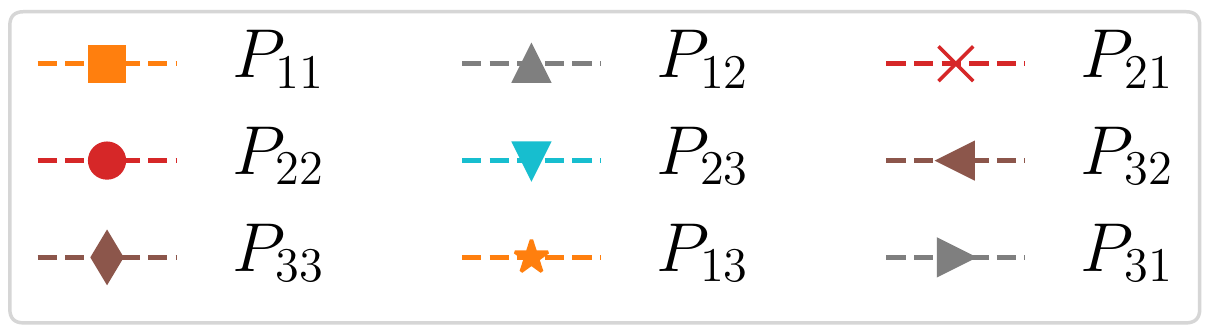}
\caption{Comparison of the predicted stress response against three uniaxial MD simulations for the energy conjugate pair $\tensor{P}-\tensor{F}$ models $\mathcal{M}_{2}$ and $\mathcal{M}_{4}$. (a,d) Uniaxial compression and extension along the $x_{1}$ axis for models $\mathcal{M}_{2}$ and $\mathcal{M}_{4}$ respectively. (b,e) Uniaxial compression and extension along the $x_{2}$ axis for models $\mathcal{M}_{2}$ and $\mathcal{M}_{4}$ respectively. (c,f) Uniaxial compression and extension along the $x_{3}$ axis for models $\mathcal{M}_{2}$ and $\mathcal{M}_{4}$ respectively.
}
\label{fig:axial_tests_PF}
\end{figure}

\begin{figure}[h!]
\newcommand\siz{.32\textwidth}
\centering
\begin{tabular}{M{.33\textwidth}M{.33\textwidth}M{.33\textwidth}}
\hspace{-1cm}\includegraphics[width=.33\textwidth ,angle=0]{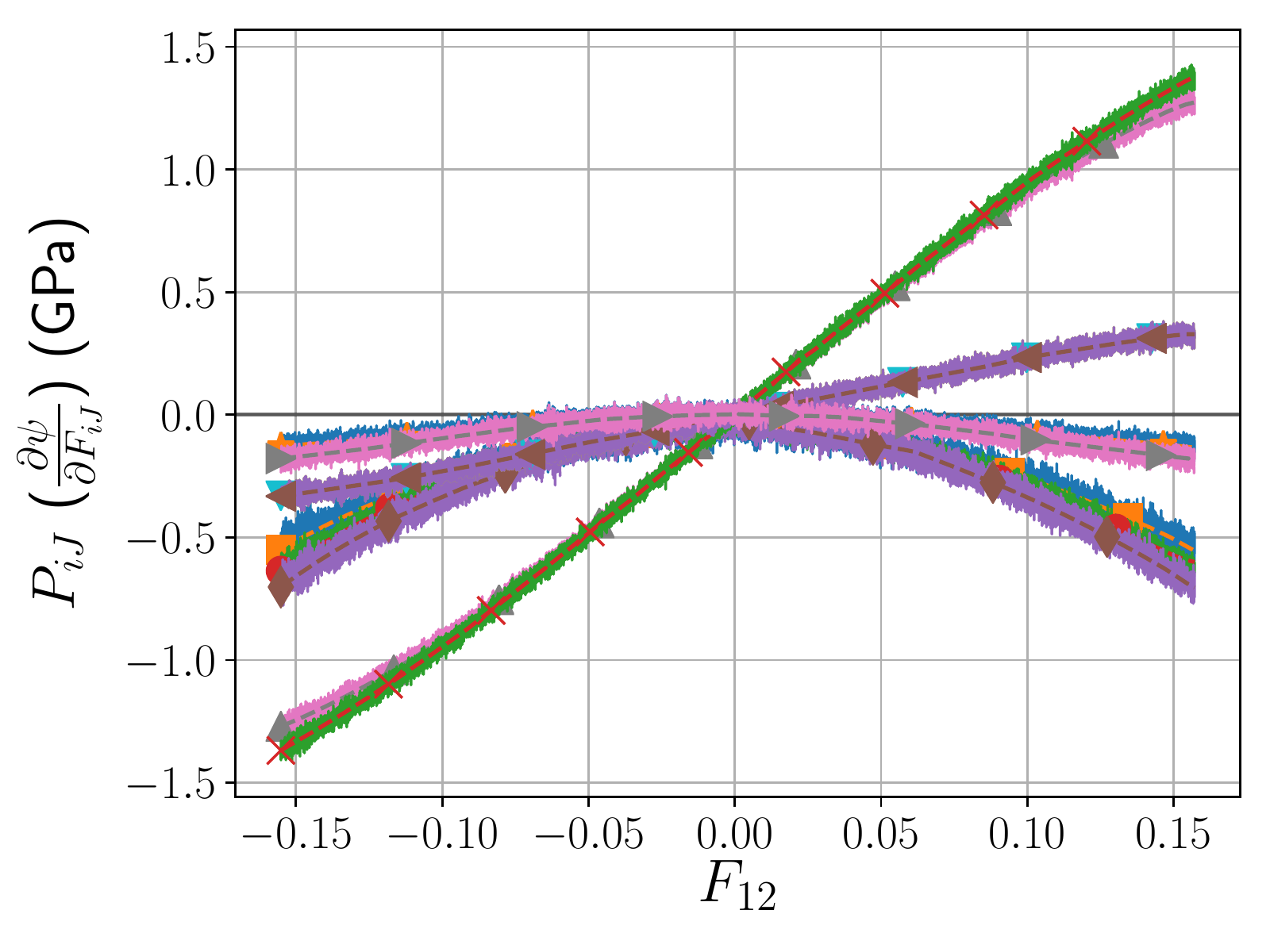} &
\hspace{-1cm}\includegraphics[width=.33\textwidth ,angle=0]{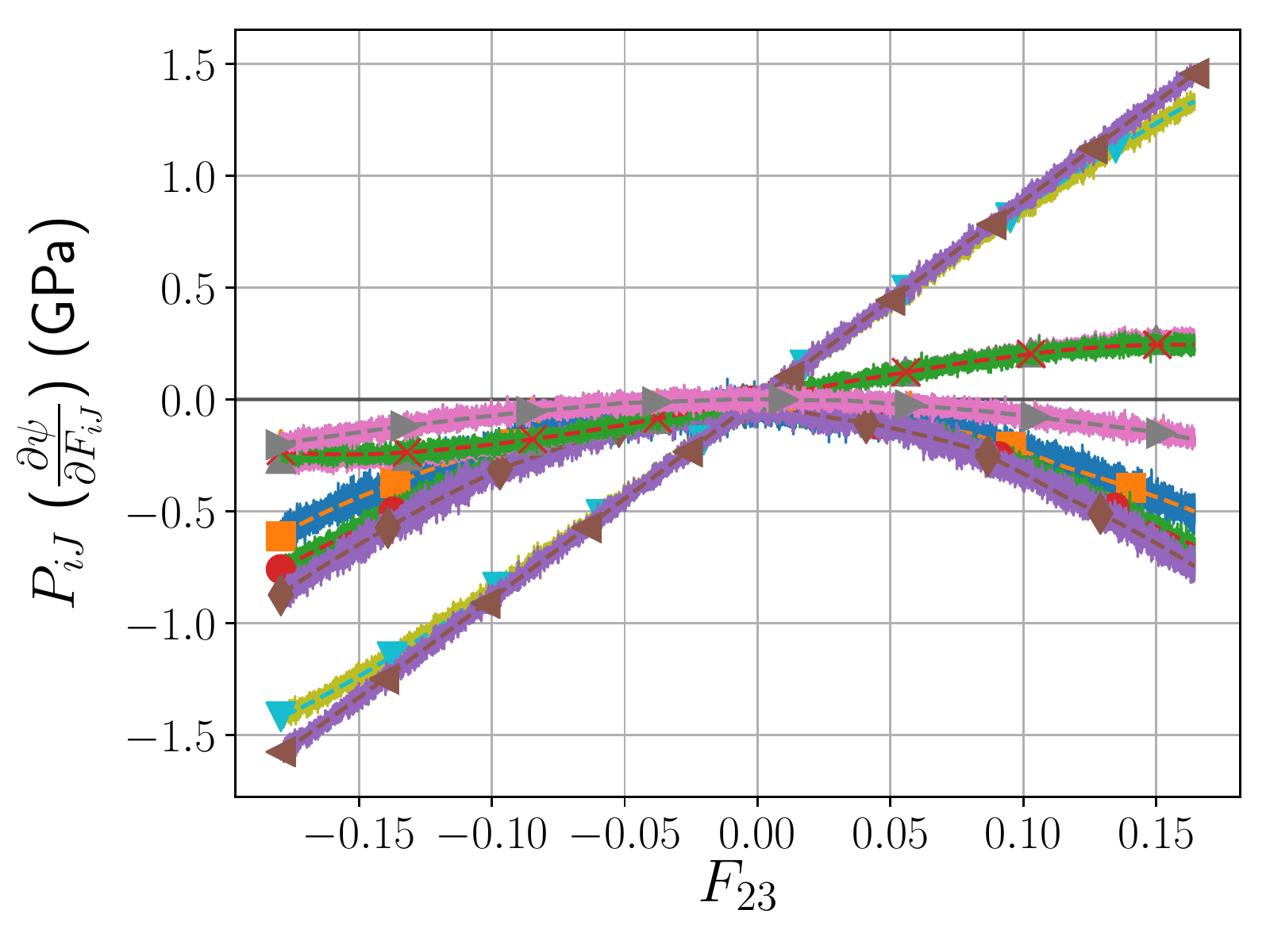} &
\hspace{-1cm}\includegraphics[width=.33\textwidth ,angle=0]{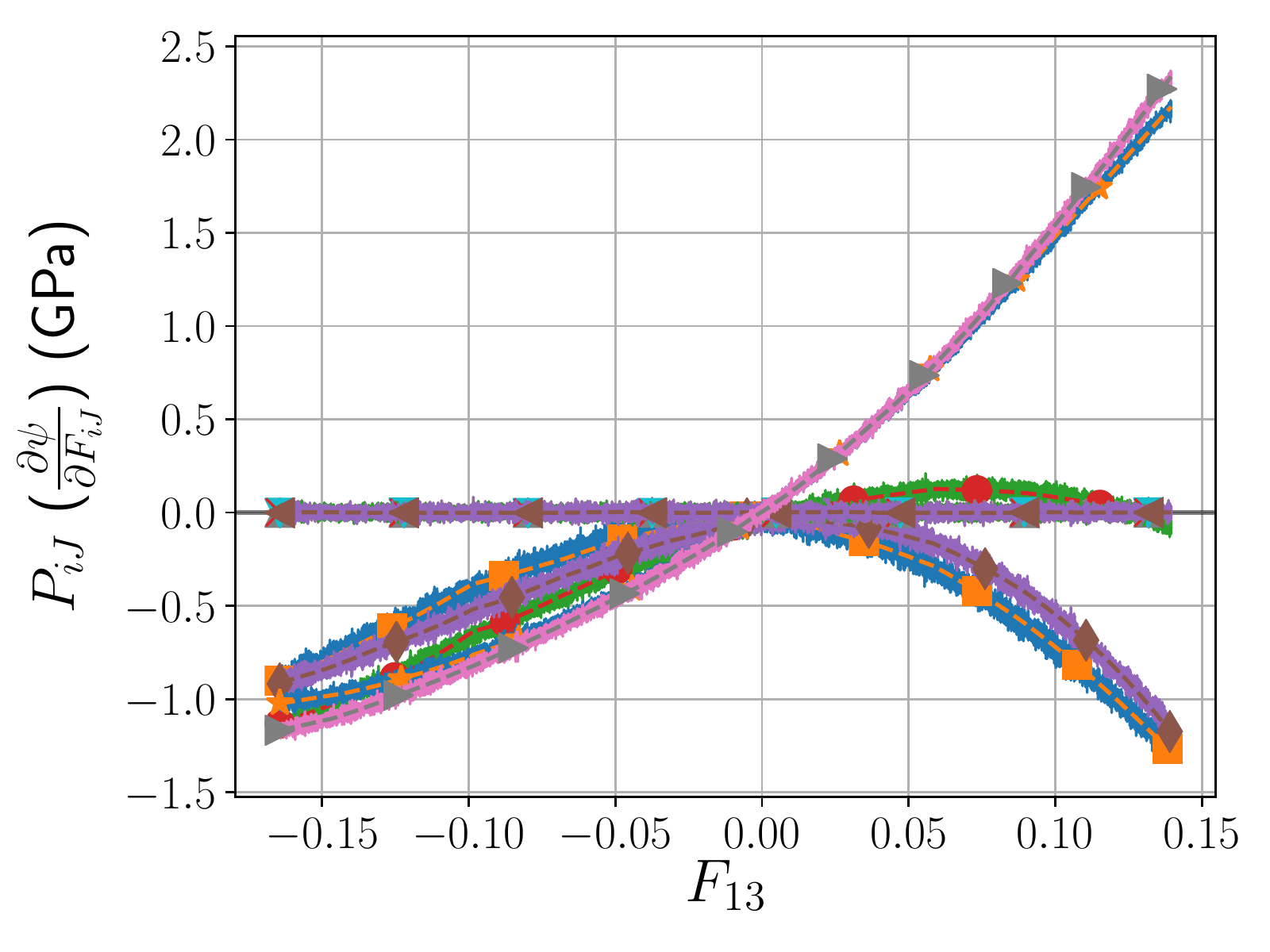} \\     

(a) & (b) & (c) \\
\end{tabular}
\includegraphics[width=.3\textwidth ,angle=0]{./figures/PF_pred_legend.pdf}
\caption{Comparison of the predicted 1st Piola-Kirchhoff stress response against three shear MD simulations for the energy conjugate pair $\tensor{P}-\tensor{F}$ model ($\mathcal{M}_{2}$). (a) Shear tests along the asymmetric positive and negative $\vec{e}_{1} \otimes \vec{e}_{2}$ direction. (b) Shear tests along the asymmetric positive and negative $\vec{e}_{2} \otimes\vec{e}_{3}$ direction. (c) Shear tests for the asymmetric positive and negative $\vec{e}_{1} \otimes \vec{e}_{3}$ direction.}
\label{fig:shear_tests_PF}
\end{figure}

\begin{figure}[h!]
\newcommand\siz{.32\textwidth}
\centering
\begin{tabular}{M{.33\textwidth}M{.33\textwidth}M{.33\textwidth}}
\hspace{-1cm}\includegraphics[width=.33\textwidth ,angle=0]{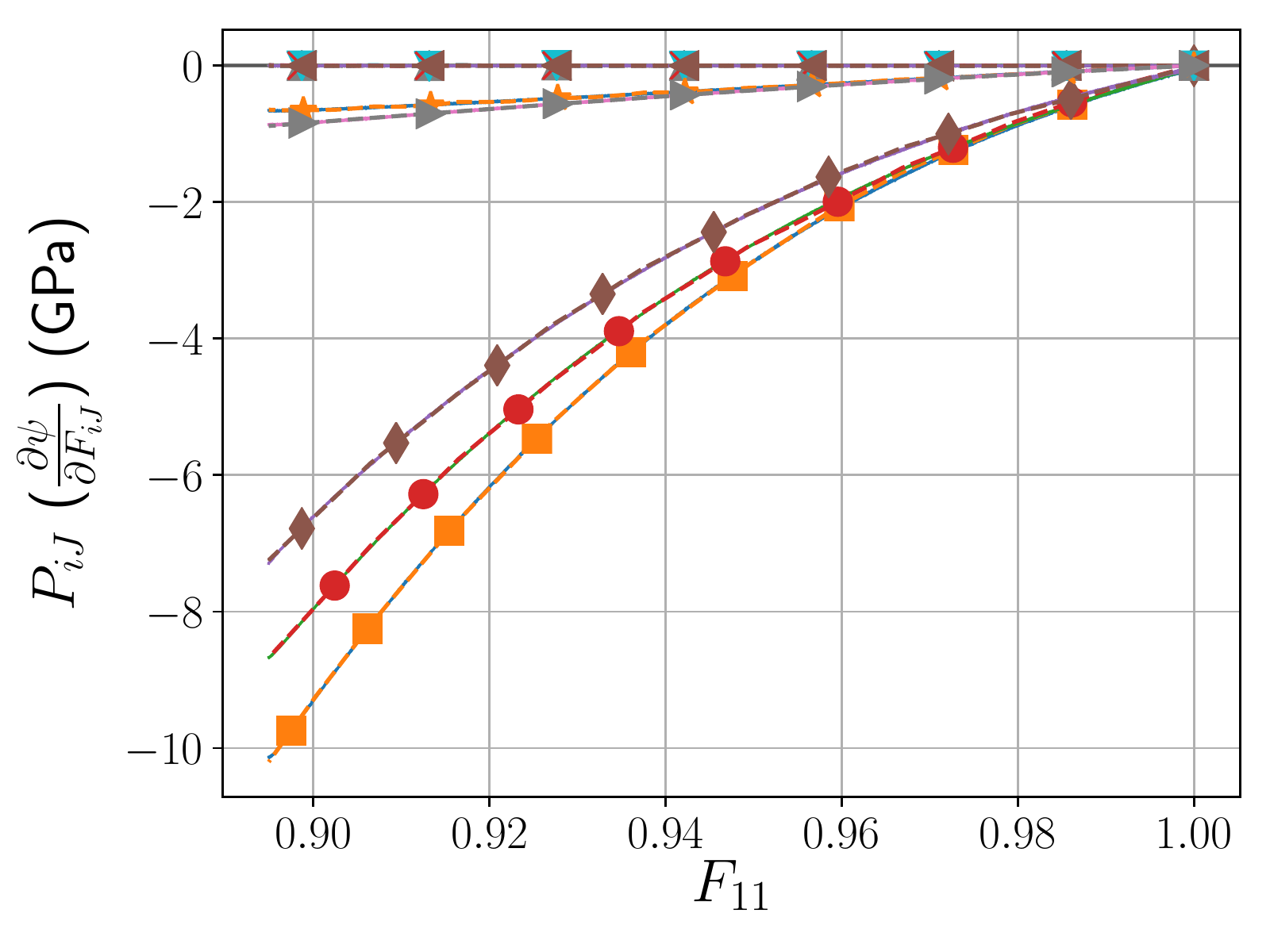} &
\hspace{-1cm}\includegraphics[width=.33\textwidth ,angle=0]{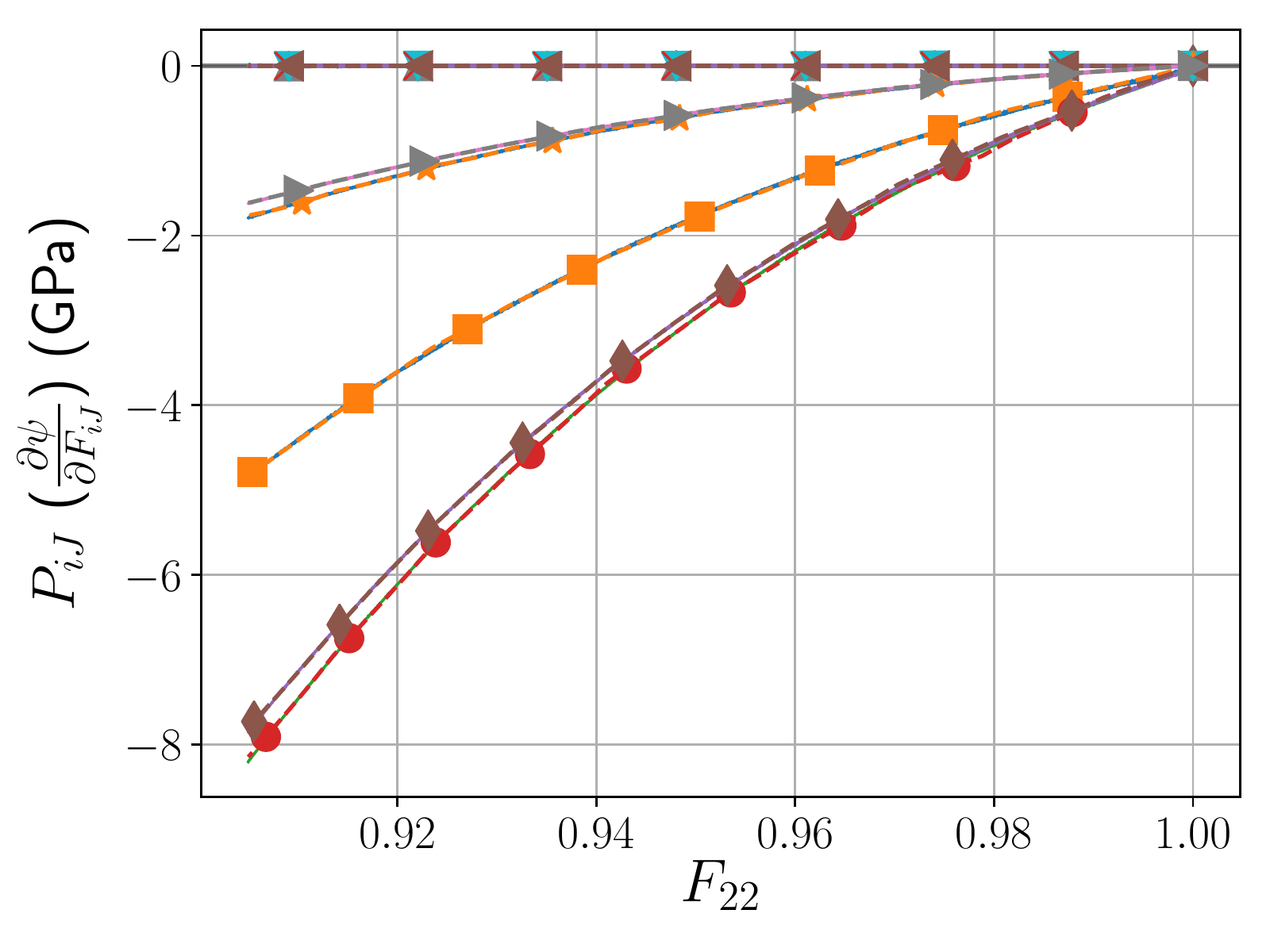} &
\hspace{-1cm}\includegraphics[width=.33\textwidth ,angle=0]{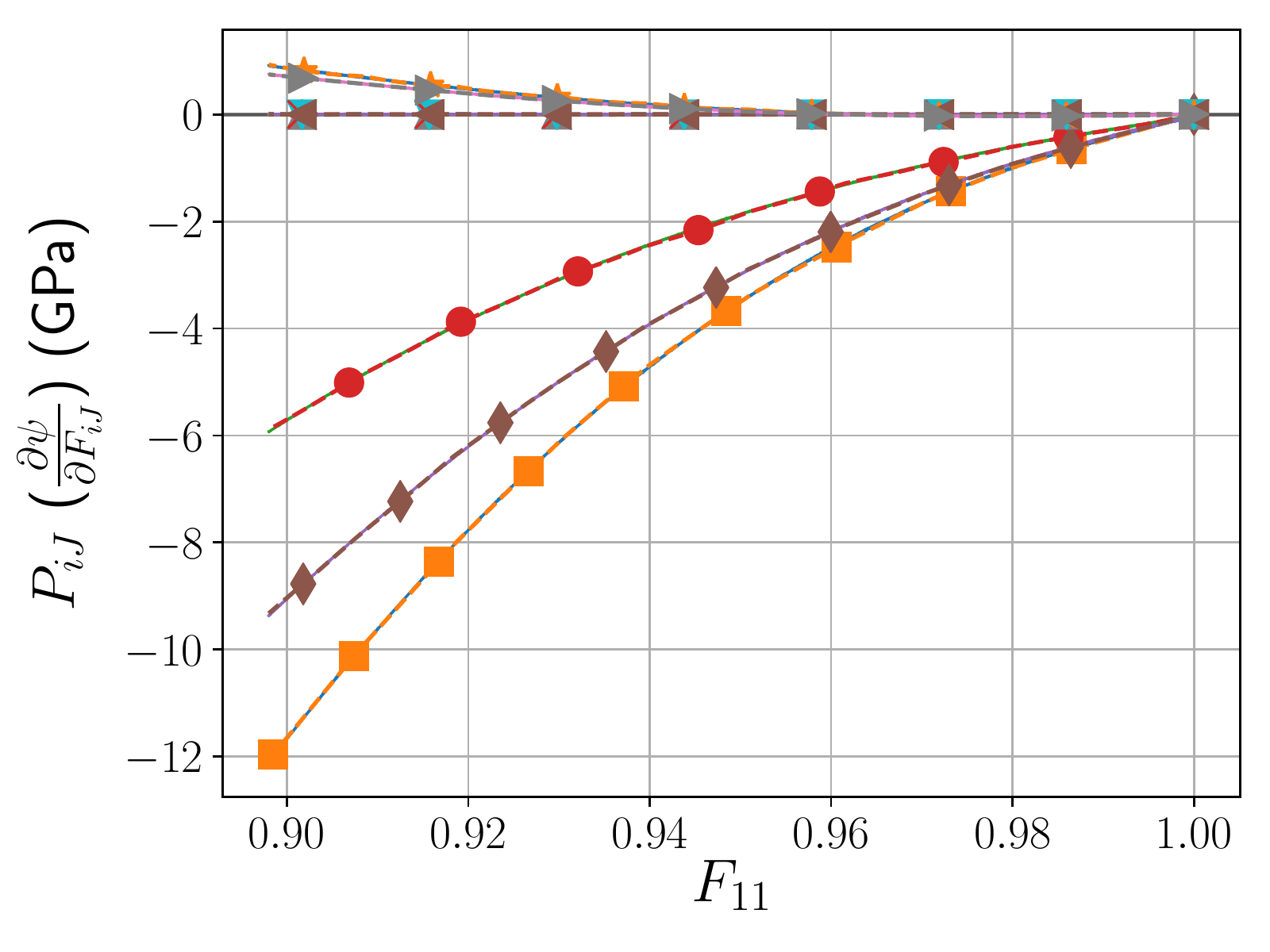} \\     

(a) & (b) & (c) \\
\end{tabular}
\includegraphics[width=.3\textwidth ,angle=0]{./figures/PF_pred_legend.pdf}
\caption{Comparison of the predicted 1st Piola-Kirchhoff stress response against three biaxial MD simulations for the energy conjugate pair $\tensor{P}-\tensor{F}$ model ($\mathcal{M}_{2}$). (a) Biaxial compression along the $x_{1}$ and $x_{2}$ axes. (b) Biaxial compression along the $x_{2}$ and $x_{3}$ axes. (c) Biaxial compression along the $x_{1}$ and $x_{3}$ axes.}
\label{fig:biaxial_compression_PF}
\end{figure}

The stress predictions of the networks against three uniaxial strains along the axes $x_{1}$, $x_{2}$, and $x_{3}$ are demonstrated in Fig.~\ref{fig:axial_tests} and Fig.~\ref{fig:axial_tests_PF}. 
All the symmetric stress tensor components are plotted against the main loading direction of the MD simulation experiment.
The predictions are compared against the raw MD simulation data before the filtering pre-processing described in Section~\ref{sec:filtering}.
The stress predictions for three pure shear MD experiments in the positive and negative $\vec{e}_{1} \otimes \vec{e}_{2}$, $\vec{e}_{2} \otimes \vec{e}_{3}$, and $\vec{e}_{1} \otimes \vec{e}_{3}$ directions are shown in Fig.~\ref{fig:shear_tests} and Fig.~\ref{fig:shear_tests_PF}.
Finally, the stress predictions for three biaxial compression tests along the $x_{1}$ and $x_{2}$, $x_{2}$ and $x_{3}$, and $x_{1}$ and $x_{3}$ axes are shown in Fig.~\ref{fig:biaxial_compression} and Fig.~\ref{fig:biaxial_compression_PF}. 
It is noted that the stress fluctuations in the MD shear data appear to have a larger magnitude than those of the axial simulations.
However, the magnitude of the fluctuations of the stress components is similar across all simulations;
it appears to be larger in the shear simulations due to the smaller scale of the stress response.

Both models are able to accurately capture the shear behavior of $\beta$-HMX,
which differs greatly in the positive vs. negative directions as seen in Fig.~\ref{fig:shear_tests} and Fig.~\ref{fig:shear_tests_PF}.
The shear stress response of the material appears to be highly non-linear and exhibits directional dependence. 
This behavior is not expected to be captured by a material model with an invariant formulation,
as it requires specific treatment of the shear response along different directions to replicate the directional dependent behavior even qualitatively.
Here, however, a more general representation of the material using the full second-order stress and strain tensors allows for the neural network to automatically recover this behavior
and rather precisely.

\rmk{}
As seen in Fig.~\ref{fig:axial_tests_PF}, the predictions of the models $\mathcal{M}_{2}$ and $\mathcal{M}_{4}$ are 
very close. We have also examined the other predictions and the discrepancies of the stress predictions 
inferred from $\mathcal{M}_{2}$ and $\mathcal{M}_{4}$ are also very minor. Hence, we do not include those comparisons in the paper for brevity. 
In the following sections, the validation tests of the energy conjugate pair $\tensor{P}-\tensor{F}$ models' properties will be performed on the model $\mathcal{M}_{2}$ as the behavior of the models $\mathcal{M}_{2}$, $\mathcal{M}_{3}$, and $\mathcal{M}_{4}$ was observed to be similar.

\subsubsection{Validation of Strong ellipticity}

In this section, we perform the strong ellipticity tests as described in Section~\ref{sec:strong_ellipticity} on the trained neural network.
The neural network architecture used in this comparison is model $\mathcal{M}_{2}$,
which uses the energy conjugate pair $\tensor{P}-\tensor{F}$.
This model was chosen for the convenience of obtaining the fourth-order elasticity tensor needed for the acoustic tensor checks.

\begin{figure}[h]
\newcommand\siz{.32\textwidth}
\centering
\begin{tabular}{M{.33\textwidth}M{.33\textwidth}M{.33\textwidth}}
\hspace{-1.5cm}\includegraphics[width=.4\textwidth ,angle=0]{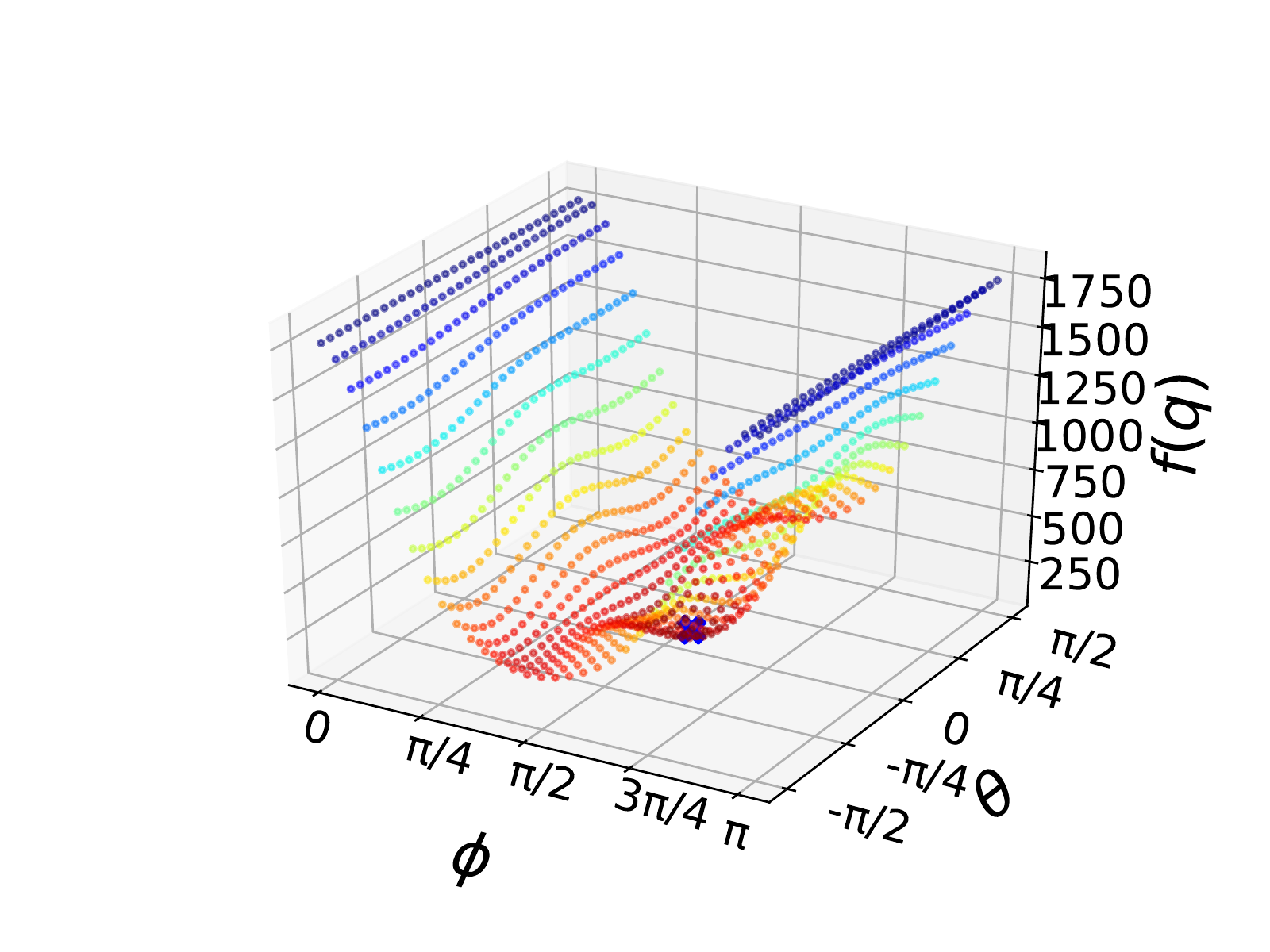} &
\hspace{-1.5cm}\includegraphics[width=.4\textwidth ,angle=0]{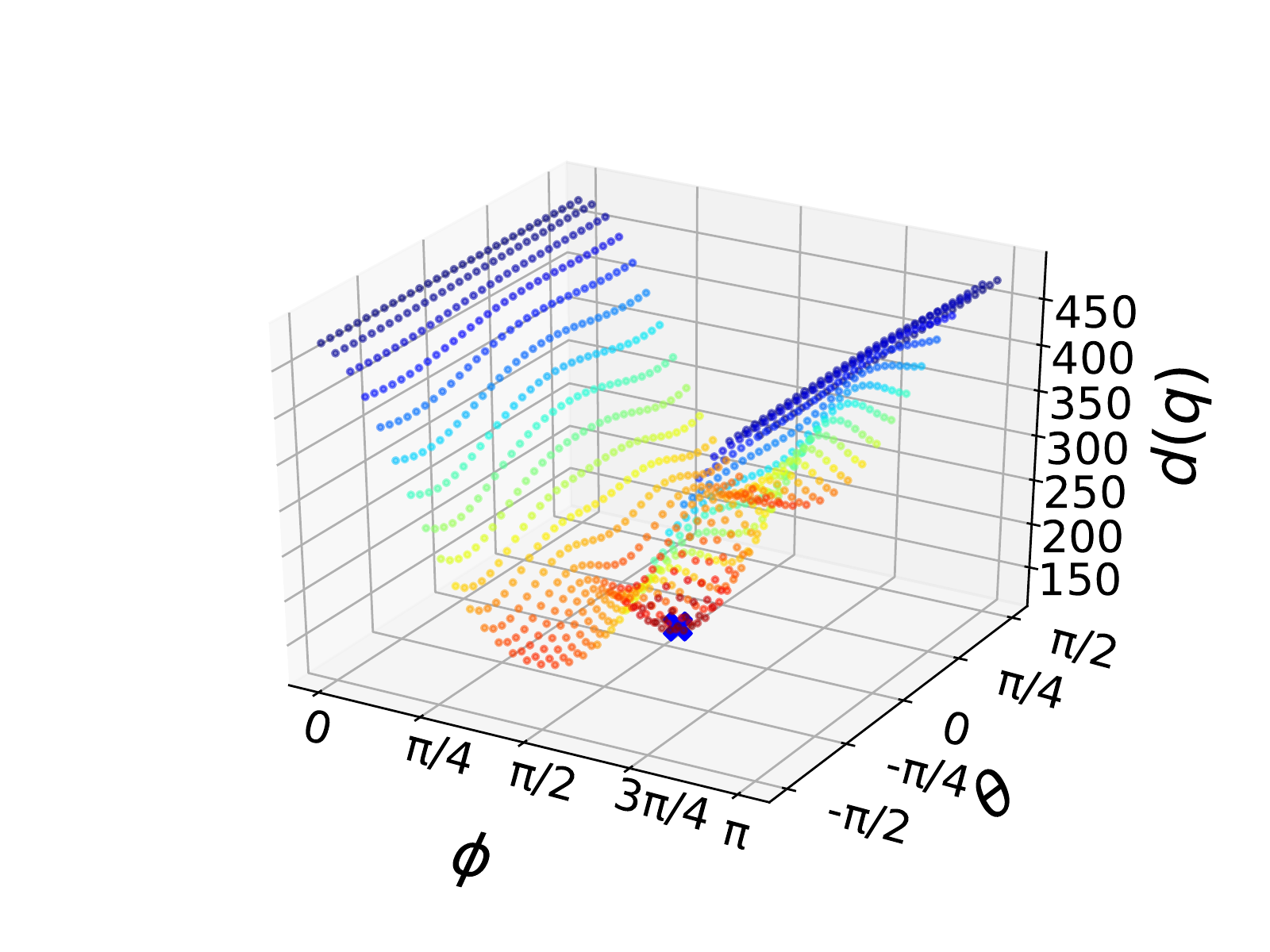} &
\hspace{-1.5cm}\includegraphics[width=.4\textwidth ,angle=0]{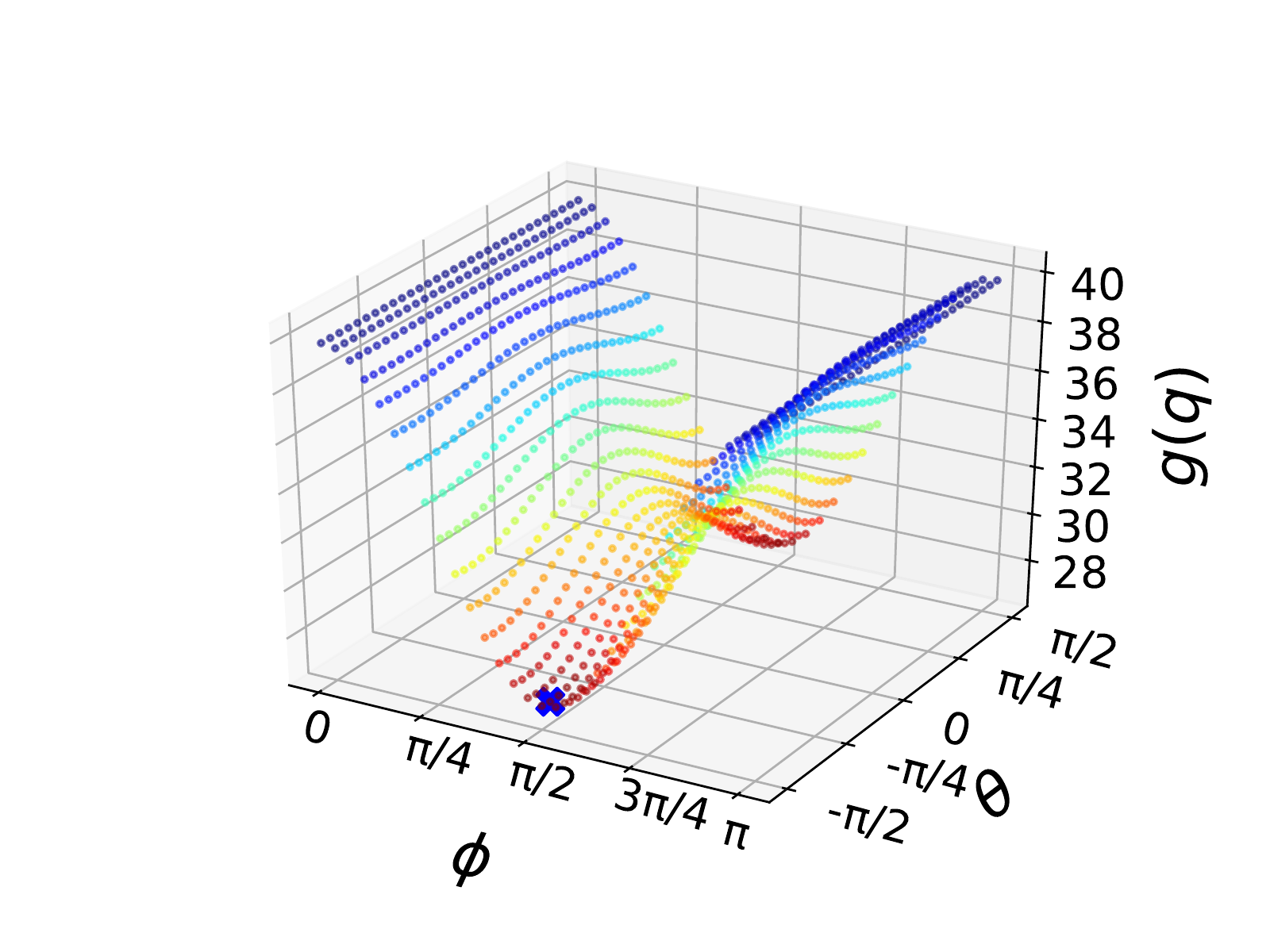} \\     

(a) & (b) & (c) \\

\hspace{-0.5cm}\includegraphics[width=.33\textwidth ,angle=0]{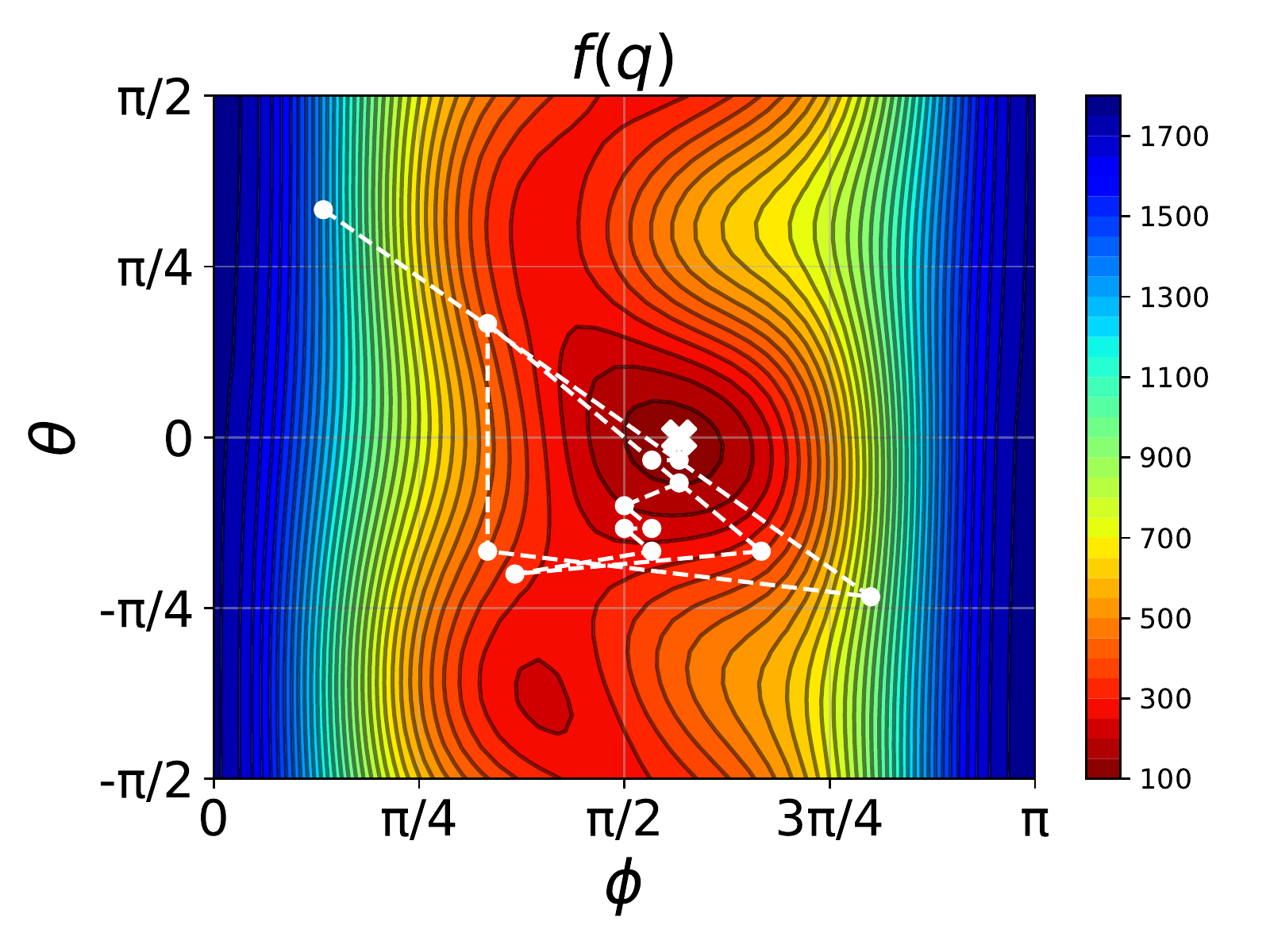} &
\hspace{-0.5cm}\includegraphics[width=.33\textwidth ,angle=0]{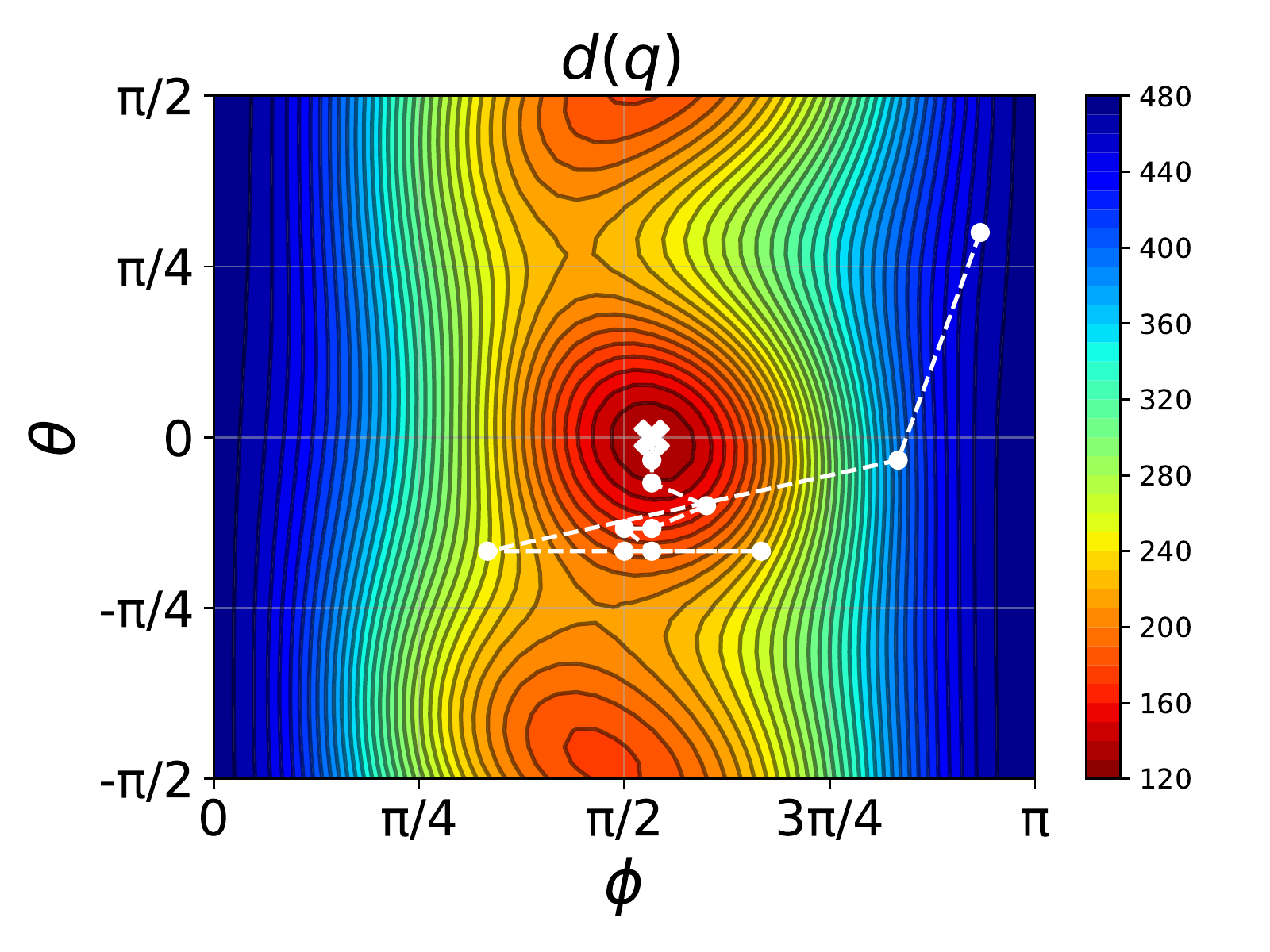} &
\hspace{-0.5cm}\includegraphics[width=.33\textwidth ,angle=0]{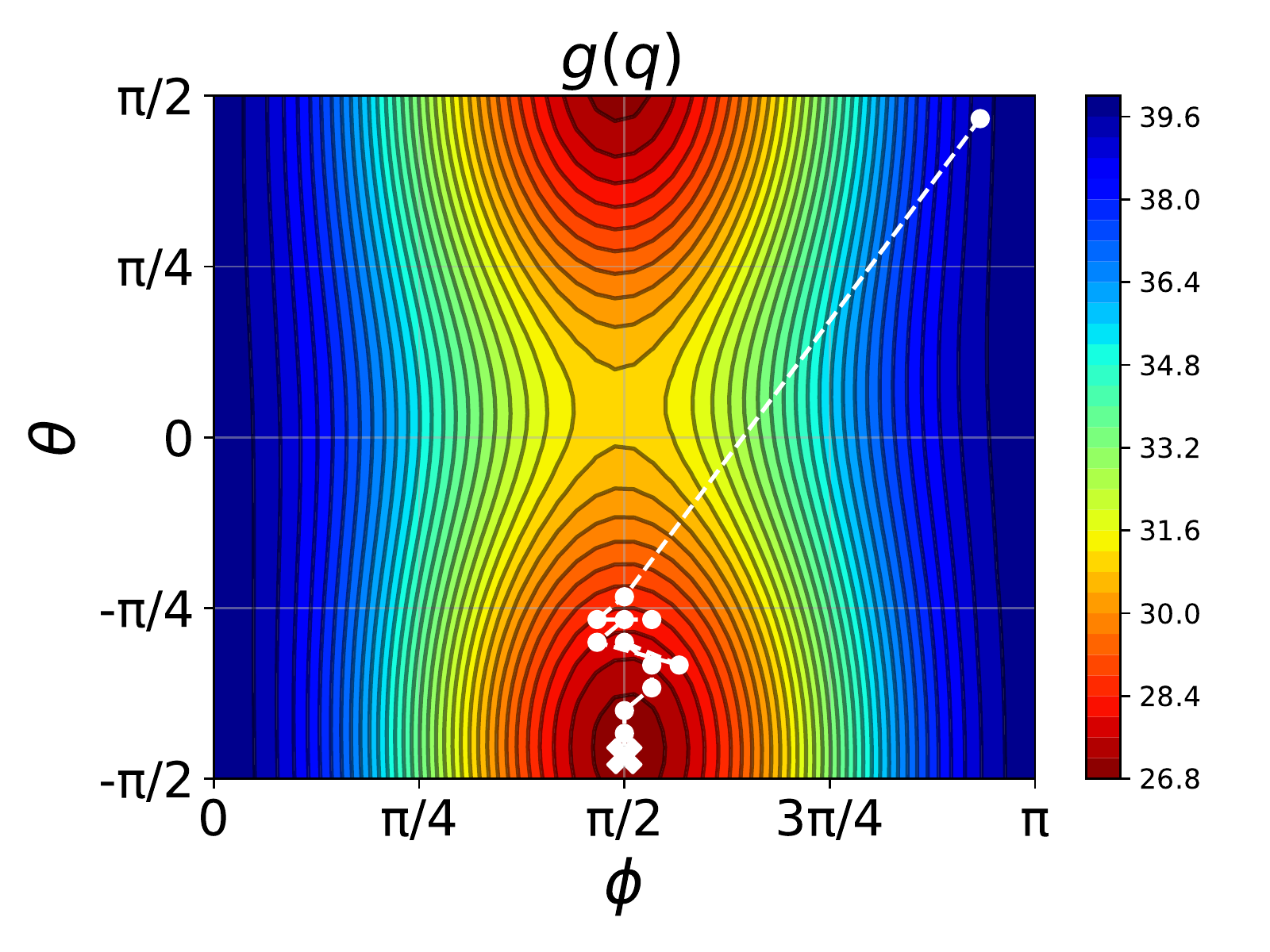} \\     

(d) & (e) & (f) \\

\end{tabular}

\caption{Validation of strong ellipticity conditions (a), (b), (c) with the criteria in Eq.~\eqref{eq:m1}, Eq.~\eqref{eq:m2}, and Eq.~\eqref{eq:m3}, respectively,
for an elasticity tensor close to the reference strain state.
The unit vectors were sampled from the surface of a unit sphere to perform the validation via a Hill Climbing gradient-free optimizer search (d, e, f).
The minimum value of the condition value discovered by the optimizer is marked.}
\label{fig:ellipticity}
\end{figure}

This check is performed by initially predicting the fourth-order elasticity tensor at a specific deformation gradient level.
We sample 1000 unit vectors $\vec{N}$ on the unit sphere $S^{2}$ in spherical coordinates by sampling the polar angle $\phi \in [0, \pi]$ and the azimuthal angle $\theta \in [0, \pi]$  
in a uniform grid, following Eq.~\eqref{eq:sample_N}. 
These vectors can be used to construct 1000 initial acoustic tensors following Eq.~\eqref{eq:acoustic_tensor}.
The acoustic tensors will be used as the initial grid landscape for a gradient-free optimizer set to discover the minimum values of the three strong ellipticity tests 
described in Eqs.~\eqref{eq:um1}, \eqref{eq:um2}, and \eqref{eq:um3}.
We use a Hill Climbing gradient-free optimizer search, using the library implemented by \citet{gfo2020},
to find the pair of $(\phi, \theta)$ that minimizes the strong ellipticity check values.
The Hill Climbing algorithm performs 10000 iterations of the search per test to discover the minimum value of the check,
which in most tests was obtained within the first 5000 iterations of the search.

\begin{figure}[h]
\newcommand\siz{.32\textwidth}
\centering
\begin{tabular}{M{.33\textwidth}M{.33\textwidth}M{.33\textwidth}}
\hspace{-1.5cm}\includegraphics[width=.4\textwidth ,angle=0]{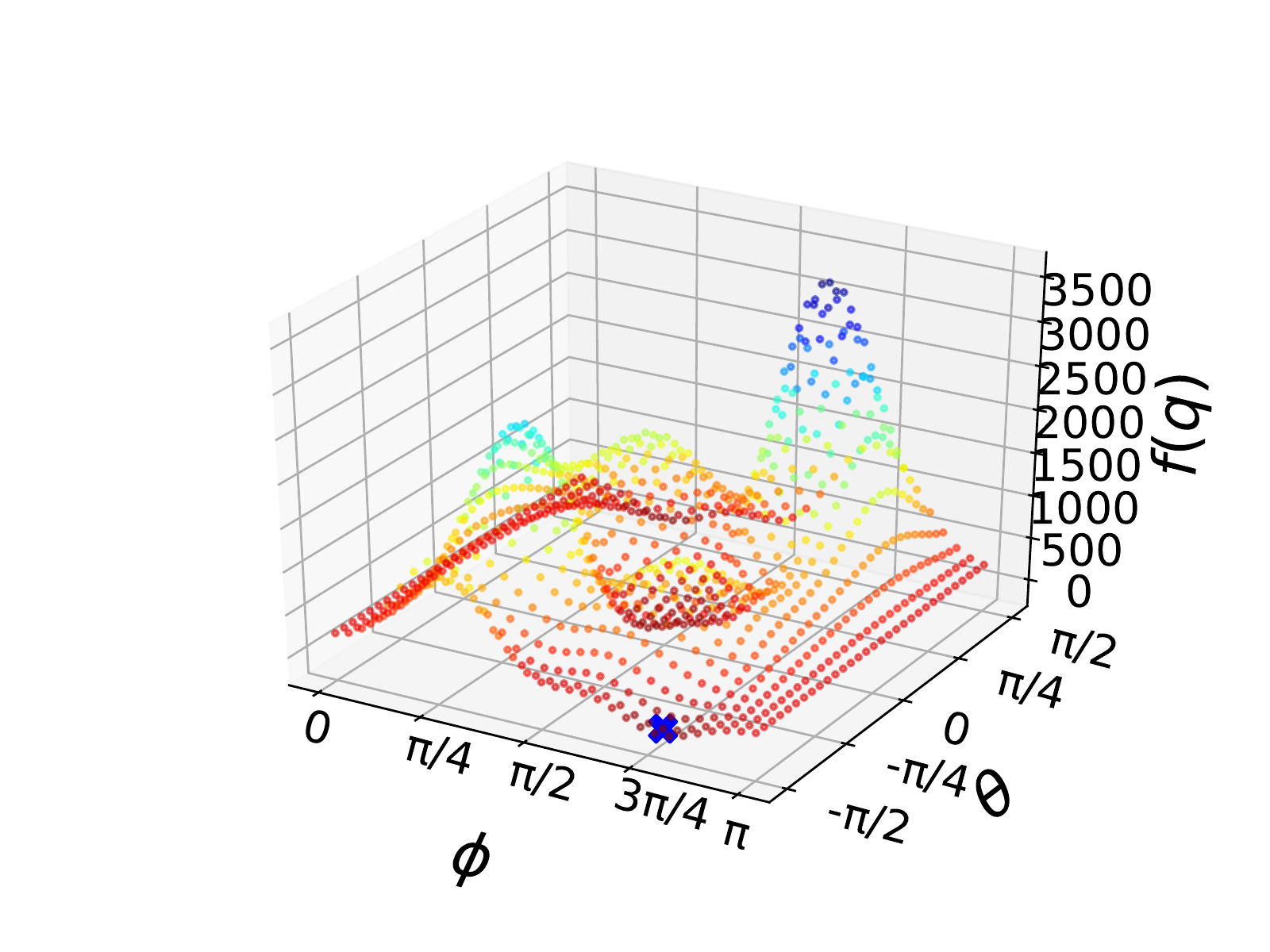} &
\hspace{-1.5cm}\includegraphics[width=.4\textwidth ,angle=0]{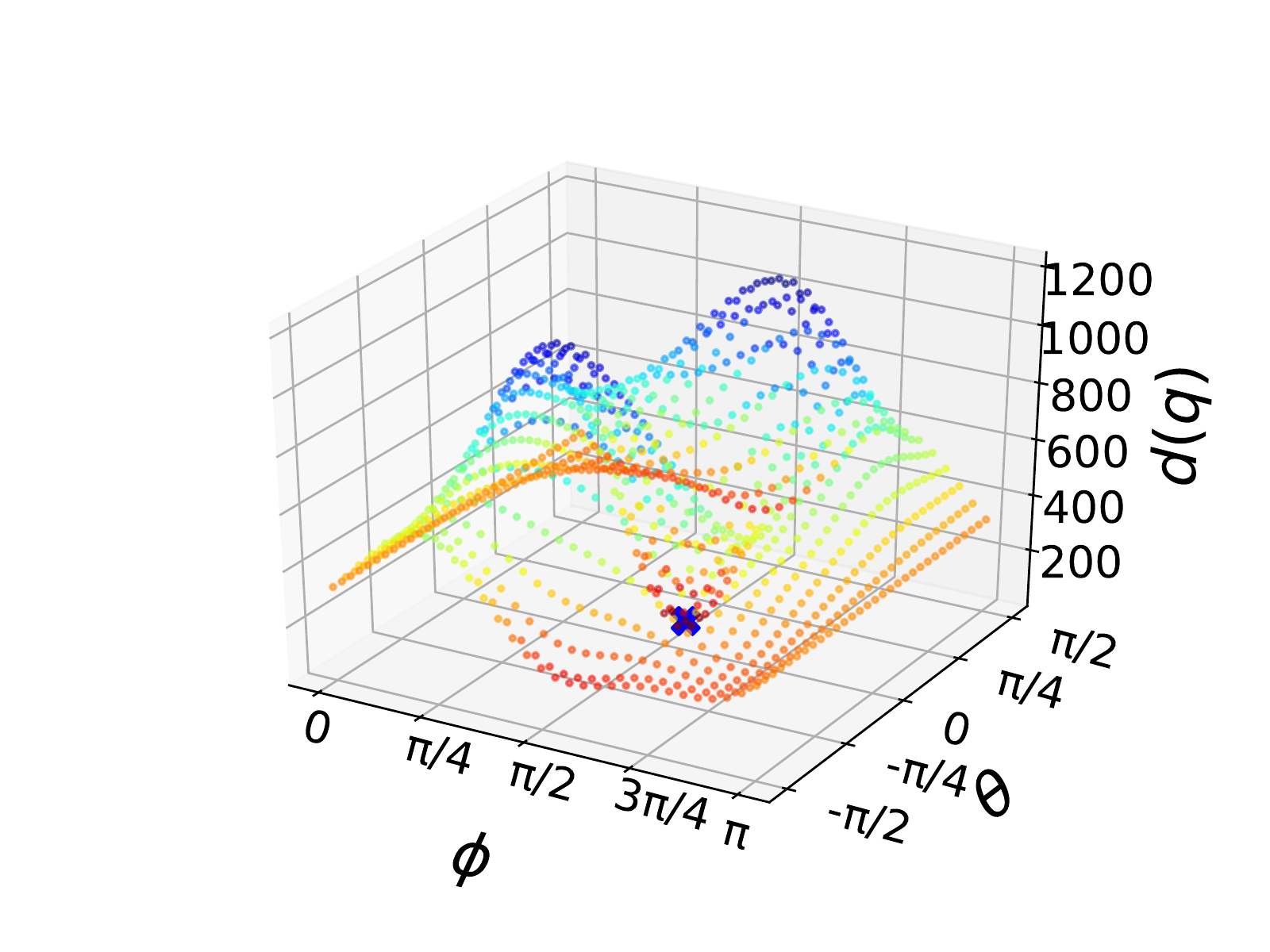} &
\hspace{-1.5cm}\includegraphics[width=.4\textwidth ,angle=0]{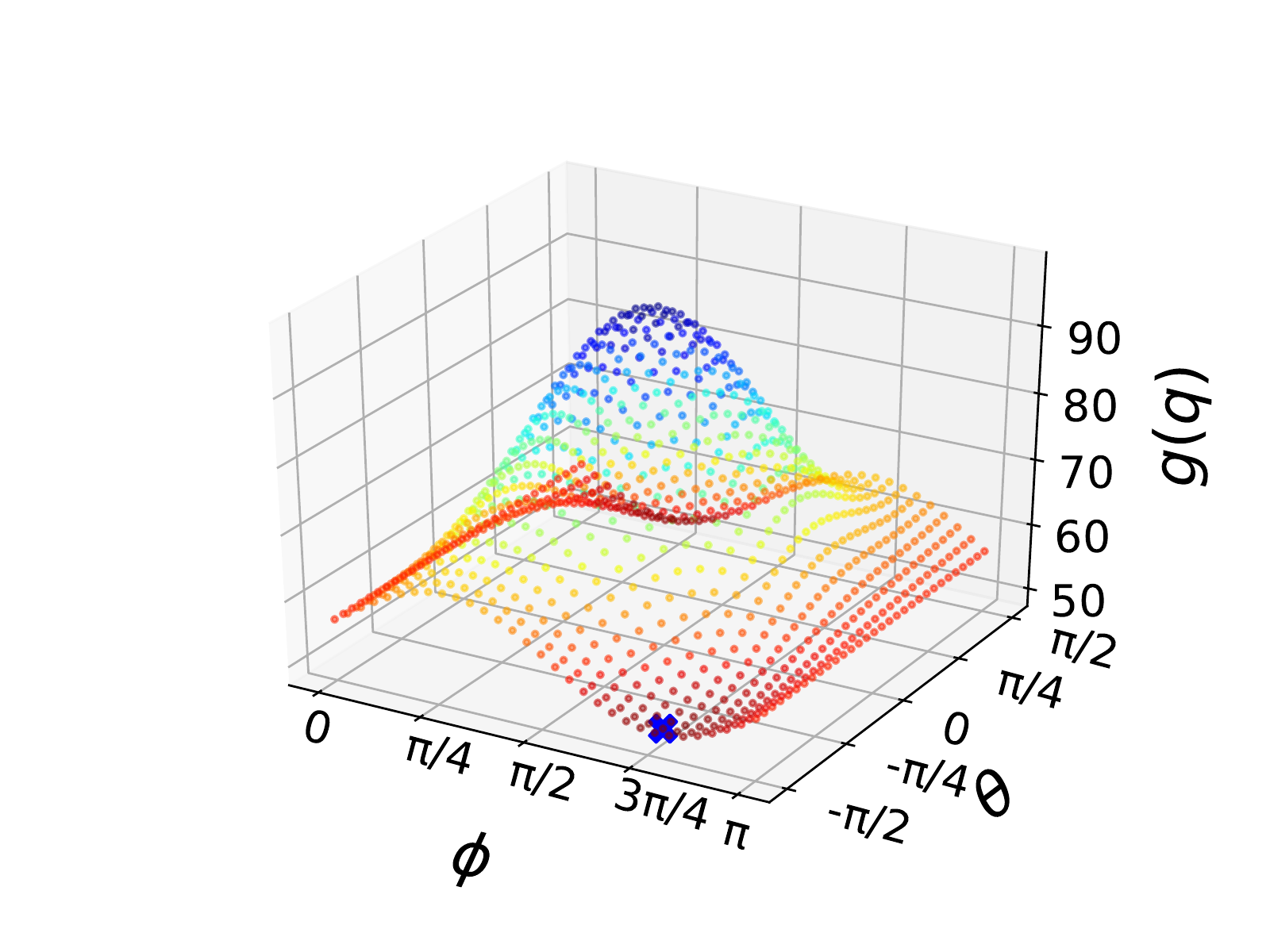} \\     

(a) & (b) & (c) \\

\hspace{-0.5cm}\includegraphics[width=.33\textwidth ,angle=0]{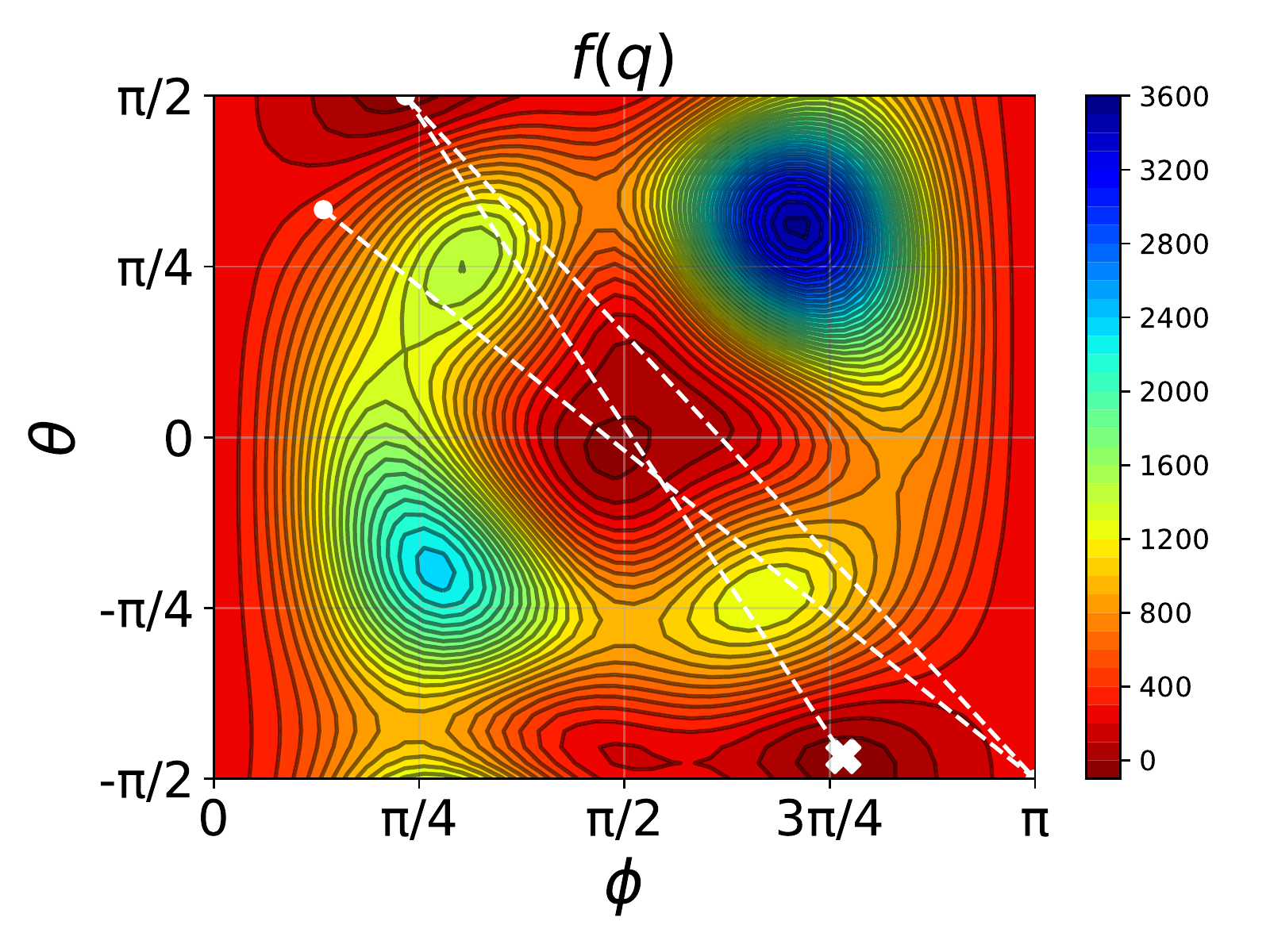} &
\hspace{-0.5cm}\includegraphics[width=.33\textwidth ,angle=0]{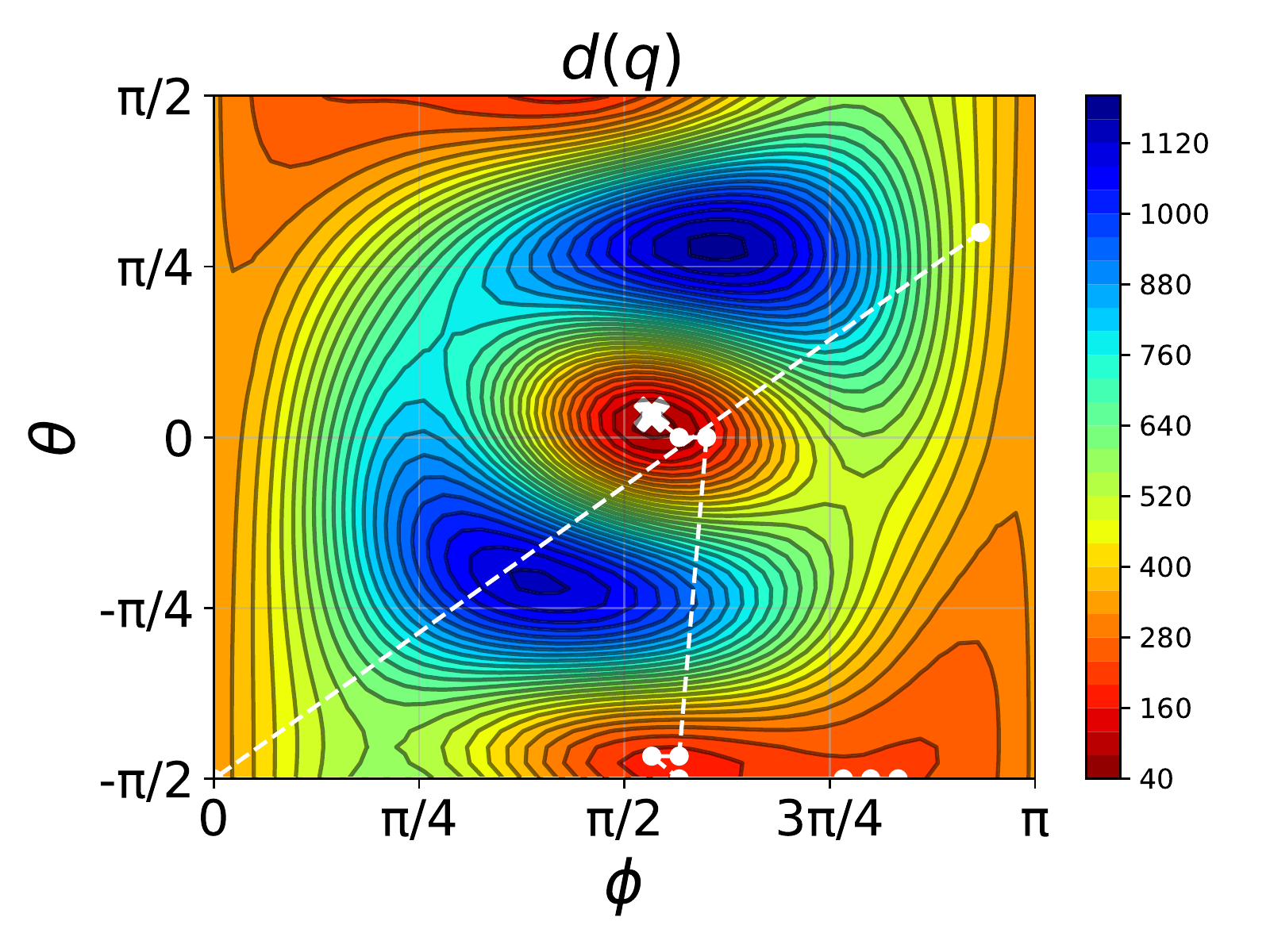} &
\hspace{-0.5cm}\includegraphics[width=.33\textwidth ,angle=0]{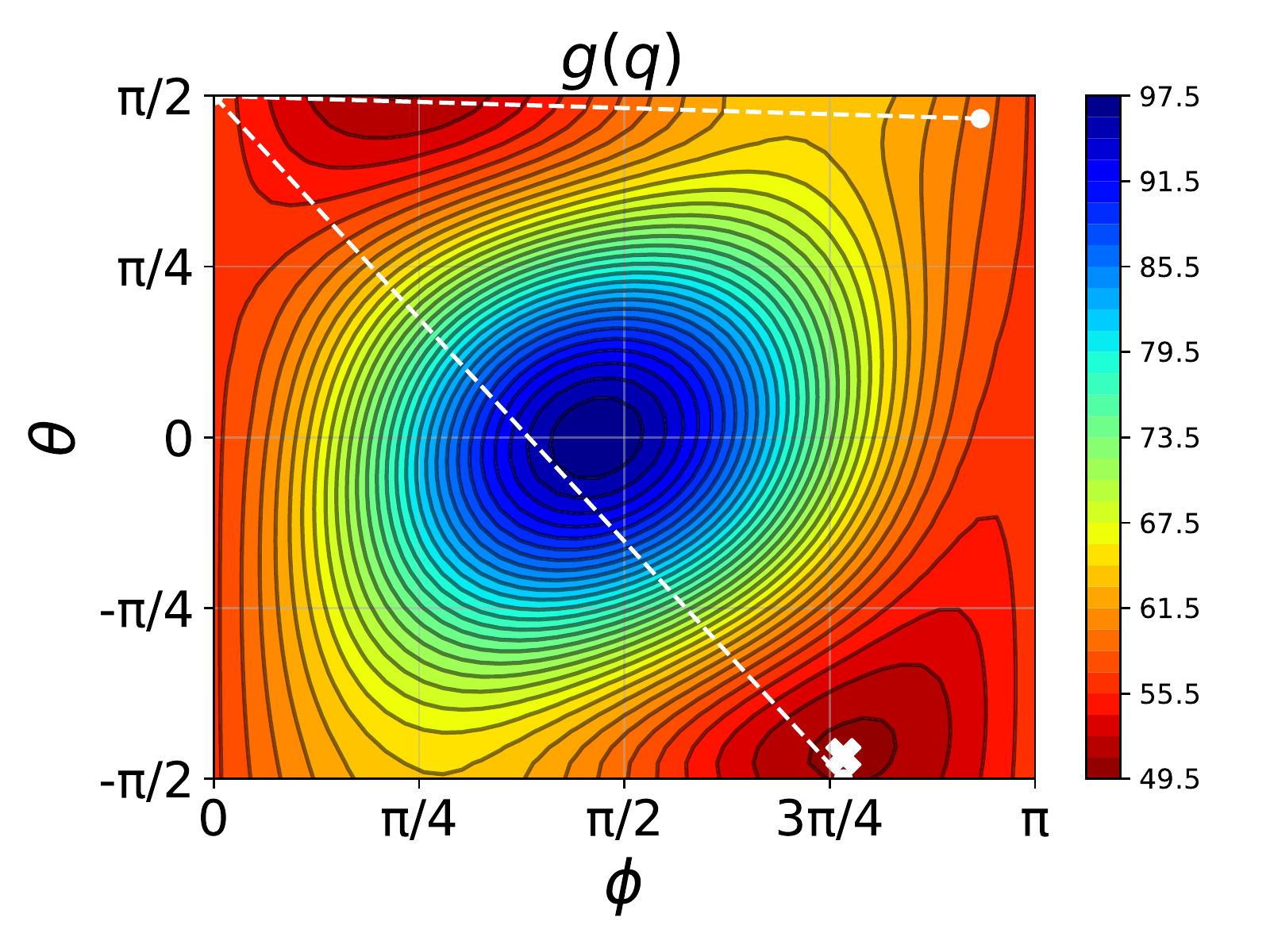} \\     

(d) & (e) & (f) \\

\end{tabular}

\caption{(a) Loss of the strong ellipticity condition in the prediction of a biaxial compression test along the $x_{1}$ and $x_{2}$ axes. The unit vectors were sampled from the surface of a unit sphere to perform the validation via a Hill Climbing gradient-free optimizer search (b). The minimum value of the condition value discovered by the optimizer is marked. }
\label{fig:ellipticity_first_zero}
\end{figure}

The predicted strong ellipticity test and the corresponding optimizer search for the minimum values
are demonstrated in Fig.~\ref{fig:ellipticity} and Fig.~\ref{fig:ellipticity_first_zero}
for two different elasticity tensors. 
In Fig.~\ref{fig:ellipticity}, we show the ellipticity test results for the elasticity tensor close to the relaxed reference state,
that is when the deformation gradient is the identity tensor. 
The neural network passes all three ellipticity tests, discovering the minimum of all tests to be greater than zero in the unit vector search space.
In Fig.~\ref{fig:ellipticity_first_zero}, we show the first strain state of a biaxial compression simulation along the $x_{1}$ and $x_{2}$ axes where the strong ellipticity test fails 
-- the acoustic tensor determinant for Eq.~\eqref{eq:um1} is found to be less than zero for the first time (compression of approximately $8\%$ along the $x_{1}$ and $x_{2}$ axes).

Given that the machine learning generated constitutive responses match very well with the filtered MD simulations (as shown in Figs. \ref{fig:axial_tests}-\ref{fig:biaxial_compression_PF}), the acoustic tensor losing positive definiteness is an indication of unstable elastic responses corresponding to the shear mode along the $\vec{N}$ direction
which could be potentially physical (C. Picu, personal communication, 2021). 

\subsubsection{Validation of energy growth for extrapolated predictions}

\begin{figure}[h]
\newcommand\siz{.32\textwidth}
\centering
\hspace{-1cm}\includegraphics[width=.45\textwidth ,angle=0]{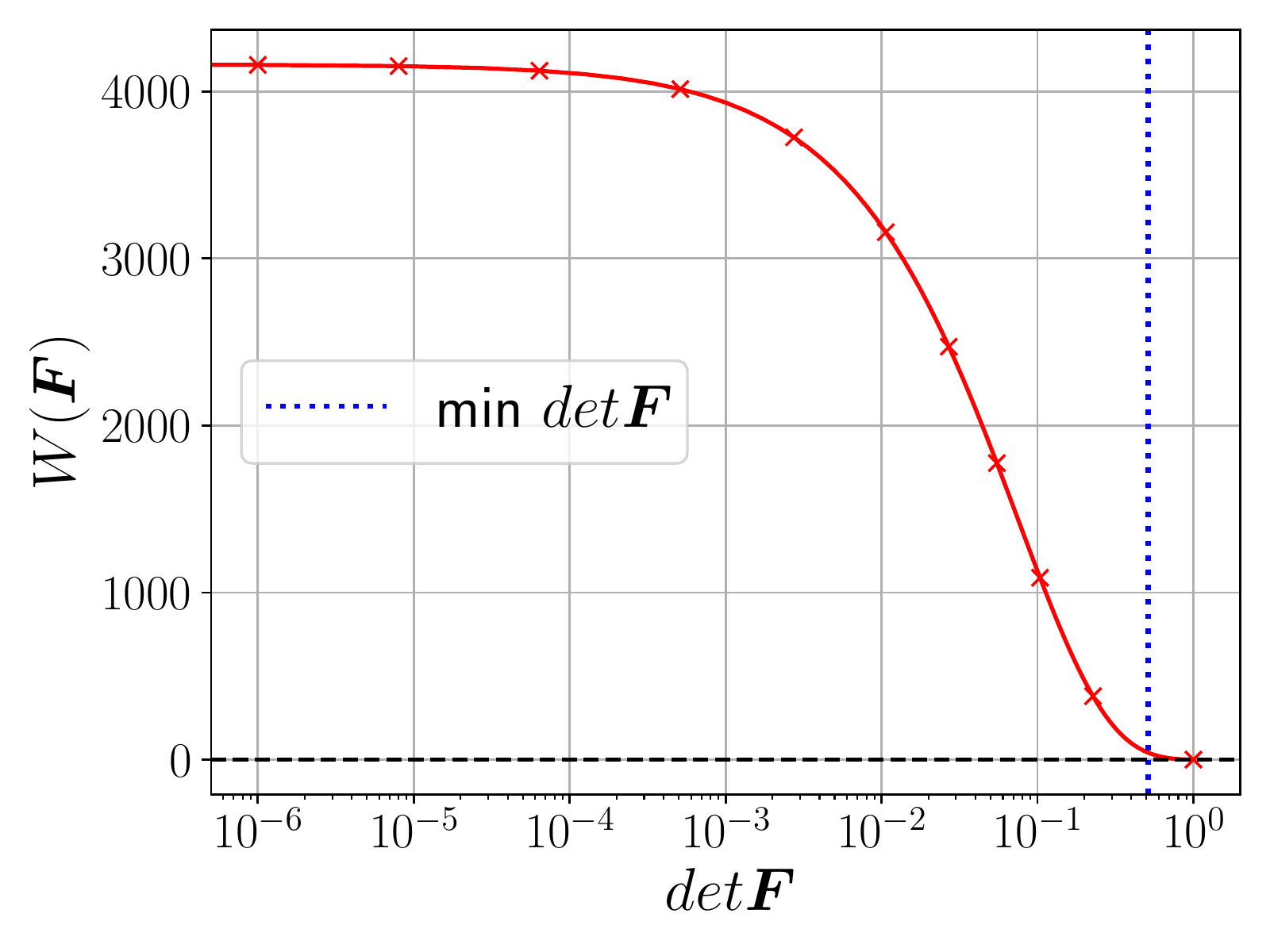} 
\caption{Results for the growth condition check, Eq. \eqref{eq:degenerated}. The predicted energy is monotonically increasing as $\det{\tensor{F}}$ approaches $0$. 
The minimum $\det{\tensor{F}}$ in the training data set is also marked.}
\label{fig:growth_condition}
\end{figure}

In this section, we perform the validation check described in Section~\ref{sec:convexity_growth} to monitor if the behavior of the predicted energy functional degenerates for very large deformations.
To test that, we impose deformation gradients on the neural network model spanning several orders of magnitude with the $\det \tensor{F}$ decreasing towards zero. 
The test is performed on the neural network architecture for the energy conjugate pair of $\tensor{P}-\tensor{F}$ (model $\mathcal{M}_2$).
As the Jacobian decreases, the energy functional values are expected to increase monotonically (Eq.~\eqref{eq:degenerated}).
Therefore, we apply a sequence of volumetric compression deformation gradients with the Jacobian approaching zero and
plot the energy against the increasing pure volumetric deformation.
The results are shown in Fig.~\ref{fig:growth_condition}. 

Note that the $\beta$-HMX may exhibit plastic yielding or damage under high pressure.
When this occurs, it is not physically feasible to have an elastic response.
Yet the continuum mechanics theory validation does require that the stored energy approach infinity
as a finite volume of HMX crystal collapses into a point \citep{rosakis1994relation}.
This does not happen in our trained neural network model even though the growth rate within the training data interval seems reasonable. 

A similar extrapolation issue has been investigated previously in \citet{versino2017data} in which the symbolic regression requires an additional artificial data point added in order to prevent an incorrect prediction of softening.
Presumably, a similar treatment can also be applied either by adding a very large artificial data point with a very large energy at the supposedly singular point or by rigorously 
enforcing the singularity in the learned energy functional.
At this point, robust ways to introduce singular data into the neural network and the formulation of the loss function are not clear,
but we intend to examine it in future studies. 
Nevertheless, the results do reveal that the energy functional trained by the neural network may only be valid within the interval of the data and that any extrapolated results outside of the data interval must be used with caution, even if a significant number of physical constraints (e.g. material symmetry) have already been applied as auxiliary objectives for the supervised learning. 

\subsubsection{Stress-dependent anisotropy of HMX crystal}

\begin{figure}[h!]
\newcommand\siz{.32\textwidth}
\centering
\begin{tabular}{M{.33\textwidth}M{.33\textwidth}M{.33\textwidth}}
\hspace{-1cm}\includegraphics[width=.33\textwidth ,angle=0]{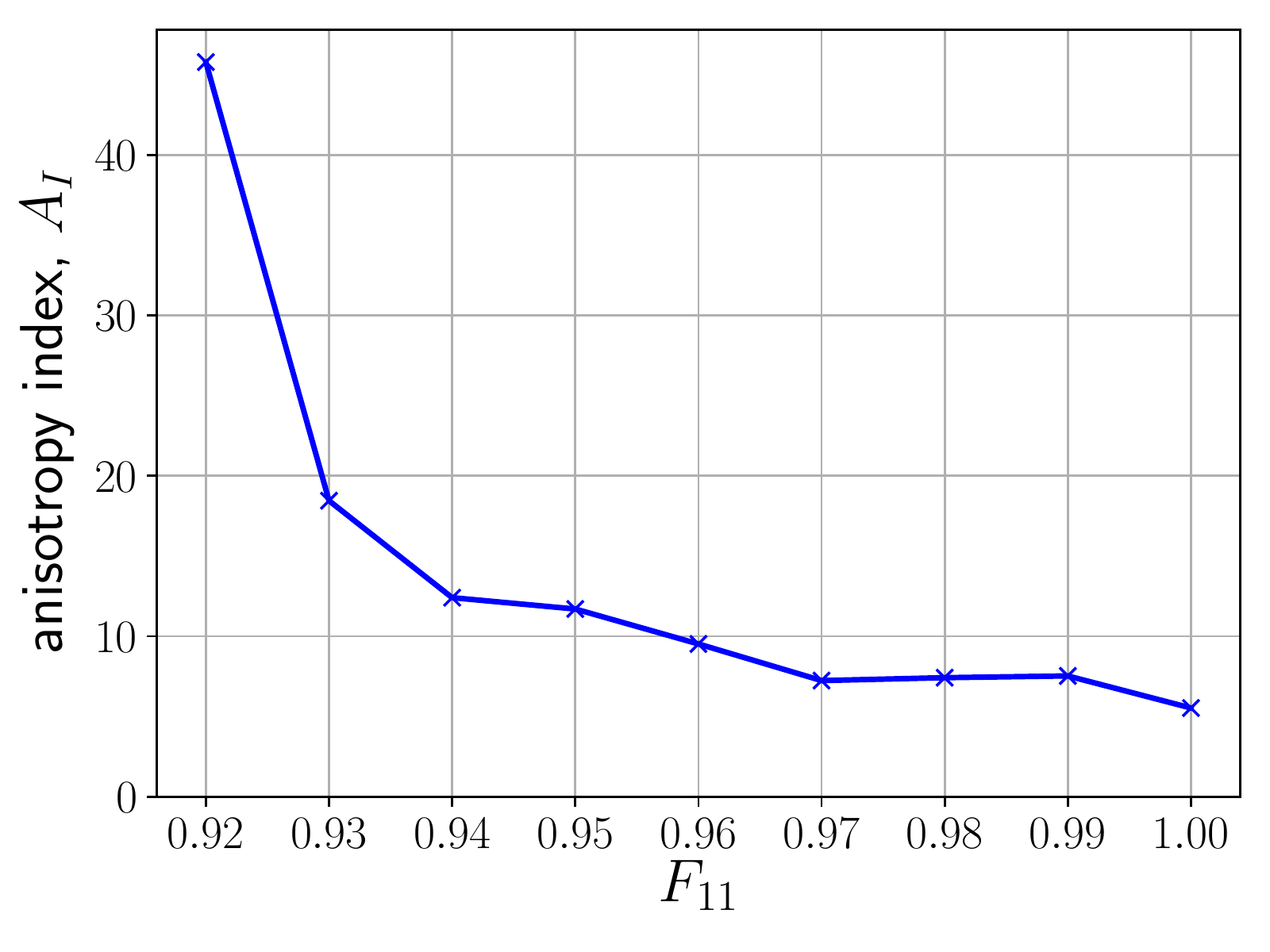} &
\hspace{-1cm}\includegraphics[width=.33\textwidth ,angle=0]{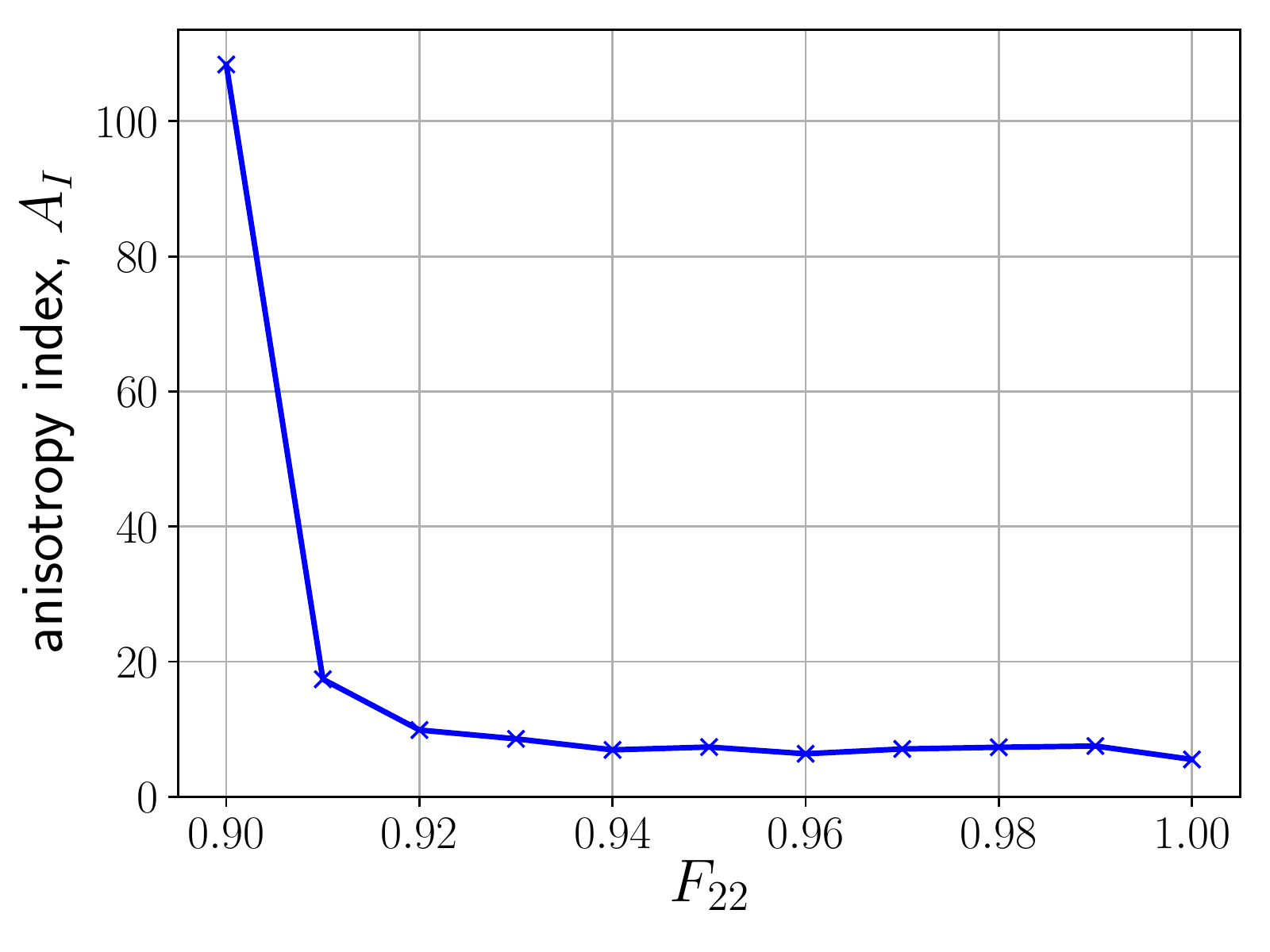} &
\hspace{-1cm}\includegraphics[width=.33\textwidth ,angle=0]{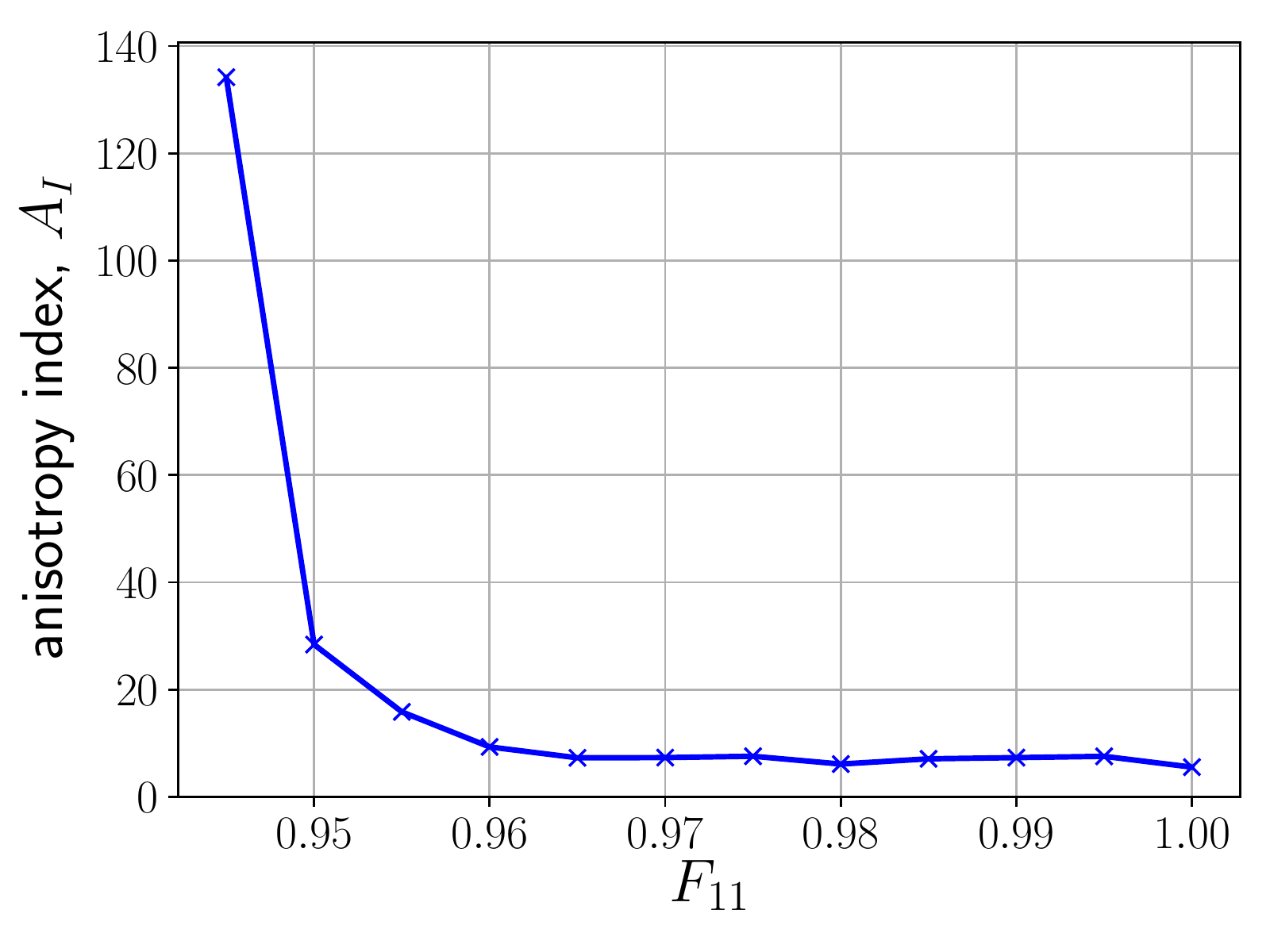} \\     

(a) & (b) & (c)\\
\end{tabular}
\caption{Anisotropy index $A_{I}$ calculated for the energy conjugate pair $\tensor{P}-\tensor{F}$ model ($\mathcal{M}_{2}$) for (a) a uniaxial compression test along the $x_{1}$ axis, (b) a uniaxial compression test along the $x_{2}$ axis, and (c) a biaxial compression test along the $x_{1}$ and $x_{2}$ axes.}
\label{fig:anisotropic_index}
\end{figure}

In this section, we recover the predicted material-response anisotropy index as described in Section~\ref{sec:anisotropic_index} to monitor the evolution of material anisotropy.
The anisotropy index is acquired for the neural network architecture of the energy conjugate pair of $\tensor{P}-\tensor{F}$ (model $\mathcal{M}_2$).
To obtain the index, the fourth-order elasticity tensor is predicted at different deformation gradients along a prescribed loading path.
We then sample 1000 unit vectors $\vec{N}$ per deformation gradient in a uniform grid, following Eq.~\eqref{eq:sample_N},
by sampling the polar angle $\phi \in [0, \pi]$ and the azimuthal angle $\theta \in [0, \pi]$. 
For each elasticity tensor, 1000 initial acoustic tensors are constructed according to Eq.~\eqref{eq:acoustic_tensor}. 
The Hill Climbing algorithm performs 10000 iterations of the search deformation gradient sample to discover the minimum $v_{1}^2$ and the maximum $v_{2}^2$ values of each acoustic tensor.
The anisotropy index $A_{I}$ is then calculated using Eq.~\eqref{eq:anisotropic_index}.
The anisotropy index calculated for three loading paths is demonstrated in Fig.~\ref{fig:anisotropic_index}.

Interestingly, for all three uniaxial compression cases, the Ledbetter-Migliori anisotropy index,
which is the ratio of the fastest and slowest shear wave speeds of the $\beta$-HMX crystal, all tend to increase significantly.
The most significant changes occur when the uniaxial deformation is more than 8\%.
In all three cases, the elastic anisotropy is not very profound when the deformation is small.
However, in all three cases, the anisotropy index jumps from less than 10 to more than 40 in the deformation
along the $\vec{x}_{1}$ direction and more than two orders in the $\vec{x}_{2}$ and $\vec{x}_{3}$  direction.
These results signify the importance of capturing the evolving anisotropy 
of the HMX materials.

\subsubsection{Comparisons with literature calculations on elasticity}

We now provide the coefficients of the elastic tangents for the $\tensor{S}-\tensor{E}$ and $\tensor{P}-\tensor{F}$ 
conjugated pairs obtained from the trained neural network energy functionals, models $\mathcal{M}_{1}$ and $\mathcal{M}_{2}$ and compare them with the elasticity tangent for the 
 $\tensor{\sigma}-\tensor{\epsilon}$ conjugated pairs previously 
 reported by \citet{pereverzev2020elastic}. 

In the present work, the strain measure is obtained differently in the sense that the models in these papers
introduce only one reference configuration such that $\tensor{F}=\tensor{I}$
when the Cauchy pressure is at $10^{-4}$ GPa. Meanwhile, the strain measure in the $\beta$-HMX in \citep{pereverzev2020elastic} is reset at different reference pressure where the MD simulation begins. This difference is minor for the atmospheric pressure case,
for which the geometrical nonlinearity is insignificant, but may lead to significant differences in the values of the elastic tangent coefficients 
for high-pressure cases. Note that, due to the anisotropic nature of the elastic responses, the imposed Cauchy pressure may also lead to isochoric deformation due to volumetric-deviatoric coupling. As such, the coordinates of the reference and current configurations $\vec{x}_{i}$ and $\vec{X}_{I}$ are not necessarily co-axial.
Hence, a direct comparison of the values of the coefficient is not productive. 

Furthermore, discrepancies may also be caused by the different data de-noising processes employed in \citet{pereverzev2020elastic}.
In this paper, we employ a de-noising algorithm to filter out the high-frequent oscillation in the constitutive responses before 
the supervised learning is conducted whereas \citet{pereverzev2020elastic} employs a finite-difference approximation with a sufficiently 
large strain increment to calculate the elasticity tensor. 

Nevertheless, a comparison of elasticity tangent operators from previous MD simulations, as well as those obtained for different conjugate 
stress-strain pairs, does indicate the significance of geometric nonlinearity in the material responses and the importance of taking it into consideration in numerical simulations. 

In the MD simulation, 
the crystal cell of $\beta$-HMX is first equilibrated at a target temperature and pressure through an isochoric-isothermal (NVT) simulation 
and then strains are imposed at different directions to obtain the stress information for the differentiation. 
The comparison between the predicted and literature reported elastic coefficients at 300 K temperature at pressure $10^{-4}$ GPa and $5$ GPa is demonstrated in Tables~\ref{tab:elast_coeff} and \ref{tab:elast_coeff_5gpa}, respectively, for the neural network models $\mathcal{M}_1$ and $\mathcal{M}_2$ as described in Section~\ref{sec:network_training}. 
It is noted that for the energy conjugate pair of $\tensor{P}-\tensor{F}$ Model $\mathcal{M}_2$, the full tangent requires  a $(9 \times 9)$ matrix to represent it in the Voigt notation. 

\begin{table}[h]
	\begin{center}
	\caption{Comparison of the predicted $\beta$-HMX elastic coefficients (GPa) at pressure $10^{-4}$ GPa for the energy conjugate pair $\tensor{S}-\tensor{E}$ (Model  $\mathcal{M}_{1}$)  and  the energy conjugate pair $\tensor{P}-\tensor{F}$ (Model  $\mathcal{M}_{2}$) to the ones reported by \citet{pereverzev2020elastic} for
	(\SI{300}{\kelvin}, \SI{e-4}{\giga\pascal}). Note that $\mathbb{C}^{\tensor{P} - \tensor{F}}$ is not symmetric but the additional terms are not shown for brevity.}
	\label{tab:elast_coeff}
\begin{tabular}{ccccc}
\hline\hline
         &            & Model $\mathcal{M}_{1}$                & Model $\mathcal{M}_{2}$                & \citet{pereverzev2020elastic}                      \\
$D_{ij}$ & $C_{ijkl}$ & $\mathbb{C}^{\tensor{S} - \tensor{E}}$ & $\mathbb{C}^{\tensor{P} - \tensor{F}}$ & $\mathbb{C}^{\tensor{\sigma} - \tensor{\epsilon}}$ \\ [0.5ex]
\hline
$D_{11}$ & $C_{1111}$ & 21.354                                 & 25.861                                 & 22.97                                              \\
$D_{22}$ & $C_{2222}$ & 22.149                                 & 18.092                                 & 22.62                                              \\
$D_{33}$ & $C_{3333}$ & 21.314                                 & 21.627                                 & 21.67                                              \\
$D_{44}$ & $C_{1212}$ & 8.616                                  & 5.8335                                 & 8.645                                              \\
$D_{55}$ & $C_{2323}$ & 10.982                                 & 8.225                                  & 10.407                                             \\
$D_{66}$ & $C_{1313}$ & 9.497                                  & 10.078                                 & 9.527                                              \\
$D_{12}$ & $C_{1122}$ & 8.789                                  & 6.898                                  & 9.2                                                \\
$D_{13}$ & $C_{1133}$ & 12.348                                 & 12.7828                                & 12.32                                              \\
$D_{23}$ & $C_{2233}$ & 15.913                                 & 13.375                                 & 12.37                                              \\
$D_{15}$ & $C_{1123}$ & -0.998                                 & -0.584                                 & -0.43                                              \\
$D_{25}$ & $C_{2223}$ & 4.247                                  & -0.877                                 & 4.47                                               \\
$D_{35}$ & $C_{3323}$ & 2.192                                  & -0.792                                 & 1.84                                               \\
$D_{46}$ & $C_{1213}$ & 2.484                                  & 1.571                                  & 2.248                            \\
\hline\hline                 
\end{tabular}
	\end{center}
\end{table}

Table~\ref{tab:elast_coeff} shows the results of different elastic tangents obtained from the neural network calculation and those obtained from \citet{pereverzev2020elastic}. While there are differences among the three tangents, they are relatively minor. This is expected as the geometrical nonlinearity is not significant. 

\begin{table}[h]
	\begin{center}
	\caption{Comparison of the predicted $\beta$-HMX elastic coefficients (GPa) at pressure $5$ GPa for the energy conjugate pair $\tensor{S}-\tensor{E}$ (model  $\mathcal{M}_{1}$)  and  the energy conjugate pair $\tensor{P}-\tensor{F}$ (model  $\mathcal{M}_{2}$) with the ones reported in \citet{pereverzev2020elastic}. Note that $\mathbb{C}^{\tensor{P} - \tensor{F}}$ is not symmetric but the additional terms are not shown for brevity.}
	\label{tab:elast_coeff_5gpa}
\begin{tabular}{ccccc}
\hline\hline
         &            & Model $\mathcal{M}_{1}$                & Model $\mathcal{M}_{2}$                & \citet{pereverzev2020elastic}                      \\
$D_{ij}$ & $C_{ijkl}$ & $\mathbb{C}^{\tensor{S} - \tensor{E}}$ & $\mathbb{C}^{\tensor{P} - \tensor{F}}$ & $\mathbb{C}^{\tensor{\sigma} - \tensor{\epsilon}}$ \\ [0.5ex]
\hline
$D_{11}$ & $C_{1111}$ & 80.157                                 & 87.556                                 & 87.71                                              \\
$D_{22}$ & $C_{2222}$ & 68.666                                 & 53.441                                 & 67.08                                              \\
$D_{33}$ & $C_{3333}$ & 71.453                                 & 72.011                                 & 62.11                                              \\
$D_{44}$ & $C_{1212}$ & 0.358                                  & 0.033                                  & 19.461                                             \\
$D_{55}$ & $C_{2323}$ & 3.71                                   & 2.813                                  & 34.08                                              \\
$D_{66}$ & $C_{1313}$ & -2.999                                 & 1.736                                  & 19.662                                             \\
$D_{12}$ & $C_{1122}$ & 44.048                                 & 26.828                                 & 36.93                                              \\
$D_{13}$ & $C_{1133}$ & 46.187                                 & 32.423                                 & 52.95                                              \\
$D_{23}$ & $C_{2233}$ & 55.267                                 & 52.603                                 & 46.49                                              \\
$D_{15}$ & $C_{1123}$ & -0.939                                 & 0.639                                  & -11.32                                             \\
$D_{25}$ & $C_{2223}$ & 10.358                                 & -6.108                                 & 11.1                                               \\
$D_{35}$ & $C_{3323}$ & -0.421                                 & -6.120                                 & 2.48                                               \\
$D_{46}$ & $C_{1213}$ & 5.546                                  & -4.082                                 & 6.06                                              \\
\hline\hline
\end{tabular}
	\end{center}
\end{table}

Table~\ref{tab:elast_coeff_5gpa}, on the other hand, shows a more significant difference in the numerical values of the coefficients  
for different energy-conjugated pairs. This is consistent with the derivation in Section \ref{sec:mapping} where 
the deformation gradient at this point is no longer infinitesimal and the incorporation of the geometrical nonlinearity is necessary to capture the elastic constitutive responses properly.

\section{Conclusions}
\label{sec:conclusion}

This paper introduce a mechanistic machine learning framework to infer anisotropic hyperelasticity 
energy functional from l from molecular dynamic simulations for $\beta$-$HMX$. 
Conventionally, machine learning constitutive laws are often formulated to match experimental data. 
As such, the discrepancy between experimental data and the predictions is often the only term
in the loss function for training and validation. 
Here we attempt to formulate the training of hyperelastic model not only to mininizing 
the discrepancy of data but also introduce additional objectives to ensure that 
the learned hyperelastic model obey the physics constraints. 
To ensure the robustness of the predictions, we also introduce a set of validation tests to examine 
the admissibility (e.g. preserving material symmetry, obeying growth conditions) and stability (convexity, strong ellipticity) of the constitutive responses generated from the trained neural networks. 
With the usage of Soblev training and automatic differentiation to facilitate the training of constitutive laws, 
the resultant model exhibit highly accurate predictions within the training data range. 
These treatments are shown to be effective in improving the accuracy and robustness of the predictions, while 
the theoretical validation may 
provide the much-needed post hoc interpretability of the neural network constitutive laws to understand 
the properties of the machine learning models. More importantly, the validation exercise may provide a 
reliable way to reveal the weakness of the models and safeguard against cherry-picking interpretation, which 
could be a key ingredient to make black-box neural network predictions more trustworthy.

\section{Data availability statements}
The code used to conduct the validation tests will be available in a Github repository upon publication of this manuscript. The datasets generated and/or analyzed during the current study are available from the authors upon reasonable request.

\section{Acknowledgements}
Fruitful discussions with Andrey Pereverzev and Bahador Bahmani are gratefully acknowledged. 
The efforts and labor hours are primarily supported by he Air Force Office of Scientific Research under grant contracts FA9550-19-1-0318, with additional support provided to WCS and NNV from the 
the NSF CAREER grant at National Science Foundation under grant contracts CMMI-1846875 and OAC-1940203.

\bibliographystyle{plainnat}
\bibliography{main}
\end{document}